\documentclass[fleqn,usenatbib]{mnras}

\usepackage[T1]{fontenc}
\usepackage{ae,aecompl}

\usepackage{subfig}
\usepackage{tikz}
\usepackage{standalone}

\usepackage{graphicx}	% Including figure files
\usepackage{amsmath}	% Advanced maths commands
\usepackage{amssymb}	% Extra maths symbols
\usepackage{booktabs}

\usepackage{mathtools}
\DeclarePairedDelimiter{\floor}{\lfloor}{\rfloor}

\usepackage{amsmath}
\usepackage{empheq}
\newcommand*\widebox[1]{\fbox{\hspace{1em}#1\hspace{1em}}}

\newcommand{\Mpch}{h^{-1}\mathrm{Mpc}}
\newcommand{\hMpc}{h\,\mathrm{Mpc}^{-1}}

\newcommand{\av}[1]{\left\langle{#1}\right\rangle}

\newcommand{\vk}{\vec k}

\newcommand{\vd}{\vec d}
\newcommand{\hd}{\hat{\vec d}}
\newcommand{\vx}{\vec x}
\newcommand{\vy}{\vec y}
\newcommand{\vr}{\vec r}

\newcommand{\ft}[1]{\mathcal{F}\left[{#1}\right]}
\newcommand{\ift}[1]{\mathcal{F}^{-1}\left[{#1}\right]}

\definecolor{darkgreen}{RGB}{0,120,0}

\definecolor{brown}{RGB}{120,60,0}
\newcommand{\resub}[1]{#1}%\textcolor{darkgreen}{#1}}

\newcommand{\hk}{\hat{\vec{k}}}
\newcommand{\hr}{\hat{\vec{r}}}
\newcommand{\hR}{\hat{\vec{R}}}
\newcommand{\hD}{\hat{\vec{\Delta}}}
\newcommand{\vD}{\vec{\Delta}}
\newcommand{\vR}{\vec{R}}
\newcommand{\vA}{\vec{A}}
\newcommand{\va}{\vec{a}}
\newcommand{\hA}{\hat{\vec{A}}}
\newcommand{\ha}{\hat{\vec{a}}}
\newcommand{\hn}{\hat{\vec n}}
\newcommand{\hx}{\hat{\vec x}}
\newcommand{\hy}{\hat{\vec y}}

\newcommand{\vz}{\vec{z}}

\newcommand{\tjo}[3]{\begin{pmatrix} {#1} & {#2} & {#3}\\ 0 & 0 & 0\end{pmatrix}}
\newcommand{\tj}[6]{\begin{pmatrix} {#1} & {#2} & {#3}\\ {#4} & {#5} & {#6}\end{pmatrix}}
\renewcommand{\max}{{\mathrm{max}}}

\def\beq{\begin{eqnarray}}
\def\eeq{\end{eqnarray}}

\numberwithin{equation}{section}

\usepackage{newtxtext,newtxmath,gensymb}

\let\vec\mathbf
\title[Beyond Yamamoto Correlators]{Beyond Yamamoto: Anisotropic Power Spectra and Correlation Functions with Pairwise Lines-of-Sight}

\author[Philcox \& Slepian]{
Oliver H.\,E. Philcox$^{1,2}$\thanks{E-mail: \href{mailto:ohep2@cantab.ac.uk}{ohep2@cantab.ac.uk}}
and Zachary Slepian$^{3,4}$\thanks{E-mail: \href{mailto:zslepian@ufl.edu}{zslepian@ufl.edu}}
\\
$^{1}$Department of Astrophysical Sciences, Princeton University, Princeton, NJ 08544, USA\\
$^{2}$School of Natural Sciences, Institute for Advanced Study, 1 Einstein Drive, Princeton, NJ 08540, USA\\
$^{3}$Department of Astronomy, University of Florida, 211 Bryant Space Science Center, Gainesville, FL 32611, USA\\
$^{4}$Lawrence Berkeley National Laboratory, 1 Cyclotron Road, Berkeley, CA 94720, USA
}
\pubyear{2021}

\begin{document}
\label{firstpage}
\pagerange{\pageref{firstpage}--\pageref{lastpage}}
 \maketitle

% Abstract of the paper
\begin{abstract}
Conventional estimators of the anisotropic power spectrum and two-point correlation function (2PCF) adopt the `Yamamoto approximation', fixing the line-of-sight of a pair of galaxies to that of just one of its members. Whilst this is accurate only to first-order in the characteristic opening angle $\theta_\max$, it allows for efficient implementation via Fast Fourier Transforms (FFTs). This work presents practical algorithms for computing the power spectrum and 2PCF multipoles using pairwise lines-of-sight, adopting either the galaxy midpoint or angle bisector definitions. Using newly derived infinite series expansions for spherical harmonics and Legendre polynomials, we construct estimators accurate to arbitrary order in $\theta_\max$, though note that the midpoint and bisector formalisms themselves differ at fourth order. Each estimator can be straightforwardly implemented using FFTs, requiring only modest additional computational cost relative to the Yamamoto approximation. We demonstrate the algorithms by applying them to a set of realistic mock galaxy catalogs, and find both procedures produce comparable results for the 2PCF, with a slight preference for the bisector power spectrum algorithm, albeit at the cost of greater memory usage. Such estimators provide a useful method to reduce wide-angle systematics for future surveys.
\end{abstract}

% Select between one and six entries from the list of approved keywords.
% Don't make up new ones.

%%%%%%%%%%%%%%%%%%%%%%%%%%%%%%%%%%%%%%%%%%%%%%%%%%

%%%%%%%%%%%%%%%%% BODY OF PAPER %%%%%%%%%%%%%%%%%%

\section{Introduction}\label{sec: intro}
%\oliver{NB: I'm now defining a $\theta_\max$ which is the characteristic size (fixed for an analysis) to differentiate from $\theta = \Delta/R$ (which is different for each pair). $\theta_\max$ is the galaxy pair opening angle in the largest bin (2PCF) or the survey size (power spectrum) - the difference arises since Pk is an integral over pairs of all separations.}
%\oliver{note Martin point?}
We have now entered the epoch of `precision cosmology'. In the coming years, the volume of cosmological data available will increase at a prodigious rate, thanks to the advent of large spectroscopic surveys such as DESI \citep{2016arXiv161100036D}, Euclid \citep{2011arXiv1110.3193L} and SPHEREx \citep{2014arXiv1412.4872D}. As the number of observed galaxies grows, so too does the precision on fundamental parameters such as the growth rate, energy densities and Hubble parameter. Given that we will soon be able to measure summary statistics at the sub-percent level, it is vital to understand also their systematics to this precision, else we risk biasing our inference or losing effective survey volume. 

Whilst there is growing interest in more complicated statistics \citep[e.g.,][]{2017MNRAS.465.1757G,2017MNRAS.469.1738S,2019JCAP...11..034C,2020PhRvD.102d3516P,2021arXiv210201696S}, the information content of future surveys will be dominated by the two-point correlator, masquerading either as the configuration-space two-point correlation function (2PCF), $\xi(\vr)$, or the Fourier-space power spectrum, $P(\vk)$ \citep[e.g.,][]{2017MNRAS.466.2242B,2017MNRAS.470.2617A,2020arXiv200708991E}. In a statistically isotropic universe, both of these will depend only the distance between galaxies, be it $r = |\vr|$ or the momentum-space equivalent $k = |\vk|$. In our Universe this is not the case, since redshift-space distortions (RSD) impart a preferred origin to the observer \citep{1987MNRAS.227....1K}, sourcing additional cosmological information \citep[e.g.,][]{2006PhR...429..307L,2013PhR...530...87W}.

To fully encapsulate RSD, two-point statistics should depend on the position vectors to the two galaxies in question, rather than just a single length. Taking into account the various rotational symmetries, such a configuration can be specified by three degrees of freedom; options include the separation $r$ (or $k$) and the angles between the separation vector and the line-of-sight (LoS) to each galaxy \citep[e.g.,][]{2008MNRAS.389..292P,2015MNRAS.447.1789Y,2018MNRAS.476.4403C} or the separation, the mean distance to the galaxy pair, and a single angle \citep[e.g.,][]{2016JCAP...01..048R,2019JCAP...03..040B}. Unless our interest lies in the largest-possible scales (for instance in $f_\mathrm{NL}$-analyses), it is usually sufficient to parametrize the two-point correlators by just \textit{two} variables; the inter-galaxy distance $r$ or $k$, and the angle of the galaxy separation vector to a joint LoS, $\mu$. In this case, the functions can be robustly expanded as a Legendre series in $\mu$, and theory and observations simply compared. Of course, any such approximation necessarily induces wide-angle effects on the largest scales, which are the subject of extensive discussion in the literature \citep[e.g.,][]{1992ApJ...385L...5H,1996MNRAS.278...73H,1998ASSL..231..185H,1996ApJ...462...25Z,1998ApJ...498L...1S,2004ApJ...614...51S,2007MNRAS.378..119D,2008MNRAS.389..292P,2008PhRvD..78j3512S,2011PhRvD..84f3505B,2014JCAP...08..022R,2015MNRAS.447.1789Y,2015arXiv151004809S,2016JCAP...01..048R,2018MNRAS.476.4403C,2019JCAP...03..040B}. In general, the error in these approaches depends on the characteristic opening angle $\theta_\max$, defined either as the galaxy pair opening angle in the maximum radial bin (2PCF) or the survey opening angle (power spectrum). The dichotomy arises since the power spectrum depends on an integral over all galaxy pairs, whilst the 2PCF only requires pairs separated by the scale of interest.

If the two-parameter formalism is adopted (as has become commonplace), an important question must be asked: how should  one choose the joint LoS to the galaxy pair? The early literature adopted a single LoS for the whole survey \citep[e.g.,][]{1987MNRAS.227....1K,1992ApJ...385L...5H}, which, whilst simple to implement, incurs significant errors if the survey is wide. A more accurate prescription is to fix the LoS to the direction vector of a single galaxy, in the `Yamamoto approximation' \citep{2006PASJ...58...93Y}. Whilst this gives an error at $\mathcal{O}(\theta_\max^2)$ for characteristic size $\theta_\max$, which becomes important for DESI volumes \citep{2019MNRAS.484..364S}, it is straightforward to implement using Fast Fourier Transforms (FFTs), thus is the approach found in most recent analyses \citep[e.g.,][]{2015PhRvD..92h3532S,2015MNRAS.453L..11B,2017JCAP...07..002H,2017MNRAS.466.2242B,2018AJ....156..160H}. Beyond this approximation, there are two appealing LoS choices: the angle between the galaxy midpoint and separation vector \citep[cf.][]{2006PASJ...58...93Y,2015PhRvD..92h3532S,2015MNRAS.452.3704S,2015MNRAS.453L..11B} and the angle bisector \citep[cf.][]{1998ApJ...498L...1S,2000ApJ...535....1M,2004ApJ...614...51S,2015MNRAS.447.1789Y}. Both are consistent to third order in the opening angle (demonstrated in \citealt{2015arXiv151004809S}), but incur an error at fourth order, and, for $\theta_\max<10\degree$, lose little information compared to the double LoS approach \citep{2012MNRAS.423.3430B,2012MNRAS.420.2102S,2015MNRAS.447.1789Y}. However, their na\"ive implementation scales as $\mathcal{O}(N^2)$ for $N$ galaxies, rather than the $\mathcal{O}(N_\mathrm{g}\log N_\mathrm{g})$ dependence enjoyed by algorithms based on FFTs with $N_\mathrm{g}$ grid cells. 

In this work, we will demonstrate that the midpoint and bisector power spectrum and 2PCF estimators may be efficiently computed in $\mathcal{O}(N_\mathrm{g}\log N_\mathrm{g})$ time using FFTs. In both cases, it is necessary to perform a series expansion of the angular dependence in the galaxy pair opening angle $\theta$ (with $\theta<\theta_\max$); however, we give the explicit form of these corrections at arbitrary order. The different LoS definitions require different mathematical treatments for greatest efficiency; for the bisector case, we implement the suggestion of \citet{2018MNRAS.476.4403C} and extend it to arbitrary order, whilst for the midpoint approach, we provide novel formulae based on a newly-derived spherical harmonic shift theorem (similar to that of \citealt{2020arXiv201103503G} for the three-point function). This paper is an extension also of \citet{2015arXiv151004809S}, which gave the lowest-order corrections for the 2PCF, but did not consider the Fourier-space counterpart (which carries somewhat more subtleties). In contrast, \citet{2015MNRAS.452.3704S} considered the power-spectrum in the midpoint formalism, but only applied to a simplified spherical cap geometry. Our work goes beyond the previous by giving a full catalog of arbitrary-order expressions for the midpoint and bisector formalism in real- and Fourier-space. We further consider their application to realistic data using the MultiDark-\textsc{patchy} mock catalogs \citep{2016MNRAS.456.4156K}, and make the analysis code publicly available.\footnote{\href{https://github.com/oliverphilcox/BeyondYamamoto}{github.com/oliverphilcox/BeyondYamamoto}}%The mathematical results proved in the appendix are fully general and may be useful beyond their application to two-point functions.

The remainder of this work is structured as follows. We begin in \S\ref{sec: defs} by recapitulating the basic two-point correlator estimators. \S\ref{sec: midpoint-pk}\,\&\,\S\ref{sec: midpoint-2pcf} present our implementations of the power spectrum and 2PCF algorithms in the midpoint formalism, before the same is done in the bisector formalism in \S\ref{sec: bisector-pk}\,\&\,\S\ref{sec: bisector-2pcf}. \S\ref{sec: application} considers the application of the algorithms to data, before we conclude in \S\ref{sec: conclusion}. A list of useful mathematical identities is given in Appendix \ref{appen: math-101}, with Appendices \ref{appen: spherical-harmonic-shift-theorem}-\ref{appen: cw-math} giving mathematical derivations of results central to this work, in particular, a shift theorem for spherical harmonics and Legendre polynomials. For the reader less interested in mathematical derivations, we recommend skipping \S\ref{subsec: mid-pk-even} and the Appendices, and note that the key equations in this work are boxed.

\section{Estimators for the Two-Point Correlators}\label{sec: defs}

We begin by stating the Fourier conventions used throughout this work. We define the Fourier and inverse Fourier transforms by
\beq
    X(\vec k) &\equiv& \ft{X}(\vk) = \int d\vec x\,e^{-i\vec k\cdot\vec x}X(\vec x),  \qquad
    X(\vec x) \equiv \ift{X}(\vx) = \int \frac{d\vec k}{(2\pi)^3}e^{i\vec k\cdot\vec x}X(\vec k),
\eeq
leading to the definition of the Dirac delta function, $\delta_{\rm D}$, as
\beq
    (2\pi)^3\delta_{\rm D}(\vec k_1-\vec k_2) = \int d\vec x\,e^{i(\vec k_1-\vec k_2)\cdot\vec x}.
\eeq
The correlation function and power spectrum of the density field, $\delta$, are defined as 
\beq
    \xi(\vec r) = \av{\delta(\vec x)\delta(\vec x+\vec r)},\qquad (2\pi)^3\delta_{\rm D}(\vec k+\vec k')P(\vec k) = 
    \av{\delta(\vec k)\delta(\vec k')},
\eeq
with the power spectrum as the Fourier transform of the correlation function. Additionally, we use the shorthand
\beq
    \int_{\vk}\equiv \int\frac{d\vk}{(2\pi)^3},\quad \int_{\Omega_k} \equiv \int \frac{d\Omega_k}{4\pi}.
\eeq

\subsection{Power Spectrum}\label{subsec: algo-motiv}

The conventional estimator for the galaxy power spectrum multipoles, $P_\ell(k)$, is defined as a Fourier transform of two density fields $\delta(\vr_1)$ and $\delta(\vr_2)$:
\beq\label{eq: Pk-def}
    \hat{P}_\ell(k) = \frac{2\ell+1}{V}\int_{\Omega_k}\int d\vr_1\,d\vr_2\,e^{-i\vk\cdot(\vr_2-\vr_1)}\delta(\vr_1)\delta(\vr_2)L_\ell(\hk\cdot\hn)
\eeq
\citep[e.g.,][]{2017JCAP...07..002H}, where $\hn$ is the joint line-of-sight (LoS) to the pair of galaxies at $(\vr_1,\vr_2)$, $L_\ell$ is the Legendre polynomial of order $\ell$, $V$ is the survey volume and hats denote unit vectors. Whilst we have not included $k$-space binning, this is a straightforward addition, requiring an additional integral over $|\vk|$. In spectroscopic surveys, we do not have access to $\delta$ directly, only a set of galaxy and random particle positions with associated density fields $n_g(\vr)$ and $n_r(\vr)$ respectively. In this context, we replace
\beq
    \delta(\vr) \rightarrow \frac{w(\vr)[n_g(\vr) - \alpha_r n_r(\vr)]}{I^{1/2}},\quad I \equiv \int d\vr\,w^2(\vr)\bar{n}^2(\vr)
\eeq
\citep{2006PASJ...58...93Y}, where $w(\vr)$ are some weights (accounting for systematics, optimality and completeness), $\bar{n}$ is the mean galaxy density and $\alpha_r$ is the ratio of randoms to galaxies. \eqref{eq: Pk-def} is strictly an estimator for the \textit{window-convolved} power spectrum,\footnote{One may remove the window function by judicious use of quadratic estimators \citep[e.g.,][]{2020arXiv201209389P}.} and includes a shot-noise term. \resub{Since the latter does not depend on the LoS, we will ignore it henceforth, and additionally denote the (windowed) density field simply by $\delta(\vr)$.}
%Since neither effect depends on the LoS choice, we will ignore them henceforth. 

In the simplest (`plane-parallel') approximation, we take $\hn$ to be fixed across the survey, thus the estimator simplifies to the form familiar from $N$-body estimators, involving the Fourier-space density field $\delta(\vk)$:\footnote{In full, $\delta(\vk)$ should be multiplied by a compensation function $M(\vk)$ to account for the mass assignment scheme window function.}
\beq
    \hat{P}^\mathrm{fix}_\ell(k) = \frac{2\ell+1}{V}\int_{\Omega_k}L_\ell(\hk\cdot\hn)\left|\delta(\vk)\right|^2.
\eeq
\citep{1987MNRAS.227....1K,1992ApJ...385L...5H,1994ApJ...426...23F}. This necessarily incurs an error at $\mathcal{O}(\theta_\max)$ where $\theta_\max$ is the survey opening angle,\footnote{Note this depends on the \textit{survey} opening angle \resub{not the pair opening angle}, since the power spectrum is an integral over all galaxy pairs. Its importance increases as the wavenumber becomes small. \resub{We note that this is a rough guide to the size of the error, rather than a formally-defined scaling.}} and is valid only for the smallest surveys. 

\begin{figure}
    \centering
    \includegraphics[width=0.9\textwidth]{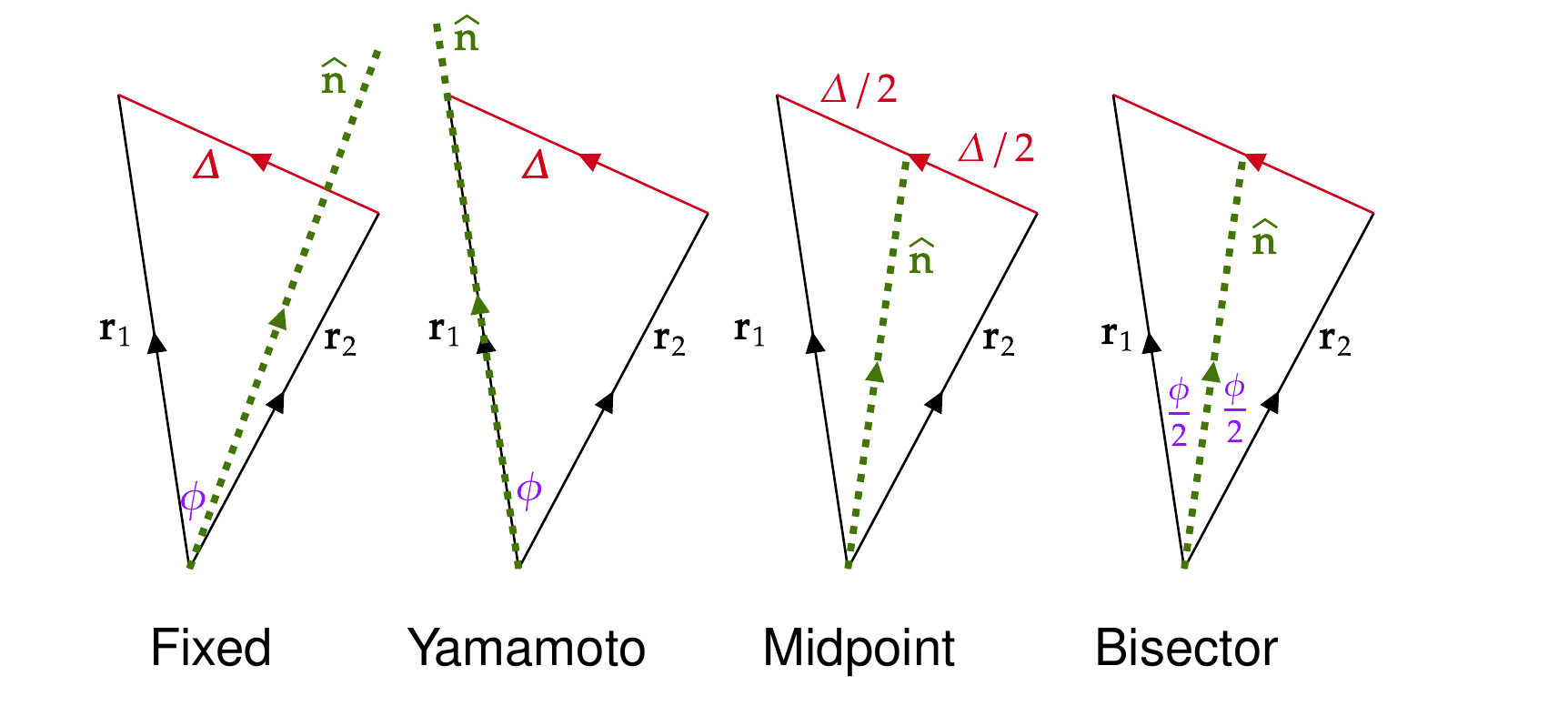}
    \caption{Comparison of the various choices of lines-of-sight (LoS) for the galaxy pair. For each example, the observer is located at the based of the triangle, with the galaxies at $\vr_1$ and $\vr_2$ with separation vector $\vD = \vr_2-\vr_1$. We define $\phi$ as the internal triangle angle with $\cos\phi = \hr_1\cdot\hr_2$. In the small-angle limit, we have $\phi\approx2\theta$, where $\theta \equiv \Delta/|\vr_1+\vr_2|$. The fixed LoS approximation ($\hn = \mathrm{const.}$) incurs an $\mathcal{O}(\theta_\max^0)$ error, whilst the Yamamoto approximation ($\hn = \hr_1$) and midpoint or bisector estimators ($\hn\propto \vr_1+\vr_2$ or $\hr_1+\hr_2$) incur $\mathcal{O}(\theta_\max^2)$ and $\mathcal{O}(\theta_\max^4)$ errors respectively, averaging over $\theta<\theta_\max$.}
    \label{fig: LoS-options}
\end{figure}

At the next order in approximation is the `Yamamoto formalism' used by most current estimators; this approximates the LoS as the position vector of a single galaxy, \textit{i.e.} $\hn = \hr_1$ or $\hn = \hr_2$, as shown in Fig.\,\ref{fig: LoS-options}. In this case, one may expand $L_\ell(\hk\cdot\hn)$ via the addition theorem for Legendre polynomials \eqref{eq: addition-theorem}, and arrive at the estimator
\beq\label{eq: Yamamoto-estimator}
    \hat{P}_\ell^\mathrm{Yama}(k) &=& \frac{4\pi}{V}\int_{\Omega_k}\sum_{m=-\ell}^\ell Y_\ell^{m*}(\hk)\int d\vr_1\,e^{i\vk\cdot\vr_1}\delta(\vr_1)\int d\vr_2\, e^{-i\vk\cdot\vr_2}Y_\ell^m(\hr_2)\delta(\vr_2)\\\nonumber
    &=& \frac{4\pi}{V}\int_{\Omega_k}\left[\sum_{m=-\ell}^\ell Y_\ell^{m*}(\hk)\ft{ Y_\ell^m\delta}(\vk)\right]\delta^*(\vk)
\eeq
\citep{2006PASJ...58...93Y,2015PhRvD..92h3532S,2015MNRAS.453L..11B,2017JCAP...07..002H,2018AJ....156..160H}. This incurs an error \resub{roughly scaling as} $\mathcal{O}(\theta_\max^2)$ in the $\ell>0$ moments (noting that the $\mathcal{O}(\theta_\max)$ part vanishes upon $\vr_1\leftrightarrow\vr_2$ symmetrization), and has been employed in almost all recent analyses \citep[e.g.,][]{2017MNRAS.466.2242B,2017MNRAS.465.1757G,2020arXiv200708991E}. The error incurred is not insignificant however; the BAO scale (much below the characteristic survey size) has opening angle $r_d/d_A(z_\mathrm{eff})$\,$\sim$\,$0.1$ for BOSS, where $d_A(z_\mathrm{eff})$ is the angular diameter distance to the mean survey redshift and $r_d$ is the sound horizon scale at decoupling. At low redshifts, this scale is at the percent level; of importance for future surveys such as Euclid and DESI. Considering the whole survey, $\theta_\max$ is significantly larger, which will affect measurements particularly on large scales.

Beyond the Yamamoto approximation, there are multiple options for how to proceed. Clearly, setting the LoS to the direction of just one galaxy not an optimal strategy on large scales, and a full treatment would include the positions of both galaxies, since their pairwise velocity cannot be simply described by a single LoS. This results in a higher-dimensional data-vector however, thus is generally disfavored. Two primary options exist for defining a single LoS correct to $\mathcal{O}(\theta_\max^2)$, both of which are shown in Fig.\,\ref{fig: LoS-options}; (a) using the position vector of the \textit{midpoint} of the two galaxies, $\hn \propto (\vr_1+\vr_2)$ \citep[cf.][]{2015PhRvD..92h3532S,2015MNRAS.452.3704S,2015MNRAS.453L..11B}, or (b) using the \textit{bisector} of the galaxy-observer-galaxy triangle, $\hn \propto (\hr_1+\hr_2)$ \citep[cf.][]{1998ApJ...498L...1S,2000ApJ...535....1M,2004ApJ...614...51S,2015MNRAS.447.1789Y}. These differ only at higher-order,\footnote{Specifically, due to the symmetry under $\vr_1\leftrightarrow\vr_2$ permutations for even $\ell$, any odd $\theta$ contribution must vanish, thus the difference between midpoint and bisector formalisms starts at $\mathcal{O}(\theta_\max^4)$.} and we will consider both in this work. In full, these are given by
\beq\label{eq: P-ell-midpoint-init}
    \hat{P}_\ell^\mathrm{midpoint}(k) &=& \frac{2\ell+1}{V}\int_{\Omega_k}\int d\vr_1\,d\vr_2\,e^{-i\vk\cdot(\vr_2-\vr_1)}\delta(\vr_1)\delta(\vr_2)L_\ell(\hk\cdot\widehat{\vr_1+\vr_2})\\\nonumber
    \hat{P}_\ell^\mathrm{bisector}(k) &=& \frac{2\ell+1}{V}\int_{\Omega_k}\int d\vr_1\,d\vr_2\,e^{-i\vk\cdot(\vr_2-\vr_1)}\delta(\vr_1)\delta(\vr_2)L_\ell(\hk\cdot\widehat{\hr_1+\hr_2}).
\eeq
Notably, neither straightforwardly factorizes into pieces depending only on $\vr_1$ and $\vr_2$, making its computation more involved than that of the Yamamoto estimator. A na\"ive implementation would involve counting all pairs of galaxies individually; this results in an estimator with $\mathcal{O}(N^2)$ complexity for $N$ galaxies; \resub{matching that of the original \citet{2006PASJ...58...93Y} estimator before the work of \citet{2015MNRAS.453L..11B}.} \resub{For upcoming galaxy surveys, such estimators will be prohibitively slow}, (though shown to be remarkably efficient on small scales in \citealt{2020MNRAS.492.1214P} and \citealt{2021MNRAS.501.4004P}). \resub{As shown below,} we can derive a \resub{separable, and hence efficient,} estimator using convergent series expansions.

\subsection{Two-Point Correlation Function}
Similar estimators may be derived for the multipoles of the two-point correlation function, $\xi_\ell(r)$. Analogous to \eqref{eq: Pk-def}, the general form is given by
\beq\label{eq: general-2pcf-form}
    \hat{\xi}_\ell(r) = \frac{2\ell+1}{V}\int d\vr_1\,d\vr_2\,\delta(\vr_1)\delta(\vr_2)L_\ell(\hD\cdot\hn)\left[\frac{\delta_{\rm D}(r-\Delta)}{4\pi r^2}\right],
\eeq
where $\hn$ is again the LoS and $\vD = \vr_2-\vr_1$ is the separation vector. Here, the square brackets pick out a particular value of the galaxy separation $\Delta = r$, and, if we substitute 
\beq
    \left[\frac{\delta_{\rm D}(r-\Delta)}{4\pi r^2}\right] \rightarrow \frac{1}{v^a}\Theta^a(r),
\eeq
where $\Theta^a(r)$ is some binning function with volume $v^a$, \eqref{eq: general-2pcf-form} becomes the estimator for the 2PCF in a finite bin $a$. Just as for the power spectrum, we cannot access the overdensity field $\delta$ directly, and must instead work with galaxies and random particle catalogs. Conventionally, this leads to the 2PCF being computed via the Landy-Szalay estimator \citep{1993ApJ...412...64L}, using $DD$, $DR$ and $RR$ counts, each of which is estimated via \eqref{eq: general-2pcf-form}. Since such complexities do not depend on our choice of $\hn$, we ignore them here, alongside the intricacies of edge-correction.

Two types of correlation function algorithms abound in the literature. Firstly, they are often computed by exhaustive pair counting, scaling as $N^2$ \citep[e.g.,][]{2020MNRAS.491.3022S}. Since we explicitly consider each pair of galaxies, any pairwise LoS can be simply included, rendering moot the analysis of this work. However, direct pair-counting can be exceedingly slow for large-datasets, thus it is commonplace to compute the 2PCF via FFT-based approaches. In the plane-parallel approximation of fixed $\hn$, the estimator may be written
\beq\label{eq: 2pcf-plane-parallel}
    \hat{\xi}_\ell^\mathrm{fix}(r) &=& \frac{2\ell+1}{V}\int d\vD\,\left[\frac{\delta_{\rm D}(r-\Delta)}{4\pi r^2}\right]L_\ell(\hD\cdot\hn) \int d\vr\,\delta(\vr)\delta(\vr+\vD)\\\nonumber
    &=& \frac{2\ell+1}{V}\int d\vD\,\left[\frac{\delta_{\rm D}(r-\Delta)}{4\pi r^2}\right]L_\ell(\hD\cdot\hn)\,\ift{|\delta(\vk)|^2}(\vD),\\\nonumber
\eeq
where we switch variables to $\vr = \vr_1, \vD = \vr_2-\vr_1$ in the first line, and write the $\vr$ integral as a convolution in the second. Following the inverse Fourier transform, \eqref{eq: 2pcf-plane-parallel} is a weighted real-space summation, which is easy to compute. This again incurs an error at $\mathcal{O}(\theta_\max^0)$, where $\theta_\max$ is now the characteristic opening of galaxy pairs separated by the maximum value of $r$ considered.

Similarly to \eqref{eq: Yamamoto-estimator}, the 2PCF may be estimated in the Yamamoto formalism by expanding the Legendre polynomial in spherical harmonics and writing the result as a convolution integral:
\beq\label{eq: 2pcf-yama}
    \hat{\xi}_\ell^\mathrm{Yama}(r) &=& \frac{4\pi}{V}\int d\vD\,\left[\frac{\delta_{\rm D}(r-\Delta)}{4\pi r^2}\right]\sum_{m=-\ell}^\ell Y_\ell^{m*}(\hD)\int d\vr\,\delta(\vr)\delta(\vr+\vD)Y_\ell^m(\widehat{\vr+\vD})\\\nonumber
    &=& \frac{4\pi}{V}\int d\vD\,\left[\frac{\delta_{\rm D}(r-\Delta)}{4\pi r^2}\right]\sum_{m=-\ell}^\ell Y_\ell^{m*}(\hD)\ift{\delta^*(\vk)\ft{Y_\ell^m\delta}(\vk)}(\vD),
\eeq
\citep[e.g.,][]{2016MNRAS.455L..31S}, which is computable in a similar manner to the plane-parallel estimator, now requiring $2(2\ell+1)$ FFTs for a given $\ell$. It again suffers an $\mathcal{O}(\theta_\max^2)$ error.

The $\mathcal{O}(\theta_\max^4)$ pairwise 2PCF estimates can be written in an analogous form to \eqref{eq: general-2pcf-form}:
\beq\label{eq: 2pcf-pair}
    \hat{\xi}_\ell^\mathrm{midpoint}(r) &=& \frac{2\ell+1}{V}\int d\vr_1\,d\vr_2\,\delta(\vr_1)\delta(\vr_2)L_\ell(\hD\cdot\widehat{\vr_1+\vr_2})\left[\frac{\delta_{\rm D}(r-\Delta)}{4\pi r^2}\right]\\\nonumber
    \hat{\xi}_\ell^\mathrm{bisector}(r) &=& \frac{2\ell+1}{V}\int d\vr_1\,d\vr_2\,\delta(\vr_1)\delta(\vr_2)L_\ell(\hD\cdot\widehat{\hr_1+\hr_2})\left[\frac{\delta_{\rm D}(r-\Delta)}{4\pi r^2}\right].
\eeq
Unlike the Yamamoto estimator, neither of these can be straightforwardly recast as a convolution, and thus evaluated via FFTs. It is however possible via series expansions, as will be discussed below.

\section{Midpoint Formalism: Power Spectrum}\label{sec: midpoint-pk}

\subsection{Series Expansion}\label{subsec: mid-pk-series}
To obtain an efficient power spectrum algorithm within the midpoint formalism, it is necessary to perform a series expansion on the angular dependence, $Y_\ell^m(\widehat{\vr_1+\vr_2})$, such that the estimator can be recast in a form conducive to FFT application. To motivate this, we start with the definition \eqref{eq: P-ell-midpoint-init} in terms of the new variables $\vD \equiv \vr_2-\vr_1$ and $\vR = \vr_1+\vr_2$, using the addition theorem \eqref{eq: addition-theorem} to write the Legendre polynomial in terms of spherical harmonics:\footnote{We note that it is formally possible to first perform the $\Omega_k$ integral analytically, which leads to the replacement $\int_{\Omega_k}e^{-i\vk\cdot\vD} L_\ell(\hk\cdot\hR) \rightarrow (-i)^\ell j_\ell(k\Delta) L_\ell(\vD\cdot\vR)$ using \eqref{eq: plane-wave}\,\&\,\eqref{eq: Omega-k-Leg-int}. However, this is not exact when the finite nature of the $k$-space grid is considered, thus will not be adopted here.} 
\beq\label{eq: Pk-pairwise}
    \hat{P}^\mathrm{pair}_\ell(k) = \frac{4\pi}{V}\sum_{m=-\ell}^\ell\int_{\Omega_k}Y_\ell^{m*}(\hk)\int d\vr_1\,d\vD\,e^{-i\vk\cdot\vD}\delta(\vr_1)\delta(\vr_1+\vD)Y_{\ell}^m(\hR).
\eeq

If the spherical harmonic factor $Y_{\ell}^m(\hR)$ can be written in a form separable in $\vD$ and $\vr_1$, the above expression can be evaluated as a convolution, allowing for acceleration by way of Fourier transforms (analogous to the 2PCF manipulations in \eqref{eq: 2pcf-plane-parallel}\,\&\,\eqref{eq: 2pcf-yama}). Such an expansion is indeed possible, via the \textit{spherical harmonic shift theorem}, which states
\begin{eqnarray}\label{eq: Ylm-shift-theorem-main-text}
    \boxed{Y_\ell^m(\widehat{\va+\vA}) = \sum_{\alpha=0}^\infty \sum_{J_1=0}^{\alpha}\sum_{J_2=\mathrm{max}(0,\ell-\alpha)}^{\ell+\alpha}\sum_{M=-J_1}^{J_1}\varphi^{\alpha,\ell m}_{J_1J_2M}\left(\frac{a}{A}\right)^{\alpha}Y_{J_1}^{M}(\ha)Y_{J_2}^{m-M}(\hA)}
\end{eqnarray}
\eqref{eq: Ylm-shift-theorem}, for arbitrary vectors $\va, \vA$ with $a\ll A$. This is proved in Appendix \ref{appen: spherical-harmonic-shift-theorem} and is a major new result of this work. Essentially, \eqref{eq: Ylm-shift-theorem-main-text} is an infinite expansion in terms of two spherical harmonics and the (small) ratio of $|\va|$ and $|\vA|$, giving an arbitrarily accurate approximation of $Y_\ell^m(\widehat{\va+\vA})$ if truncated at sufficiently large $\alpha$. This uses the numerical coefficients $\varphi^{\alpha,\ell m}_{J_1J_2M}$ given in \eqref{eq: shift-coeff-def}, which may be pre-computed and obey the relations
\beq\label{eq: varphi-relations-main-text}
    &&\varphi^{\alpha,\ell m}_{J_1J_2M} = 0 \quad\text{ if } J_1+\alpha \text{ or } J_2+\alpha+\ell \text{ is odd},\\\nonumber
    &&\,\varphi^{0,\ell m}_{J_1J_2M} = \sqrt{4\pi}\delta^\mathrm{K}_{J_10}\delta^\mathrm{K}_{J_2\ell}\delta^\mathrm{K}_{M0}, \quad \varphi^{\alpha,00}_{J_1J_2M} = \sqrt{4\pi}\delta^\mathrm{K}_{\alpha 0}\delta^\mathrm{K}_{J_10}\delta^\mathrm{K}_{J_20}\delta^\mathrm{K}_{M0}\,,
\eeq
(where $\delta^K_{ij}$ is the Kronecker delta, equal to unity if $i=j$ and zero else), 
%are non-zero unless both $J_1+\alpha$ and $J_2+\alpha+\ell$ are even (proved in Appendix \ref{appen: shift-coeff-prop}). 
as proved in Appendix \ref{appen: shift-coeff-prop}. 

In our context, we may use \eqref{eq: Ylm-shift-theorem-main-text} to expand $Y_{\ell}^m(\hR)$ using $\vR = 2\vr_1+\vD$ or $\vR = 2\vr_2-\vD$, \resub{recalling $\Delta \ll R$, \textit{i.e.} that $r_1\approx r_2$.}\footnote{An alternative approach would be to find an expansion of $Y_\ell^m(\hR)$ that is separable in $\vr_1$ and $\vr_2$ rather than $\vr_i$ and $\vD$. In this case, the power spectrum could be computed in the same manner as in the Yamamoto approximation, however, this is more difficult to obtain since $r_1/r_2$ is order unity, so one cannot simply apply \eqref{eq: Ylm-shift-theorem-main-text}. An approach similar to this will prove useful for the bisector formalism however (\S\ref{sec: bisector-pk}).} Suppressing summation limits for clarity, these lead to
\beq
    Y_{\ell}^m(\hR) = \sum_{\alpha J_1J_2M}\varphi^{\alpha,\ell m}_{J_1J_2M}\left(\frac{\Delta}{2r_1}\right)^{\alpha}Y_{J_1}^{M}(\hD)Y_{J_2}^{m-M}(\hr_1) = \sum_{\alpha J_1J_2M}\varphi^{\alpha,\ell m}_{J_1J_2M}(-1)^{J_1}\left(\frac{\Delta}{2r_2}\right)^{\alpha}Y_{J_1}^{M}(\hD)Y_{J_2}^{m-M}(\hr_2),
\eeq
using $Y_{J_1}^{M}(-\hD) = (-1)^{J_1}Y_{J_1}^M(\hD)$. The two may be combined to give the symmetrized form:
\beq\label{eq: YlmR-symm}
    Y_{\ell}^m(\hR) = \frac{1}{2}\sum_{\alpha J_1J_2M}\varphi^{\alpha,\ell m}_{J_1J_2M} Y_{J_1}^M(\hD)\left[\epsilon_1^\alpha Y_{J_2}^{m-M}(\hr_1)+(-1)^{J_1}\epsilon_2^\alpha Y_{J_2}^{m-M}(\hr_2)\right],
\eeq
where $\epsilon_i \equiv \Delta/(2r_i)$. Each term is fully separable in $\hD$ and $\hr_i$, and the expansion is simply a power series in $\epsilon_i$, which is closely related to the pair opening angle $\theta \equiv \Delta/R$.\footnote{Note the distinction between $\theta$, the opening angle for a particular pair, and $\theta_\max$, the characteristic opening angle, which is fixed for a particular analysis.} Indeed, including all terms up to $\alpha = K$ gives an approximation incurring an error only at $\mathcal{O}(\theta^{K+1})$. At lowest order ($\alpha=0$), we have $\varphi^{0,\ell m}_{J_1J_2M} = \delta^\mathrm{K}_{J_10}\delta^\mathrm{K}_{J_2\ell}\delta^\mathrm{K}_{M0}\times\sqrt{4\pi}$ \eqref{eq: varphi-relations-main-text}, thus
\beq
    Y_{\ell}^m(\hR) \rightarrow \frac{1}{2}\left[Y_{\ell}^m(\hr_1)+(-1)^\ell Y_{\ell}^m(\hr_2)\right];
\eeq
this simply yields the Yamamoto estimator in symmetrized form.

As an example, we consider the expansion of $Y_2^1(\hR)$, including all terms up to $\alpha=2$. From the above expressions, we obtain
\beq\label{eq: Y21-all}
    2\,Y_2^1(\hR) &=& \left\{\,\textcolor{red}{Y_2^1(\hr_1)}\right.\\\nonumber
    &&\,+\textcolor{orange}{\sqrt{\frac{4\pi}{105}}\epsilon_1\left[3\sqrt{7} Y_1^1(\hD)Y_1^0(\hr_1)+2\sqrt{3}Y_1^1(\hD)Y_3^0(\hr_1)+3\sqrt{7}Y_1^0(\hD)Y_1^1(\hr_1)-4\sqrt{2}Y_1^0(\hD)Y_{3}^1(\hr)+2\sqrt{10}Y_1^{-1}(\hD)Y_{3}^2(\hr_1)\right]}\\\nonumber
   &&+\textcolor{darkgreen}{\frac{\sqrt{4\pi}}{105}\epsilon_1^2\left[21Y_2^1(\hD)Y_0^0(\hr_1)-6\sqrt{5}Y_2^1(\hD)Y_2^0(\hr_1)-32Y_2^1(\hD)Y_4^0(\hr_1)-105Y_0^0(\hD) Y_2^1(\hr_1)-6 \sqrt{5} Y_2^0(\hD) Y_2^1(\hr_1)\right.}\\\nonumber
   &&\textcolor{darkgreen}{\quad+6 \sqrt{30} Y_2^2(\hD)
   Y_{2,-1}(\hr_1)+6 \sqrt{30} Y_2^{-1}(\hD) Y_2^2(\hr_1)+8 \sqrt{5} Y_2^2(\hD) Y_4^{-1}(\hr_1)+8
   \sqrt{30} Y_2^0(\hD) Y_4^1(\hr_1)-16 \sqrt{10} Y_2^{-1}(\hD) Y_4^2(\hr_1)}\\\nonumber
   &&\textcolor{darkgreen}{\quad\left.+8 \sqrt{35}
   Y_2^{-2}(\hD) Y_4^3(\hr_1)\right]}\\\nonumber
   &&\left.+\textcolor{black}{\mathcal{O}(\epsilon_1^3)}\right\} + (\vr_1\leftrightarrow\vr_2),
\eeq
with zeroth-, first- and second-order pieces shown in red, orange and green. Whilst this is lengthy, it is nonetheless computationally tractable. 

\subsection{Implementation}\label{subsec: mid-pk-impl}
The above series expansion may be used to write the pairwise power spectrum estimator as a series of convolutions. Inserting \eqref{eq: YlmR-symm} into \eqref{eq: Pk-pairwise} gives
\beq
    \hat{P}^\mathrm{midpoint}_\ell(k) &=& \frac{4\pi}{V}\sum_{m=-\ell}^\ell\int_{\Omega_k}Y_{\ell}^{m*}(\hk)\sum_{\alpha J_1J_2M}\varphi^{\alpha,\ell m}_{J_1J_2M}\int d\vD\,e^{-i\vk\cdot\vD}Y_{J_1}^{M}(\hD)\Delta^\alpha\\\nonumber
    &&\,\times\,\frac{1}{2}\left[\int d\vr_1\,\delta(\vr_1)\delta(\vr_1+\vD)\frac{Y_{J_2}^{m-M}(\hr_1)}{(2r_1)^\alpha}+(-1)^{J_1}\int d\vr_2\,\delta(\vr_2-\vD)\delta(\vr_2)\frac{Y_{J_2}^{m-M}(\hr_2)}{(2r_2)^\alpha}\right].
\eeq
Relabelling $\vD\rightarrow-\vD$, $\vr_2\rightarrow\vr_1$ and $\vk\rightarrow-\vk$ in the second term shows that the two are equivalent up to a factor $(-1)^{\ell}$. Here and henceforth we will assume even $\ell$, allowing us to absorb the symmetrization:\footnote{We replace $(-1)^{J_1}$ with $(-1)^{J_2}$ for later convenience; for even $\ell$, these must have the same sign, due to the parity-rules on $\varphi^{\alpha,\ell m}_{J_1J_2M}$.}
\beq\label{eq: P-ell-k-symm}
    \hat{P}^\mathrm{midpoint}_\ell(k) &=& \frac{4\pi}{V}\sum_{m=-\ell}^\ell\int_{\Omega_k}Y_{\ell}^{m*}(\hk)\sum_{\alpha J_1J_2M}\varphi^{\alpha,\ell m}_{J_1J_2M}(-1)^{J_2}\int d\vD\,e^{-i\vk\cdot\vD}Y_{J_1}^{M}(\hD)\Delta^\alpha\int d\vr_2\,\delta(\vr_2)\delta(\vr_2-\vD)\frac{Y_{J_2}^{m-M}(\hr_2)}{(2r_2)^\alpha}.
\eeq
Next, we note that the $\vr_2$ integral can be written as a convolution:
\beq\label{eq: g-JM-form}
    g_{J_2(m-M)}^{\alpha}(\vD) &\equiv& (-1)^{J_2}\int d\vr_2\,\left[\delta(\vr_2)(2r_2)^{-\alpha}Y_{J_2}^{m-M}(\hr_2)\right]\delta(\vr_2-\vD)\\\nonumber
    &=& (-1)^{J_2}\ift{\delta^*(\vk)\ft{(2r)^{-\alpha}Y_{J_2}^{m-M}\delta}(\vk)}(\vD),
\eeq
and the $\vD$ integral is then just a Fourier transform:
\begin{eqnarray}\label{eq: FFT-midpoint}
    \boxed{\hat{P}^\mathrm{midpoint}_\ell(k) = \frac{4\pi}{V}\sum_{m=-\ell}^\ell\int_{\Omega_k}Y_{\ell}^{m*}(\hk)\sum_{\alpha J_1J_2M}\varphi^{\alpha,\ell m}_{J_1J_2M}\ft{Y_{J_1}^{M}(\hD)\Delta^\alpha g_{J_2(m-M)}^\alpha(\vD)}(\vk).}
\end{eqnarray}
Calculation of the spectra is thus reduced to computing a convolution for each $\{\alpha, J_2, (m-M)\}$ triplet, then performing a Fourier transform for each $(\ell m)$ pair. We note that the summation over $\{\alpha, J_1,J_2\}$ can be moved inside the $\vk$-space Fourier transform; we separate it here for clarity. For $\alpha=0$, we require $J_1=0$, $J_2=\ell$, $M=0$ as before; this implies
\beq
    g^0_{\ell m}(\vD) = (-1)^\ell \ift{\delta^*(\vk)\ft{Y_{\ell}^{m}\delta}(\vk)}(\vD),
\eeq
and, since $Y_{J_1}^{M}(\hD)\rightarrow(4\pi)^{-1/2}$, the estimator is equal to that of \citet{2017JCAP...07..002H}, as expected. 

When implementing \eqref{eq: FFT-midpoint}, we must be aware of a certain subtlety. Our formalism requires the Fourier transform of a function of $\vD$ weighted by $\Delta^{\alpha}$ where $\alpha\geq0$. If $g^\alpha_{J_2(m-M)}(\vD)$ decays slower than $\Delta^{-\alpha}$, we will obtain an integrand that, in the limit of infinite survey volume, is not square integrable, \textit{i.e.} it diverges at large $\Delta$. This is particularly clear when the mean survey distance, $\bar{r}$ is large; in this case $\Delta/(2r_i)\approx \Delta/(2\bar{r})$, which is a monotonically increasing function of $\Delta$. Whilst the infinite volume limit is somewhat academic (since our expansion parameter $\theta_\max$ cannot be assumed small), for finite volumes, we note that the magnitude of $\Delta^\alpha g^\alpha_{J_2(m-M)}$ increases towards the survey edges for $\alpha>0$. As such, it is important that we consider the full extent of the $\vD$-space function. For a survey of characteristic width $L$, the convolved width is $2L$, thus this requires us to use a grid at least twice the survey width when painting particles. This restriction reduces the efficiency of the algorithm, since we must double the number of grid cells per dimension to obtain the same Nyquist frequency.

The selection rules on $\{J_1,J_2,M\}$ allow us to quantify the method's complexity. In general, if one wishes to compute all even power spectrum multipoles up to $\ell_\mathrm{max}$ using even (odd) $\alpha=\alpha_0$, we must compute all $g^\alpha_{J'M'}$ functions with even (odd) $J'$ up to $J' = \ell_\mathrm{max}+\alpha$; a total of $\left[1+(\ell_\mathrm{max}+\alpha_0)/2\right]^2$, each of which requires two Fourier transforms. The Fourier transform over $\vD$ can then be performed just once per $(\ell,m)$ pair (\textit{i.e.} $\left[1+\ell_\mathrm{max}/2\right]^2$ times). As a concrete example, computing the spectra up to $\ell_\mathrm{max} = 4$ requires $21$ $g^\alpha_{J'M'}$ coefficients for $\alpha=1$, and 28 for $\alpha=2$. 

\subsection{Parity-Even Form}\label{subsec: mid-pk-even}
Whilst the above estimator is mathematically valid, closer inspection reveals a curious property; it contains terms both odd and even in the small angles $\epsilon_i$. Since (for even $\ell$) the power-spectrum definition is symmetric under permutation of $\vr_1$ and $\vr_2$, we would expect any contributions of $\mathcal{O}(\theta^K)$ to vanish for odd $K$ (recalling $\theta \equiv \Delta/R$). In fact, this is the case, and the above expansion can be recast in a manner to make this manifest. Below, we consider a straightforward way to achieve this, based on iterated infinite sequences. An alternative method, which is less obvious \textit{a priori}, but simpler to implement, is described in Appendix \ref{appen: parity-even-alternate}.

In order to demonstrate that the odd terms in $\theta$ vanish, we first consider the term
\beq
    \left[\epsilon_1^\alpha Y_{J_2}^{m-M}(\hr_1) + (-1)^{J_1}\epsilon_2^\alpha Y_{J_2}^{m-M}(\hr_2)\right]
\eeq
appearing in \eqref{eq: YlmR-symm}. At lowest order in $\theta$, $\epsilon_1\approx\epsilon_2\approx\theta$ and $\vr_1\approx\vr_2\approx\vr$, giving
\beq
    \left[\epsilon_1^\alpha Y_{J_2}^{m-M}(\hr_1) + (-1)^{J_1}\epsilon_2^\alpha Y_{J_2}^{m-M}(\hr_2)\right] = \theta^\alpha Y_{J_2}^{m-M}(\hr)\left[1+(-1)^{J_1}\right]+\mathcal{O}(\theta^{\alpha+1}).
\eeq
For even $J_1$, the contribution is $\mathcal{O}(\theta^\alpha)$, yet for odd $J_1$, the term starts only at $\mathcal{O}(\theta^{\alpha+1})$. Since $J_1+\alpha$ must be even for $\varphi^{\alpha,\ell m}_{J_1J_2M}$ to be non-zero, we find that even $\alpha$ terms in \eqref{eq: YlmR-symm} (\textit{i.e.} those with even powers of $\epsilon_i$) begin to contribute at $\mathcal{O}(\theta^{\alpha})$, whilst the leading-order piece of odd $\alpha$ terms vanishes, and their contribution starts at $\mathcal{O}(\theta^{\alpha+1})$. This is not sufficient to demonstrate that there are \textit{no} terms odd in $\theta$ however, since, the term containing $\epsilon_i^3$ could include a non-cancelling $\epsilon_i^5$ contribution for example.

To obtain a manifestly parity-even expansion, we first split the $\alpha$ summation of \eqref{eq: YlmR-symm} into even and odd pieces, noting that the latter contributions must start at $\mathcal{O}(\theta^{\alpha+1})$, \textit{viz.} the above discussion:
\beq\label{eq: YlmR-tmp}
    Y_{\ell}^m(\hR) &=& \frac{1}{2}\sum_{\mathrm{even}\,\alpha}\sum_{ J_1J_2M}\varphi^{\alpha,\ell m}_{J_1J_2M} Y_{J_1}^M(\hD)\left[\epsilon_1^\alpha Y_{J_2}^{m-M}(\hr_1)+\epsilon_2^\alpha Y_{J_2}^{m-M}(\hr_2)\right]\\\nonumber
    &&+\frac{1}{2}\sum_{\mathrm{odd}\,\alpha}\sum_{ J_1J_2M}\varphi^{\alpha,\ell m}_{J_1J_2M} Y_{J_1}^M(\hD)\left[\epsilon_1^\alpha Y_{J_2}^{m-M}(\hr_1)-\epsilon_2^\alpha Y_{J_2}^{m-M}(\hr_2)\right].
\eeq
Next, we rewrite the odd-parity piece by expanding the $\vr_2$ term around $\vD = \vec 0$, which allows us to explicitly cancel the lowest-order piece. To do so, we require a generalized version of the spherical harmonic shift theorem, proved in appendix \ref{appen: generalized-shift-theorem}:
\beq
    \left(\frac{|\va|}{|\vA+\va|}\right)^\alpha Y_{J_1}^{M}(\ha)Y_{J_2}^{m-M}(\widehat{\va+\vA}) &=& \sum_{\beta S_1S_2T}\Omega^{\beta,\alpha J_1J_2mM}_{S_1S_2T}\left(\frac{a}{A}\right)^{\alpha+\beta}Y_{S_1}^T(\ha)Y_{S_2}^{m-T}(\hA),
\eeq
where the $\Omega$ coefficients are given in \eqref{eq: generalized-shift-theorem}, and the $S_i$ summation is limited to the range $[\mathrm{max}(0,J_i-\beta),J_i+\beta]$. Notably, $\Omega^{0,\alpha J_1J_2mM}_{S_1S_2T} = \delta^\mathrm{K}_{S_1J_1}\delta^\mathrm{K}_{S_2J_2}\delta^\mathrm{K}_{TM}$ from \eqref{eq: Omega-alpha-0-coeff}. Applying this to $\epsilon_2^\alpha Y_{J_1}^{M}(\hD)Y_{J_2}^{m-M}(\hr_2)$ with $\va = \vD$, $\vA = \vr_1$ yields
\beq
    \epsilon_2^\alpha Y_{J_1}^{M}(\hD)Y_{J_2}^{m-M}(\hr_2)  = \epsilon_1^\alpha Y_{J_1}^{M}(\hD)Y_{J_2}^{m-M}(\hr_1) + \sum_{\beta>0}\sum_{S_1S_2T}2^\beta \Omega^{\beta,\alpha J_1J_2mM}_{S_1S_2T}\epsilon_1^{\alpha+\beta} Y_{S_1}^T(\hD)Y_{S_2}^{m-T}(\hr_1),
\eeq
separating out the $\beta = 0$ term. This is now an expansion in $\Delta/r_i\approx 2\theta$, thus the radius of convergence is somewhat diminished compared to the all-parity form. Inserting this in the parity-odd part of \eqref{eq: YlmR-tmp} and symmetrizing over $\vr_1\leftrightarrow\vr_2$ gives
\beq\label{eq: Ylm-even-tmp}
    \left.Y_\ell^m(\hR)\right|_\mathrm{odd} = -\frac{1}{4}\sum_{\mathrm{odd}\,\alpha}\sum_{J_1J_2M}\varphi^{\alpha,\ell m}_{J_1J_2M}\sum_{\beta>0}\sum_{ S_1S_2T}2^\beta \Omega^{\beta,\alpha J_1J_2mM}_{S_1S_2T}Y_{S_1}^{T}(\hD)\left[\epsilon_1^{\alpha+\beta}Y_{S_2}^{m-T}(\hr_1)-(-1)^{J_1+S_1}\epsilon_2^{\alpha+\beta}Y_{S_2}^{m-T}(\hr_2)\right].
\eeq
Importantly the $\beta = 0$ part vanishes, thus the summand at order $\alpha$ contains terms only starting at $\mathcal{O}(\epsilon_i^{\alpha+1})$. We have therefore succeeded in removing the asymmetric piece at lowest order. In practice, this means that we have removed all terms proportional to $\epsilon_i^1$, reducing the total of convolutions that need to be performed, since there is no longer a requirement to compute $\alpha=1$.

The pudding is not yet proved however, since we have not removed terms with, for example, $\epsilon_i^{3}$. This is possible via a similar prescription, first noting that the lowest-order piece of the summand in \eqref{eq: Ylm-even-tmp} (\textit{i.e.} that with $\vr_1=\vr_2)$ is non-zero only for odd $\beta$; a consequence of the parity rules given in Appendix \ref{appen: generalized-shift-theorem}, restricting $\beta+J_1+S_1$ to be even. For even $\beta$, we again have a cancellation between terms involving $\vr_1$ and $\vr_2$. As before, we may expand the relevant $\vr_2$ piece in terms of $\vr_1$, and cancel the lowest-order contribution. This procedure of ``split according to parity, expand, symmetrize'' may be iterated, and leads to the following form:
\beq
    Y_\ell^m(\hR) &=&
    \frac{1}{2}\sum_{\mathrm{even}\,\alpha}\sum_{J_1J_2M}\varphi^{\alpha,\ell m}_{J_1J_2M}Y_{J_1}^{M}(\hD)\left[\epsilon_1^{\alpha}Y_{J_2}^{m-M}(\hr_1)+\epsilon_2^{\alpha}Y_{J_2}^{m-M}(\hr_2)\right]\\\nonumber
    &&-\frac{1}{4}\sum_{\mathrm{odd}\,\alpha}\sum_{J_1J_2M}\varphi^{\alpha,\ell m}_{J_1J_2M}\sum_{\mathrm{odd}\,\beta}\sum_{S_1S_2T}2^\beta \Omega^{\beta,\alpha J_1J_2mM}_{S_1S_2T}Y_{S_1}^{T}(\hD)\left[\epsilon_1^{\alpha+\beta}Y_{S_2}^{m-T}(\hr_1)+\epsilon_2^{\alpha+\beta}Y_{S_2}^{m-T}(\hr_2)\right]\\\nonumber
    &&+\frac{1}{8}\sum_{\mathrm{odd}\,\alpha}\sum_{J_1J_2M}\varphi^{\alpha,\ell m}_{J_1J_2M}\sum_{\mathrm{even}\,\beta>0}\sum_{S_1S_2T}2^\beta \Omega^{\beta,\alpha J_1J_2mM}_{S_1S_2T}\sum_{\mathrm{odd}\,\gamma}\sum_{Q_1Q_2R}2^\gamma \Omega^{\gamma,(\alpha+\beta) S_1S_2mT}_{Q_1Q_2R}Y_{Q_1}^{R}(\hD)\\\nonumber
    &&\quad\times\quad\left[\epsilon_1^{\alpha+\beta+\gamma}Y_{Q_2}^{m-R}(\hr_1)+\epsilon_2^{\alpha+\beta+\gamma}Y_{Q_2}^{m-R}(\hr_2)\right]\\\nonumber
    &&+\,...\,,
\eeq
which contains only even powers of $\epsilon_i$. Whilst each successive line requires more work to evaluate, we note that the only terms up to the second (third) line are required for an expansion correct to third (fifth) order in $\theta$. Furthermore, we may simplify the above by introducing redefined coefficients $\Phi^{\alpha,\ell m}_{J_1J_2M}$, such that
\beq\label{eq: YlmR-symm-even}
    \boxed{Y_\ell^m(\hR) =
    \frac{1}{2}\sum_{\mathrm{even}\,\alpha}\sum_{J_1J_2M}\Phi^{\alpha,\ell m}_{J_1J_2M}Y_{J_1}^{M}(\hD)\left[\epsilon_1^{\alpha}Y_{J_2}^{m-M}(\hr_1)+\epsilon_2^{\alpha}Y_{J_2}^{m-M}(\hr_2)\right],}
\eeq
for even $J_1, J_2,\alpha$, using
\beq
    \Phi^{\alpha,\ell m}_{J_1J_2M} &=& \varphi^{\alpha,\ell m}_{J_1J_2M}\\\nonumber
    &&- \frac{1}{2}\sum_{\mathrm{odd}\,\beta<\alpha}\sum_{J_1'J_2'M'}2^{\alpha-\beta}\varphi^{\beta,\ell m}_{J_1'J_2'M'}\Omega^{(\alpha-\beta),\beta J_1'J_2'mM'}_{J_1J_2M}\\\nonumber
    %&&+\frac{1}{4}\sum_{\mathrm{odd}\,\beta<\alpha}\sum_{J_1'J_2'M'}\varphi^{\beta,\ell m}_{J_1'J_2'M'}\sum_{\mathrm{even}\,\gamma<\alpha-\beta}\sum_{J_1''J_2''M''}2^{\alpha-\beta}\Omega^{(\alpha-\beta),\beta J_1'J_2'mM'}_{J_1J_2M}\\\nonumber
    &&+\frac{1}{4}\sum_{\mathrm{odd}\,\beta<\alpha}\sum_{J_1'J_2'M'}\varphi^{\beta,\ell m}_{J_1'J_2'M'}2^{\alpha-\beta}\sum_{\mathrm{even}\,\gamma<\alpha-\beta,\gamma>0}\sum_{ J_1''J_2''M''}\Omega^{\gamma,\beta J_1'J_2'mM'}_{J_1''J_2''M''} \Omega^{(\alpha-\beta-\gamma),(\beta+\gamma) J_1''J_2''mM''}_{J_1J_2M} + ...\,.
\eeq
As an example, for $\ell = 2$, $m = 1$, the parity-even expansion truncating at $\alpha=2$ gives 
\beq\label{eq: Y21-even}
    2Y_2^1(\hR) &=& \left\{\,\textcolor{red}{Y_2^1(\hr_1)}\right.\\\nonumber
    &&\,-\,\textcolor{darkgreen}{\frac{\sqrt{4\pi}}{105}\epsilon_1^2\left[21Y_2^1(\hD)Y_0^0(\hr_1)-6\sqrt{5}Y_2^1(\hD)Y_2^0(\hr_1)-32Y_2^1(\hD)Y_4^0(\hr_1)-105Y_0^0(\hD) Y_2^1(\hr_1)-6 \sqrt{5} Y_2^0(\hD) Y_2^1(\hr_1)\right.}\\\nonumber
   &&\textcolor{darkgreen}{\quad\,+\,6 \sqrt{30} Y_2^2(\hD)
   Y_{2,-1}(\hr_1)+6 \sqrt{30} Y_2^{-1}(\hD) Y_2^2(\hr_1)+8 \sqrt{5} Y_2^2(\hD) Y_4^{-1}(\hr_1)+8
   \sqrt{30} Y_2^0(\hD) Y_4^1(\hr_1)-16 \sqrt{10} Y_2^{-1}(\hD) Y_4^2(\hr_1)}\\\nonumber
   &&\textcolor{darkgreen}{\quad\left.\,+\,8 \sqrt{35}
   Y_2^{-2}(\hD) Y_4^3(\hr_1)\right]}\\\nonumber
   &&\left.\,+\,\mathcal{O}(\epsilon_1^4)\right\} + (\vr_1\leftrightarrow\vr_2),
\eeq
marking zeroth- (second-)order terms in red (green). At fixed maximum order in $\epsilon_i$, this contains significantly fewer terms than \eqref{eq: Y21-all}, since all odd powers vanish. We note that $\Phi^{2,\ell m}_{J_1J_2M} = -\varphi^{2,\ell m}_{J_1J_2M}$ here.\footnote{This arises naturally in the alternate derivation given in Appendix \ref{appen: parity-even-alternate}.}

Adopting this notation, the full parity-even power spectrum estimator is given by
\beq
    \hat{P}^\mathrm{midpoint}_\ell(k) &=& \frac{4\pi}{V}\sum_{m=-\ell}^\ell\int_{\Omega_k}Y_{\ell}^{m*}(\hk)\sum_{\mathrm{even}\,\alpha}\sum_{J_1J_2M}\Phi^{\alpha,\ell m}_{J_1J_2M}\int d\vD\,e^{-i\vk\cdot\vD}Y_{J_1}^{M}(\hD)\Delta^\alpha\int d\vr_1\,\delta(\vr_1)\delta(\vr_1+\vD)\frac{Y_{J_2}^{m-M}(\hr_1)}{(2r_1)^\alpha}\nonumber
\eeq
\beq
    \Rightarrow \boxed{\hat{P}_\ell^\mathrm{midpoint}(k)= \frac{4\pi}{V}\sum_{m=-\ell}^\ell\int_{\Omega_k}Y_{\ell}^{m*}(\hk)\sum_{\mathrm{even}\,\alpha}\sum_{J_1J_2M}\Phi^{\alpha,\ell m}_{J_1J_2M}\ft{Y_{J_1}^{M}(\hD)\Delta^\alpha g_{J_2(m-M)}^\alpha(\vD)}(\vk),}
\eeq
analogous to \eqref{eq: P-ell-k-symm}, but now requiring only even $\alpha$ (and thus even $J_1$, $J_2$), significantly reducing the necessary number of $g^\alpha_{J'M'}$ fields \eqref{eq: g-JM-form}. For $\ell_\mathrm{max} = 2$, we require 28 (45) functions for $\alpha = 2$ ($\alpha = 4$). It is important to note that the parity-even estimator will exhibit somewhat slower convergence than the all-parity form, since we capture only a subset of the pieces containing odd $\alpha$ (\textit{i.e.} those that contribute to even $\alpha<\alpha_\mathrm{max}$), and the expansion formally requires $2\epsilon_{i,\max}\approx 2\theta_\max\ll 1$, rather than $\epsilon_{i,\max}\approx \theta_\max\ll 1$. If the latter conditions are met, both forms are fully convergent.

\section{Midpoint Formalism: 2PCF}\label{sec: midpoint-2pcf}

\subsection{Series Expansion}\label{subsec: mid-2pcf-series}
Just as for the power spectrum, we may construct an efficient 2PCF estimator via series expansions coupled with FFTs. In this instance, the angular dependence appears through $L_\ell(\hD\cdot\hR)$; our goal therefore is to expand this in a form separable in $\vD$ and $\vr_1$. To this end, we use the \textit{Legendre polynomial shift theorem}, which states
\beq\label{eq: Legendre-shift-theorem-all-parity-main-text}
    \boxed{L_\ell(\ha\cdot\widehat{\va+\vA}) = \sum_{\alpha=0}^\infty \sum_{J=\mathrm{max}(\ell-\alpha,0)}^{\ell+\alpha}\frac{2J+1}{2\ell+1}f_{J}^{\alpha,\ell}\left(\frac{a}{A}\right)^\alpha L_{J}(\ha\cdot\hA)}
\eeq
(cf.\,\ref{eq: Legendre-shift-theorem-all-parity}), for $a \ll A$. This is proved in Appendix \ref{appen: leg-shift-theorem-deriv}, and uses the coefficients $f^{\alpha,\ell}_J$ presented in \eqref{eq: f-alpha-def}. In brief, the derivation proceeds by noting that $L_\ell(\ha\cdot\widehat{\va+\vA})$ can be written as a sum of spherical harmonics in $\ha$ and $\widehat{\va+\vA}$ using the addition theorem \eqref{eq: addition-theorem}, then expanding $Y_\ell^m(\widehat{\va+\vA})$ using the spherical harmonic shift theorem of Appendix\,\ref{appen: spherical-harmonic-shift-theorem}, and simplifying the resulting coefficients. An alternative approach would be to write the Legendre polynomial as a power series then perform a Taylor expansion; this gives the same results.

In our context, we set $\va = \vD$, $\vA = 2\vr_1$ or $\va = -\vD$, $\vA = 2\vr_2$ in \eqref{eq: Legendre-shift-theorem-all-parity-main-text} to yield the symmetrized form
\beq\label{eq: Legendre-shift-theorem-applied}
    L_\ell(\hD\cdot\hR) = \frac{1}{2}\sum_{\alpha=0}^\infty \sum_{J=\mathrm{max}(\ell-\alpha,0)}^{\ell+\alpha}\frac{2J+1}{2\ell+1}f_{J}^{\alpha.\ell}\left[\epsilon_1^\alpha L_{J}(\mu_1)+(-1)^{\ell+J}\epsilon_2^\alpha L_{J}(\mu_2)\right]\,,
\eeq
where $\mu_i = \hD\cdot\hr_i$ and $\epsilon_i = \Delta/(2r_i)$. This is markedly simpler than the result for the spherical harmonic shift theorem, and can equivalently be derived by an expansion of $L_\ell(\hD\cdot\hR)$ in powers of $\hD\cdot\hR$, which is then expanded via the binomial theorem. Key properties of this expansion are discussed in Appendix \ref{appen: leg-shift-theorem-deriv}; here we note that the coefficients are non-zero only for even $\alpha+\ell+J$ and $f^{0,\ell}_J = \delta^\mathrm{K}_{J\ell}$, such that $L_\ell(\hD\cdot\hR) = L_\ell(\mu_1)=L_\ell(-\mu_2)$ at lowest order, \textit{i.e.} the Yamamoto approximation. 

As an example, we consider the expansion of the $\ell = 2$ moment. At fourth order in $\epsilon_i$ (and hence the opening angle $\theta$):
\beq\label{eq: L2-all}
    2\,L_2(\hD\cdot\hR) &=& \textcolor{red}{L_2(\mu_1)}+\textcolor{orange}{\frac{6}{5} \epsilon_1 \left[L_1(\mu_1 )-L_3(\mu_1
   )\right]}+\textcolor{darkgreen}{\frac{1}{35} \epsilon_1^2 \left[7 L_0(\mu_1 )-55 L_2(\mu_1 )+48
   L_4(\mu_1 )\right]}\\\nonumber
   &&\,-\,\textcolor{blue}{\frac{4}{105} \epsilon_1^3 \left[9 L_1(\mu_1 )-49 L_3(\mu_1
   )+40 L_5(\mu_1 )\right]}+\textcolor{purple}{\frac{1}{385} \epsilon_1 ^4 \left[11 L_0(\mu_1 )+165
   L_2(\mu_1 )-816 L_4(\mu_1 )+640 L_6(\mu_1 )\right]}\\\nonumber
   &&\,+\,\mathcal{O}\left(\epsilon_1^5\right)+ (\vr_1\leftrightarrow-\vr_2),
\eeq
where colors separate the different orders, as before. For even (odd) $\alpha$, the expansion simply consists of Legendre polynomials of even (odd) order up to $\ell+\alpha$, weighted by powers of $\epsilon_i$.

\subsection{Implementation}\label{subsec: mid-2pcf-imp}
Inserting \eqref{eq: Legendre-shift-theorem-applied} into the 2PCF estimator \eqref{eq: 2pcf-pair} yields
\beq\label{eq: xi-midpoint-tmp}
    \hat{\xi}^\mathrm{midpoint}_\ell(r) = \frac{1}{V}\sum_{\alpha J}(2J+1)f^{\alpha,\ell}_{J}\int d\vr_1\,d\vr_2\,\delta(\vr_1)\delta(\vr_2)\left(\frac{\Delta}{2r_2}\right)^\alpha L_J(-\hD\cdot\hr_2)\left[\frac{\delta_{\rm D}(r-\Delta)}{4\pi r^2}\right],
\eeq
suppressing summation indices for clarity and absorbing the $\vr_1\leftrightarrow-\vr_2$ symmetrization (for even $\ell$). Expanding the Legendre polynomial via the addition theorem \eqref{eq: addition-theorem}, we may express the integrand in separable form:
\beq
    \hat{\xi}^\mathrm{midpoint}_\ell(r) = \frac{4\pi}{V}\sum_{\alpha J}\sum_{M=-J}^J f^{\alpha,\ell}_{J}\int d\vD\,Y_J^{M*}(\hD)\Delta^\alpha \left[\frac{\delta_{\rm D}(r-\Delta)}{4\pi r^2}\right] \int d\vr_2\,\delta(\vr_2)\delta(\vr_2-\vD)\frac{Y_J^{M}(-\hr_2)}{(2r_2)^\alpha}.
\eeq
As for the power spectrum, the $\vr_2$ integral is simply a convolution:
\beq
    g^\alpha_{JM}(\vD) &=& (-1)^J\int d\vr_2\,\left[\delta(\vr_2)(2r_2)^{-\alpha}Y_J^M(\hr_2)\right]\delta(\vr_2-\vD)\\\nonumber
    &=& (-1)^J\ift{\delta^*(\vk)\ft{(2r)^{-\alpha}Y_J^M\delta}(\vk)}(\vD),
\eeq
where the $g^\alpha_{JM}(\vD)$ functions are the same as those in \eqref{eq: g-JM-form}. In this notation, the 2PCF estimator is given by
\beq\label{eq: xi-parity-all-estimator}
    \boxed{\hat{\xi}^\mathrm{midpoint}_\ell(r) = \frac{4\pi}{V}\sum_{\alpha JM}f^{\alpha,\ell}_{J}\int d\vD\,Y_J^{M*}(\hD)\Delta^\alpha g^\alpha_{JM}(\vD) \left[\frac{\delta_{\rm D}(r-\Delta)}{4\pi r^2}\right],}
\eeq
which may be computed via a simple summation over real-space pixels, given some $r=|\vD|$ binning scheme. This is analogous to the estimator obtained within the Yamamoto scheme:
\beq
    \hat{\xi}^\mathrm{Yama}_\ell(r) = \frac{4\pi}{V}\sum_{M=-\ell}^{M=\ell} \int d\vD\,Y_\ell^{M*}(\hD) g^0_{\ell M}(\vD) \left[\frac{\delta_{\rm D}(r-\Delta)}{4\pi r^2}\right],
\eeq 
and simply requires additional 2PCF multipoles to be computed, weighted by powers of $\Delta/(2r_2)$. Since algorithm this does not require a Fourier transform in $\vD$ space, we do not need to use an increased boxsize, contrary to the power spectrum case (cf.\,\S\ref{subsec: mid-pk-impl}). Instead, we require only that the width of the box is at least the maximum galaxy dimension plus the maximum separation of interest, to avoid particle overlap on periodic wrapping. To obtain a 2PCF estimate in multipoles up to $\ell_\mathrm{max}$ including $\mathcal{O}(\theta_\max^K)$ corrections, we need to compute the $g^\alpha_{JM}$ functions with $\alpha \leq K$, $J\leq\ell_\mathrm{max}+\alpha$, $|M|\leq J$, each of which requires a forward and inverse Fourier transform. For $\ell_\mathrm{max} = 4$ case requires 21 (28) $g^\alpha_{JM}$ coefficients for $\alpha = 1$ ($\alpha = 2$) just as for $\hat{P}_\ell(k)$, each of which is computed in $\mathcal{O}(N_\mathrm{g}\log N_\mathrm{g})$ time for $N_\mathrm{g}$ grid points.

\subsection{Parity-Even Form}\label{subsec: 2pcf-mid-even}
We may further simplify the estimator by writing it as a sum over only \textit{even} $\alpha$ (and thus even multipoles $J$). This is possible via recasting \eqref{eq: Legendre-shift-theorem-applied} in a manifestly parity-even form:
\beq\label{eq: Legendre-shift-theorem-even-parity-main-text}
    \boxed{L_\ell(\hD\cdot\hR) = \frac{1}{2}\sum_{\mathrm{even}\,\alpha=0}^\infty \sum_{J=\mathrm{max}(\ell-\alpha,0)}^{\ell+\alpha}F_{J}^{\alpha.\ell}\left[\epsilon_1^\alpha L_{J}(\mu_1)+(-1)^{\ell+J}\epsilon_2^\alpha L_{J}(\mu_2)\right]}
\eeq
\eqref{eq: Legendre-shift-theorem-even-parity}, where the corresponding coefficients, $F^{\alpha,\ell}_J$, are given in \eqref{eq: F-coeff-def}. The corresponding estimator is thus
\beq
    \boxed{\hat{\xi}^\mathrm{midpoint}_\ell(r) = \frac{4\pi}{V}\sum_{\mathrm{even}\,\alpha}\sum_{JM} F^{\alpha,\ell}_{J}\int d\vD\,Y_J^{M*}(\hD)\Delta^\alpha g^\alpha_{JM}(\vD) \left[\frac{\delta_{\rm D}(r-\Delta)}{4\pi r^2}\right].}
\eeq
The derivation of \eqref{eq: Legendre-shift-theorem-even-parity-main-text} is analogous to that of \S\ref{subsec: mid-pk-even} and sketched in Appendix \ref{appen: leg-shift-theorem-even}. As an example, the even-parity expansions of $\ell=2$ and $\ell = 4$ become
\beq\label{eq: L24-even}
    2\,L_2(\hD\cdot\hR) &=& \textcolor{red}{L_2(\mu_1)}+\textcolor{darkgreen}{\frac{1}{35}\epsilon_1^2 \left[-7 L_0(\mu_1 )+55 L_2(\mu_1 )-48
   L_4(\mu_1 )\right]}\\\nonumber
   &&\,+\,\textcolor{purple}{\frac{1}{77} \epsilon_1^4 \left[11 L_0(\mu_1 )+165 L_2(\mu_1
   )-816 L_4(\mu_1 )+640 L_6(\mu_1 )\right]}+\mathcal{O}\left(\epsilon_1^6\right)+ (\vr_1\leftrightarrow-\vr_2)\\\nonumber
   2\,L_4(\hD\cdot\hR) &=& \textcolor{red}{L_4(\mu_1)}-\textcolor{darkgreen}{\frac{10}{77} \epsilon_1^2 \left[11 L_2(\mu_1)-39 L_4(\mu_1)+28
   L_6(\mu_1)\right]}\\\nonumber
   &&\,-\,\textcolor{purple}{\frac{5}{9009}\epsilon_1^4 \left[143 L_0(\mu_1)+4420 L_2(\mu_1)-52083 L_4(\mu_1)+110240 L_6(\mu_1)-62720 L_8(\mu_1)\right]}+\mathcal{O}\left(\epsilon_1^6\right)+(\vr_1\leftrightarrow-\vr_2),
\eeq
correct to fifth-order in $\theta$. These results are consistent with the second-order calculation of \citet[Eqs.\,19\,\&\,20]{2015arXiv151004809S}, and may be extended to arbitrarily high order in $\ell$ and $\alpha$, with each order just requiring computation of more $g^\alpha_{JM}(\vD)$ functions.

\section{Bisector Formalism: Power Spectrum}\label{sec: bisector-pk}
We now consider the estimators for the power spectrum in the bisector formalism. The discussion in this section closely follows that of \citet[Appendix E]{2018MNRAS.476.4403C}, but is extended to give closed-form estimators at arbitrary order in $\theta_\max$.

\subsection{Series Expansion}\label{subsec: pk-bis-series}
To derive the midpoint power spectrum estimator (\S\ref{sec: midpoint-pk}), we began by expanding the spherical harmonic $Y_\ell^m(\hn)$ as a separable function in $\hr_1$ and $\hD$. This was both tractable, since one can write $\hn \propto (2\vr_1 + \vD)$, and favorable, since it led to an explicit expansion in powers of $\epsilon = \Delta/(2r)\approx \theta$. When adopting the bisector LoS, we have $\hn \propto \hr_1+\hr_2$, which does not facilitate as straightforward separation into $\vD$ and $\vr_1$ pieces due to the additional normalization by $|\vr_1|$ and $|\vr_2|$. For this reason, we take a different approach, proposed by \citet{2018MNRAS.476.4403C}, whereupon we first expand instead in $\vr_1$ and $\vr_2$, then in the small angle $\theta$.

Whilst the explicit details of this can be found in Appendix \ref{appen: cw-math}, for now we note that the derivation proceeds by writing the bisector vector $\vec d = t\vr_1+(1-t)\vr_2$, where $t = r_2/(r_1+r_2)$. Expanding the spherical harmonic $Y_\ell^m(\hat{\vec d})$ in terms of $t\vr_1$ and $(1-t)\vr_2$ via the solid harmonic addition theorem \eqref{eq: solid-harmonic-addition} gives the finite series
\beq\label{eq: bisector-solid}
    Y_\ell^m(\hat{\vec d}) = \sum_{\lambda=0}^\ell \sum_{\mu=-\lambda}^\lambda A_{\ell m}^{\lambda\mu} \left(\frac{tr_1}{d}\right)^\ell Y_{\lambda}^\mu(\hr_1)Y_{\ell-\lambda}^{m-\mu}(\hr_2),
\eeq
with $d = |\vd|$ and coupling coefficient $A_{\ell m}^{\lambda\mu}$ defined in \eqref{eq: solid-harm-application}. Due to the $(tr_1/d)^{-\ell}$ factor, this is not directly separable in $\vr_1,\vr_2$; however, we can write
\beq
    \frac{tr_1}{d} = \left[2(1+\cos\phi)\right]^{-1/2},
\eeq
and expand in the small parameter $(1-\cos\phi)$, where $\cos\phi = \hr_1\cdot\hr_2$. From the geometry of Fig.\,\ref{fig: LoS-options}, it is clear that $\theta\approx\phi/2$ for $\Delta\ll r_1,r_2$, thus this expansion is effectively one in $\theta^2$ (since $1-\cos\phi\approx 2\theta^2$). Omitting lengthy algebra, this leads to the series expansion for $Y_\ell^m(\hat{\vec d})$:
\beq\label{eq: bisector-ylm-expan}
    \boxed{Y_\ell^m(\hat{\vec d}) = \sum_{\beta=0}^\infty \sum_{J_1=0}^{\ell+\beta}\sum_{J_2=0}^{\ell+\beta}\sum_{M=-J_1}^{J_1}B^{\beta,\ell m}_{J_1J_2M}Y_{J_1}^M(\hr_1)Y_{J_2}^{m-M}(\hr_2)}
\eeq
(cf.\,\ref{eq: Castorina-expan}), where the coupling coefficients $B^{\beta,\ell m}_{J_1J_2M}$ are defined in \eqref{eq: B-def}. These obey the properties
\beq\label{eq: Bk-properties}
    B^{\beta,\ell m}_{J_1J_2M} &=& 0 \text{ if }\ell+J_1+J_2 \text{ is odd}\\\nonumber
    B^{0,\ell m}_{J_1J_2M} =  2^{-\ell}A_{\ell m}^{J_1M}\delta^\mathrm{K}_{J_2(\ell-J_1)} &,& \quad B^{\beta, 00}_{J_1J_2M} = \sqrt{4\pi}\delta^\mathrm{K}_{J_10}\delta^\mathrm{K}_{J_20}\delta^\mathrm{K}_{M0}
\eeq
(cf.\,Appendix \ref{appen: cw-math}), with the latter indicating that $Y_0^0(\hd) = 1/\sqrt{4\pi}$, as expected.

Notably, \eqref{eq: bisector-ylm-expan} is an infinite expansion involving only separable functions of $\hr_1$ and $\hr_2$ (and no powers of $|\vr_i|$), which will allow for an efficient power spectrum estimator to be wrought. The approximation order is controlled by the value of $\beta$; fixing to $\beta \leq K$ expands the internal angle up to $(1-\cos\phi)^K$, which scales as $(\sqrt{2}\theta)^{2K}$. Unlike in the midpoint formalism, the expansion is automatically symmetric in $\phi$ and hence $\theta$, thus there is no need for the parity-even manipulations of \S\ref{subsec: mid-pk-even}. In this case, the symmetries of $B^{\beta,\ell m}_{J_1J_2M}$ require both odd and even $J_1, J_2$ (albeit with the restriction $J_1+J_2=\mathrm{even}$ for even $\ell$), unlike for the midpoint estimator.

As an example, we consider the expansion of $Y_2^1(\hd)$ at $\beta=0$ and $\beta=1$, \textit{i.e.} the $\mathcal{O}(\theta^0)$ and $\mathcal{O}(\theta^2)$ contributions. This yields
\beq
    \left.Y_2^1(\hat{\vec d})\right|_{\beta=0} &=& \textcolor{red}{\frac{\sqrt{\pi}}{6}\left[3Y_2^1(\hr_1)Y_0^0(\hr_2)+\sqrt{15}Y_1^1(\hr_1)Y_1^0(\hr_2)\right]}+(\vr_1\leftrightarrow\vr_2)\\\nonumber
    \left.Y_2^1(\hat{\vec d})\right|_{\beta=1} &=& \textcolor{darkgreen}{\frac{\sqrt{\pi}}{420}\left[35Y_2^1(\hr_1)Y_0^0(\hr_2)+5\sqrt{42}Y_3^2(\hr_1)Y_1^{-1}(\hr_2)+21\sqrt{15}Y_1^1(\hr_1)Y_1^0(\hr_2)-2\sqrt{210}Y_3^1(\hr_1)Y_1^0(\hr_2)\right.}\\\nonumber
    &&\,\textcolor{darkgreen}{\left.\quad\quad+3\sqrt{35}Y_3^0(\hr_1)Y_1^1(\hr_2)+7\sqrt{30}Y_2^2(\hr_1)Y_2^{-1}(\hr_2)-7\sqrt{5}Y_2^{1}(\hr_1)Y_2^0(\hr_2)\right]}+(\vr_1\leftrightarrow\vr_2)\,.
\eeq
Several points are of note. Firstly, in this formalism we have the same terms appearing at $\beta=0$ and $\beta=1$; for instance there is a $Y_2^1(\hr_1)Y_0^0(\hr_2)$ term sourced both at $\beta=0$ and all higher orders. This differs from the midpoint expansion of \S\ref{sec: midpoint-pk} in which the $\mathcal{O}(\theta^K)$ part contains a factor $\epsilon_i^K$. Secondly, setting $\beta=0$ does not recover the Yamamoto formalism (which would have $2\,Y_2^1(\hd) = Y_2^1(\hr_1)+Y_2^1(\hr_2)$). This is clear from the non-trivial $\beta=0$ coefficients in \eqref{eq: Bk-properties}, and indicates that the scheme includes a resummation of higher-order terms, due to the non-perturbative solid harmonic expansion carried out in \eqref{eq: bisector-solid}. Importantly, this implies that the midpoint formalism with $\alpha=2$ will not equal that for the bisector method with $\beta=1$, even though both are $\mathcal{O}(\theta_\max^2)$ (the order to which the two LoS definitions agree). The difference is due to different higher-order terms (and exists also between the midpoint expansion and its even-parity equivalent).

Given the simple nature of \eqref{eq: bisector-ylm-expan} (containing only spherical harmonics in $\hr_1$ and $\hr_2$), it is interesting to consider whether the above approach may be applied also to the midpoint estimator. In this case, we use $\vR = \vr_1+\vr_2$, with \eqref{eq: bisector-solid} becoming
\beq\label{eq: Castorina-tmp-midpoint}
    Y_\ell^m(\hat{\vec R}) &=& \sum_{\lambda=0}^\ell \sum_{\mu=-\lambda}^\lambda A_{\ell m}^{\lambda\mu} \left(\frac{r_1}{R}\right)^\ell \left(\frac{r_1}{r_2}\right)^{\lambda-\ell}Y_{\lambda}^\mu(\hr_1)Y_{\ell-\lambda}^{m-\mu}(\hr_2)\\\nonumber
    &=& \left[1+s^2+2s\cos\phi\right]^{-\ell/2}\sum_{\lambda=0}^\ell \sum_{\mu=-\lambda}^\lambda A_{\ell m}^{\lambda\mu}s^{\ell-\lambda} Y_{\lambda}^\mu(\hr_1)Y_{\ell-\lambda}^{m-\mu}(\hr_2),
\eeq
where $s = r_1/r_2$, using the definition $R^2 = r_1^2+r_2^2+2r_1r_2\cos\phi$. As before, this can be expanded around the point $\cos\phi = 1$, via
\beq
    \left[1+s^2+2s\cos\phi\right]^{-\ell/2} = \sum_{\beta=0}^\infty \binom{-\ell/2}{\beta}(1+s)^{-\ell-2\beta}(2s)^\beta(\cos\phi-1)^\beta,
\eeq    
provided that $(1+s)^2>4s$, \textit{i.e.} $s>0$. The difficulty arises from the additional factor of $s$; we require a separable perturbative expansion of $(1+s)^{-\ell-2\beta}$ yet $s$ is order unity. It is this complexity (not found in the bisector approach) that yields the $(\vD, \vr_1)$ separation of \S\ref{subsec: mid-pk-series} more useful for the midpoint LoS definition.

\subsection{Implementation}\label{subsec: pk-bis-imp}
By inserting the expansion of \eqref{eq: bisector-ylm-expan} into \eqref{eq: Pk-pairwise}, we obtain an estimator for the power spectrum in the bisector formalism:
\beq\label{eq: bisector-pk-fft}
    \hat{P}^\mathrm{bisector}_\ell(k) &=& \frac{4\pi}{V}\sum_{m=-\ell}^\ell \int_{\Omega_k}Y_\ell^{m*}(\hk)\int d\vr_1\,d\vr_2\,e^{-i\vk\cdot(\vr_2-\vr_1)}\delta(\vr_1)\delta(\vr_2)\sum_{kJ_1J_2M}B^{\beta, \ell m}_{J_1J_2M}Y_{J_1}^{M}(\hr_1)Y_{J_2}^{m-M}(\hr_2)\\\nonumber
    &=& \frac{4\pi}{V}\sum_{m=-\ell}^\ell \int_{\Omega_k}Y_{\ell}^{m*}(\hk)\sum_{\beta J_1J_2M}B^{\beta, \ell m}_{J_1J_2M}\left(\int d\vr_1\,e^{i\vk\cdot\vr_1}\delta(\vr_1)Y_{J_1}^M(\hr_1)\right)\left(\int d\vr_2\,\delta(\vr_2)Y_{J_2}^{m-M}(\hr_2)\right)\\\nonumber
\eeq
\beq
    \Rightarrow \boxed{\hat{P}^\mathrm{bisector}_\ell(k) = \frac{4\pi}{V}\sum_{m=-\ell}^{\ell}\int_{\Omega_k}Y_{\ell}^{m*}(\hk)\sum_{\beta J_1J_2M}B^{\beta, \ell m}_{J_1J_2M}\ft{Y_{J_1}^M\delta}(-\vk)\ft{Y_{J_2}^{m-M}\delta}(\vk).\nonumber}
\eeq
In the last line, we have written the estimator in terms of the Fourier transforms of $Y_{J}^M\delta$, resulting in a simple-to-implement estimator. This bears strong similarities to the Yamamoto approximation (but now involves two density-weighted spherical harmonics), and does not include any factors of $\epsilon_i=\Delta/(2r_i)$. In comparison to the midpoint estimator, the lack of $\epsilon_i$ terms has the advantage that we do not need to use a wider box-size relative to the standard estimators (cf.\,\S\ref{subsec: mid-pk-impl}).

Practically, one estimates the power spectra by first  computing $\ft{Y_J^M\delta}(\vk)$ for all (odd and even) $J$ up to $\ell_\mathrm{max}+\beta_\mathrm{max}$, giving a total of $(1+\ell_\mathrm{max}+\beta_\mathrm{max})(2+\ell_\mathrm{max}+\beta_\mathrm{max})/2$ terms, using the symmetry $\ft{Y_J^M\delta}(-\vk) = (-1)^M\ft{Y_J^{-M}\delta}^*(\vk)$. For $\ell_\mathrm{max} = 4$, this requires 15 (21) terms for $\beta_\mathrm{max} = 0$ ($\beta_\mathrm{max} = 1)$. For modest computational resources, it may be impractical to store all possible $\ft{Y_J^M\delta}$ grids, since each requires substantial memory. If these are computed separately, we note that the total number of relevant $Y_\ell^{m*}Y_{J_1}^MY_{J_2}^{m-M}$ combinations is significant; 20 for $\beta=0$ and 68 for $\beta=1$ at $\ell=2$. We caution that this approach may thus be quite computationally intensive.

\section{Bisector Formalism: 2PCF}\label{sec: bisector-2pcf}
We finally turn to the two-point correlation function in the bisector formalism. In this case, the angular dependence of the estimator \eqref{eq: 2pcf-pair} is given by $L_\ell(\hD\cdot\hd)$ where $\vd$ is the bisector vector as before. For this, we do not require a specialized perturbative expansion, instead expanding the Legendre polynomial via the addition theorem \eqref{eq: addition-theorem} and utilizing the expansion of the previous section \eqref{eq: bisector-ylm-expan}:
\beq\label{eq: bisector-leg-expan}
    \boxed{L_\ell(\hD\cdot\hd) = \frac{4\pi}{2\ell+1}\sum_{m=-\ell}^\ell Y_\ell^{m*}(\hD)\sum_{\beta=0}^\infty \sum_{J_1=0}^{\ell+\beta}\sum_{J_2=0}^{\ell+\beta}\sum_{M=-J_1}^{J_1}B^{\beta,\ell m}_{J_1J_2M}Y_{J_1}^M(\hr_1)Y_{J_2}^{m-M}(\hr_2).}
\eeq
Since we expand $Y_\ell^m(\hd)$ in terms of $\hr_1$ and $\hr_2$, contracting with $Y_\ell^m(\hD)$ does not simplify the coefficients (unlike that seen for the midpoint case in \S\ref{subsec: mid-2pcf-series}). Whilst this form is appealing since it involves no new derivations or powers of $\Delta/(2r_i)$, we note that, at least at leading order, it is possible to obtain a series expansion of $L_\ell(\hD\cdot\hd)$ in powers of $\epsilon_i$ and $L_\ell(\hD\cdot\hr_i)$ as in the midpoint formalism. This can be done by writing the Legendre polynomial as a power series then performing a Taylor expansion (as in \citealt{2015arXiv151004809S}). %, though it is more difficult to find a perturbative closed-form solution in this case.

To obtain an efficient 2PCF estimator, we insert \eqref{eq: bisector-leg-expan} into \eqref{eq: 2pcf-pair}, giving
\beq\label{eq: CW-2pcf}
    \hat{\xi}^\mathrm{bisector}_\ell(r) &=& \frac{4\pi}{V}\sum_{m=-\ell}^\ell\int d\vD\,Y_\ell^{m*}(\hD)\left[\frac{\delta_{\rm D}(r-\Delta)}{4\pi r^2}\right]\sum_{\beta J_1J_2M}B_{J_1J_2M}^{\beta, \ell m}\int d\vr_1\,\left[Y_{J_1}^M(\hr_1)\delta(\vr_1)\right]\left[Y_{J_2}^{m-M}(\widehat{\vr_1+\vD})\delta(\vr_1+\vD)\right],
\eeq
writing $\vr_2 = \vr_1+\vD$. As in \S\ref{sec: defs}, the $\vr_1$ integral is a convolution and can thus be computed via Fourier methods; specifically
\beq
    h_{J_1J_2M}^m(\vD) &\equiv& \int d\vr_1\,\left[Y_{J_1}^M(\hr_1)\delta(\vr_1)\right]\left[Y_{J_2}^{m-M}(\widehat{\vr_1+\vD})\delta(\vr_1+\vD)\right]\\\nonumber
    &=& \ift{\ft{Y_{J_1}^M\delta}(-\vk)\ft{Y_{J_2}^{m-M}\delta}(\vk)}(\vD).
\eeq
This gives the full form:
\beq
    \boxed{\hat{\xi}^\mathrm{bisector}_\ell(r) =\frac{4\pi}{V}\sum_{m=-\ell}^\ell\int d\vD\,Y_\ell^{m*}(\hD)\left[\frac{\delta_{\rm D}(r-\Delta)}{4\pi r^2}\right]\sum_{\beta J_1J_2M}B_{J_1J_2M}^{\beta, \ell m}h_{J_1J_2M}^m(\vD).}
\eeq
Whilst this does not require powers of $\epsilon_i$ (as in \S\ref{sec: midpoint-2pcf}), it is quite computationally expensive to implement since we require separate convolutions for each $\{J_1, J_2,m,M\}$ quartet; a total of $20$ ($68$) for $\beta = 0$ ($\beta=1$) at $\ell=2$, as before. This may be contrasted from the midpoint case \eqref{eq: xi-parity-all-estimator}, which requires estimation of the $g^\alpha_{JM}$ functions, depending only on a single total angular momentum. However, note that the summation over $\beta,J_1,J_2,M$ can be performed before the inverse transform, leading to significant expedition.

\section{Results}\label{sec: application}
Having established the existence of efficient pairwise power spectrum and 2PCF estimators, we are now ready to implement them. Below, we will consider their application on realistic survey, including a comparison between the two LoS schemes, before first discussing the convergence of the aforementioned series expansions.

\subsection{Convergence Tests}\label{subsec: convergence}
Three series expansions have been introduced in this work; one for each of $Y_\ell^m(\hR)$ and $L_\ell(\hD\cdot\hR)$ (\S\ref{subsec: mid-pk-series}\,\&\,\S\ref{subsec: mid-2pcf-series}), relevant to the midpoint estimators, and one for $Y_\ell^m(\hd)$ (\S\ref{subsec: pk-bis-series}) for the bisector estimators. To test these, we consider a simple scenario where we fix $\vR$ and vary $\vD$ such that we scan over the convergence parameter $\theta = \Delta/R$. We compute the three angular statistics using the associated $\vr_1,\vr_2$ vectors as a function of $\theta$, storing both the true value and the approximation at a given value of $\alpha$ or $\beta$. For the midpoint estimators, we consider both the standard expansions and those after the even-parity transformations of \S\ref{subsec: mid-pk-even}\,\&\,\S\ref{subsec: 2pcf-mid-even}.

The results are shown in Fig.\,\ref{fig: convergence-plots} for $\ell = 2$ and $4$. Notably, the fractional error in the spherical harmonic or Legendre polynomial generally decreases as we include more terms in the infinite series, indicating convergence for both the midpoint and bisector formalisms. As expected, the fractional error becomes larger as $\theta$ increases (and thus the survey becomes wider), and, for $\theta = 0.2$ (around twice that of the BOSS BAO scale), the even-parity midpoint expansion is non-convergent. The bisector expansions are found to be highly convergent here, as a consequence of their inherent partial resummations of higher-order terms (cf.\,\S\ref{subsec: mid-2pcf-series}). For the midpoint series, we find that the even parity expansions (involving only spherical harmonics of even order) give somewhat larger fractional errors than their all parity equivalents; this is unsurprising since (a) the even parity scheme has a stricter convergence criterion ($2\theta\ll1$ rather than $\theta\ll1$) and (b) each odd-parity term contributes to an infinite number of even-parity terms; our formalism only captures their contributions to even terms up to the given $\alpha_\mathrm{max}$, and will thus incur a larger error. That being said, we find the even parity expansions (which are cheaper to compute) to converge fairly well for $\theta\lesssim0.1$. As a point of comparison, the mock data-sets used in \S\ref{subsec: mock-surveys} have $\theta$\,$\sim$\,$0.1$ at the BAO peak. Finally, we consider the $\alpha_\mathrm{max} = 0$ intercept of midpoint plots. This is simply the Yamamoto approximation, and gives a useful indication of its intrinsic error. In particular, for $\ell = 2$ we find a fractional error of $0.5\%$ ($2.1\%$) for $\theta = 0.1$ ($\theta =0.2$) or $1.8\%$ ($7.5\%$) for $\ell=4$; not an insignificant error!

\begin{figure}%
    \centering
    \includegraphics[width=0.95\textwidth]{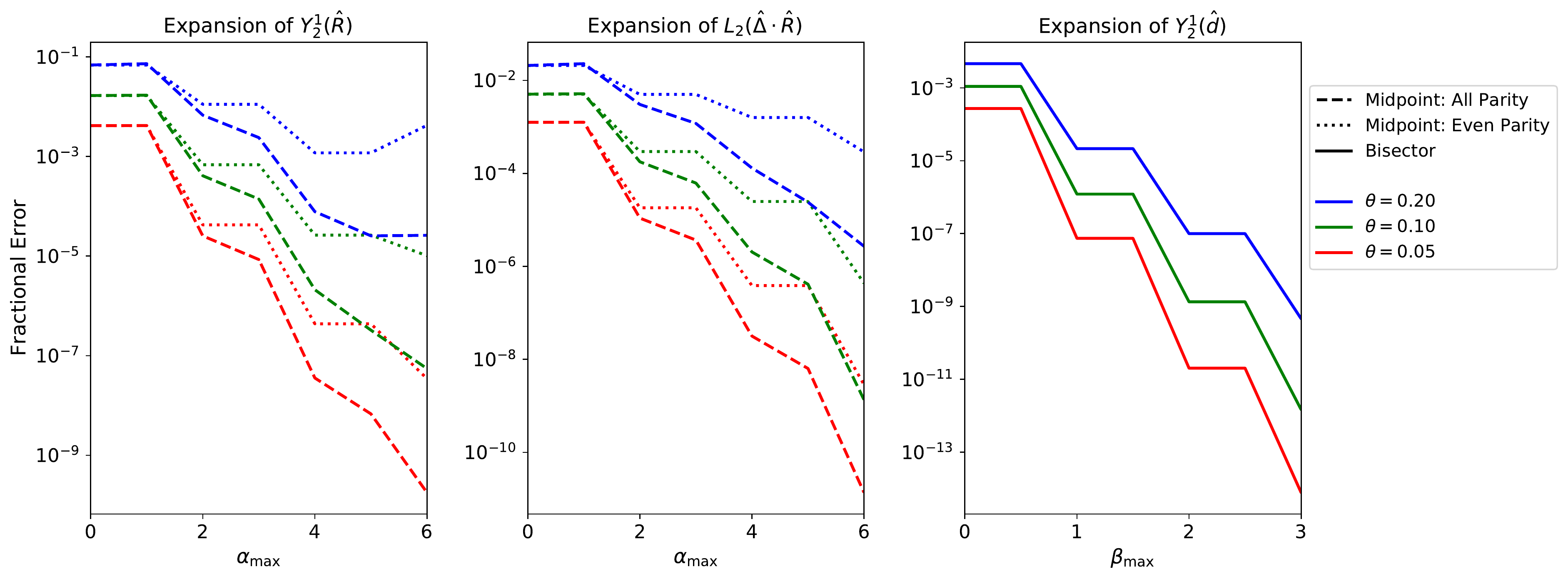}\\
    \includegraphics[width=0.95\textwidth]{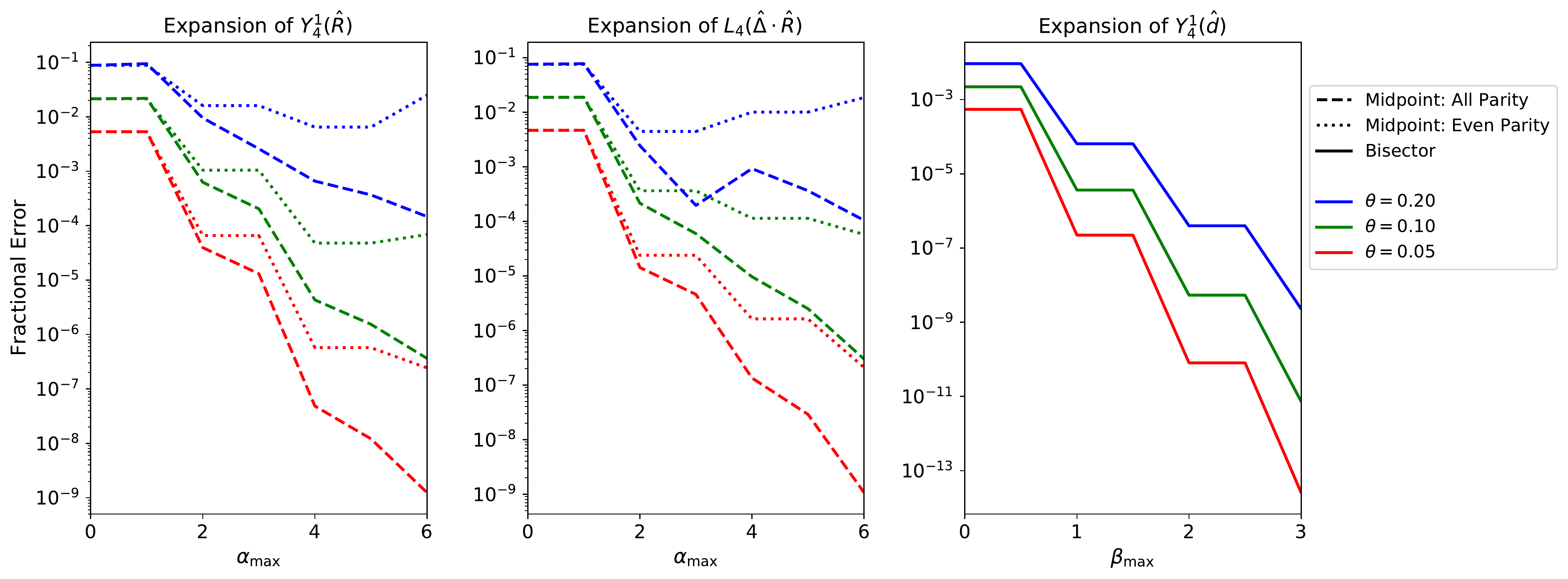}%
    \caption{Convergence plots for the series expansions used in this work. The left, center and right panels show the expansions of $Y_\ell^m(\hR)$, $L_\ell(\hD\cdot\hR)$ and $Y_\ell^m(\hd)$ respectively (\S\ref{subsec: mid-pk-series},\,\S\ref{subsec: mid-2pcf-series}\,\&\,\S\ref{subsec: pk-bis-series}), where $\hR$ is the direction vector of the galaxy midpoint, $\hd$ is the angle bisector and $\hD$ is the separation vector. Results are shown for both $\ell=2$ (top) and $\ell=4$ (bottom), and, for the midpoint expansions, include both the full and even-parity results of \S\ref{subsec: mid-pk-series}\,\&\,\S\ref{subsec: mid-pk-even}. In all cases, we plot the fractional error, defined as $|A^\mathrm{true}-A^\mathrm{approx}|/|A^\mathrm{true}|$ for (possibly complex) statistic $A$. Whilst we pick a single value of $m$ for the spherical harmonic plots, we have checked that this is representative of all $Y_\ell^m$ functions for the stated $\ell$. The horizontal axis shows the approximation order, with $\alpha=K$ ($\beta=K$) corresponding to $\theta^{K}$ ($\theta^{2K}$) corrections, where $\theta = \Delta/R$; the galaxy pair opening angle (with $\theta$\,$\sim$\,$ 0.1$ for the BOSS DR12 sample at the BAO scale). As the expansion parameter $\theta_\max$ becomes large, the convergence is reduced, and divergences occur. This is more prominent for the even-parity expansions, which are formally valid only for small $2\theta$. For the bisector cases, the $\alpha=0$ results give the Yamamoto approximation, whose error is significant at large $\theta$.}%
    \label{fig: convergence-plots}%
\end{figure}

\subsection{Application to Mock Galaxy Surveys}\label{subsec: mock-surveys}

To provide a practical demonstration of above algorithms, we apply them to a set of realistic mock galaxy catalogs. For this purpose, we use 24 MultiDark-\textsc{patchy} (hereafter `\textsc{patchy}') simulations\footnote{Publicly available at \href{https://data.sdss.org/sas/dr12/boss/lss/}{data.sdss.org/sas/dr12/boss/lss}.} \citep{2016MNRAS.460.1173R,2016MNRAS.456.4156K}, created for the analysis of the twelfth data-release (DR12) \citep{2017MNRAS.470.2617A} of the Baryon Oscillation Spectroscopic Survey (BOSS), part of SDSS-III \citep{2011AJ....142...72E,2016arXiv161100036D}. The data is split according to the criteria discussed in \citep{2017MNRAS.466.2242B}; here we specialize to the patch with the largest number density (and volume); the north Galactic cap in the redshift range $0.2<z<0.5$. This has total volume $1.46h^{-3}\mathrm{Gpc}^3$ and mean redshift $z = 0.38$. The simulations are generated with the cosmology $\{\Omega_m = 0.307115, \Omega_b = 0.048, \sigma_8 = 0.8288, h = 0.6777\}$, and each contains $\sim$\,$4.8\times 10^5$ simulated galaxies, alongside a random catalog $50\times$ larger. 

For each simulation, the mock galaxy positions are painted to a cuboidal grid using \textsc{nbodykit} \citep{2018AJ....156..160H}, using FKP weights \citep{1994ApJ...426...23F} and triangle-shaped-cell interpolation, with a fiducial value $\Omega_m = 0.31$ used to convert redshifts and angles into comoving coordinates. We use the same gridding parameters as in the final BOSS data release but double the box-size (as discussed in \S\ref{subsec: mid-pk-impl}), giving a Nyquist frequency $k_\mathrm{Nyq} = 0.3\hMpc$. All further computations are carried out in Python, making use of the \textsc{pyfftw} library to implement the algorithms of \S\ref{subsec: mid-pk-impl},\,\S\ref{subsec: mid-2pcf-imp},\,\S\ref{subsec: pk-bis-imp}\,\&\,\S\ref{sec: bisector-2pcf}. When binning spectra, we adopt the parameters $\{k_\mathrm{min}=0.01\hMpc, k_\mathrm{max}=0.25\hMpc, \Delta k = 0.01\hMpc\}$, $\{r_\mathrm{min} = 10\Mpch, r_\mathrm{max} = 190\Mpch, \Delta r = 10\Mpch\}$ in Fourier- and configuration-space respectively. In both cases, we use a maximum Legendre multipole of $\ell_\mathrm{max} = 4$.\footnote{\textsc{Jupyter} notebooks containing our analysis pipeline are publicly available at \href{https://github.com/oliverphilcox/BeyondYamamoto}{github.com/oliverphilcox/BeyondYamamoto}. Our implementation takes $\sim$\,$3$ hours to analyze the 2PCF and $P_\ell(k)$ of a single BOSS-like simulation on 4 Intel Skylake processors using $\alpha_\mathrm{max}=4$, $\beta_\mathrm{max}=2$, and $\ell_\mathrm{max}=4$. The analysis requires $\sim$\,$150\,\mathrm{GB}$ of memory, though this can be significantly reduced at the expense of longer computation time.}

\begin{figure}
    \centering
    \includegraphics[width=0.8\textwidth]{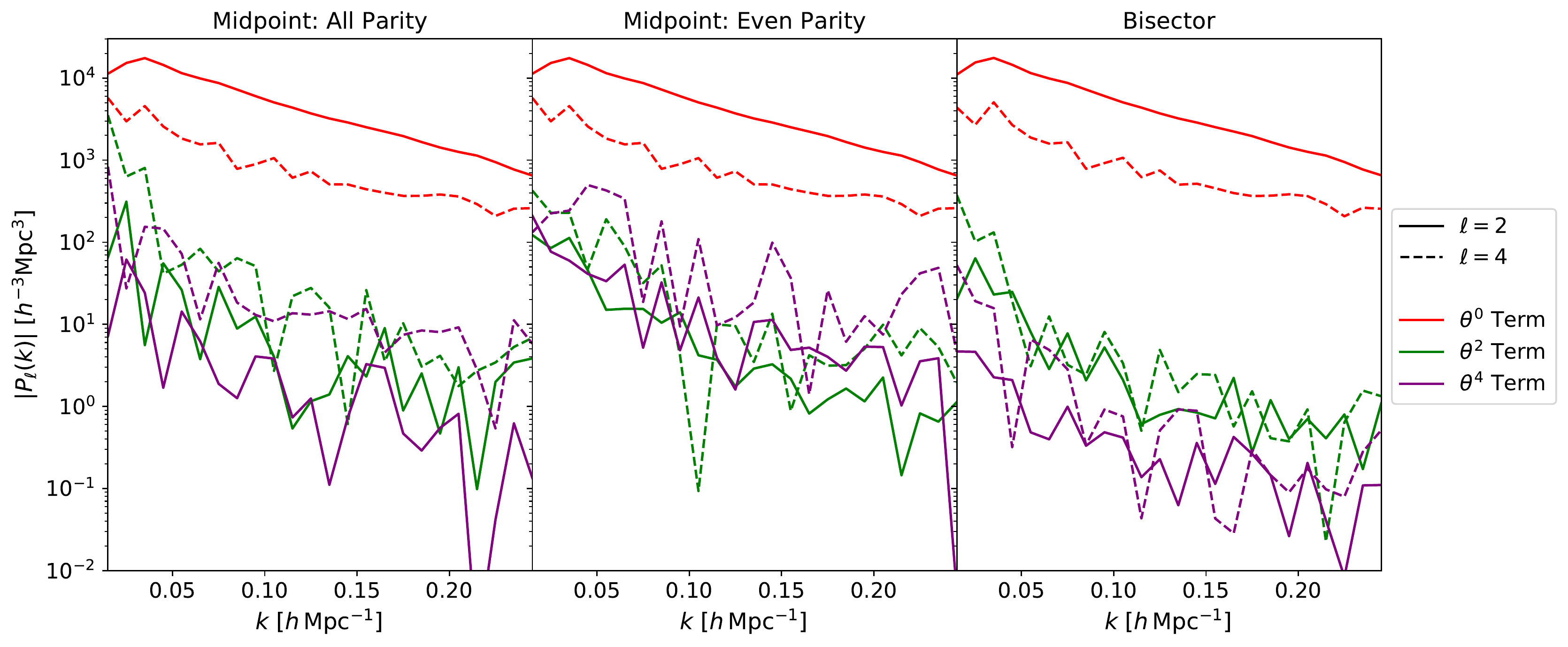}\\
    \includegraphics[width=0.8\textwidth]{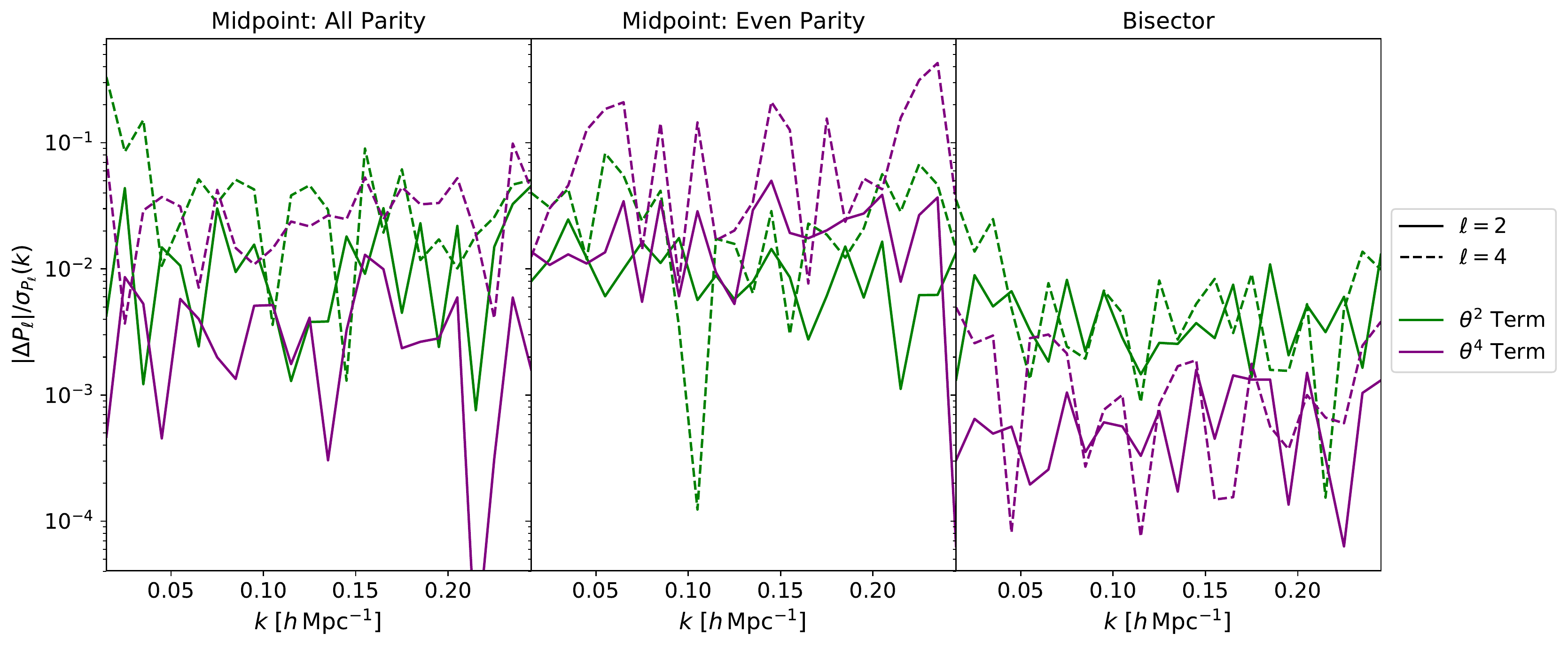}
    \caption{Contributions to the pairwise power spectra from the midpoint and bisector formalisms for a BOSS-like sample. The left (middle) panel shows the all parity (even parity) midpoint estimator of \S\ref{subsec: mid-pk-impl} (\S\ref{subsec: mid-pk-even}), whilst the right gives the bisector approach of \S\ref{subsec: pk-bis-imp}, based on \citet{2018MNRAS.476.4403C}. We show the absolute magnitudes of the relevant terms in the top panel, whilst the bottom gives the ratio to the standard deviation of the Yamamoto spectra. In all cases, we plot the mean of 24 \textsc{patchy} mocks and display the quadrupole (full lines) and hexadecapole (dashed lines), noting that the monopole receives no wide-angle corrections. The red, green and blue lines correspond to corrections of order $\theta_\max^0$, $\theta_\max^2$ and $\theta_\max^4$ in the characteristic angle $\theta_\max$ (\textit{i.e.} $\theta_\max^2$ corresponds to $\alpha=1$ plus $\alpha=2$ in the all parity midpoint formalism, $\alpha=2$ in the even parity midpoint approach, or $\beta=1$ for the bisector algorithm). For the midpoint spectra, the $\theta_\max^0$ term is the Yamamoto approximation, whilst for the bisector it is the sum of this and higher-order corrections, as discussed in \S\ref{subsec: pk-bis-series}. In general the corrections are a small fraction of the errorbars, and we find some convergence issues for the midpoint approximations, indicating that the convergence criterion $\theta_\max\ll1$ is not well-satisfied.}
    \label{fig: pk-corrections}
\end{figure}

\subsubsection{Power Spectrum}
Contributions to the power spectrum multipoles in the midpoint and bisector formalisms are shown in Fig.\,\ref{fig: pk-corrections}. The figure displays the contributions as a function of their order in the (assumed small) parameter $\theta_\max$, with, for example, the $\beta=1$ bisector piece giving a contribution which starts at $\mathcal{O}(\theta_\max^2)$. For the midpoint, the same order is obtained by summing the $\alpha=1$ and $\alpha=2$ pieces, or just from the $\alpha=2$ piece in the even parity formalism of \S\ref{subsec: mid-pk-even}. As previously mentioned, the $\mathcal{O}(\theta_\max^0)$ piece is equal to the Yamamoto spectrum in the midpoint formalism, but includes higher-order corrections for the bisector LoS definition. Our first note is that, in all cases, the leading order contribution is subdominant, indicating that the Yamamoto approximation is, in general, a fair approximation for BOSS. Relative to the errorbars, any wide-angle corrections represent a few-percent correction at best for this survey.\footnote{\resub{The fractional error is roughly scale-invariant; the reason for this is not obvious, due to the inherent scale mixing for the power spectrum compared to the 2PCF. This observation is consistent with the conclusions of \citet{2018MNRAS.476.4403C} however.}} Whilst they seem somewhat smaller for the bisector LoS, this is primarily due to the additional $\mathcal{O}(\theta_\max^2)$ terms absorbed into the $\beta=0$ estimator. For a larger volume survey, the statistical errors shrink, thus these wide-angle effects will grow in importance. Secondly, it is clear that, in the bisector formalism, the quartic terms are of significantly reduced magnitude compared to those at quadratic order; this indicates that the underlying perturbative expansion is well convergent, and matches that found in Fig.\,\ref{fig: convergence-plots}. For the all parity midpoint case, the convergence is more tenuous; though the results appear convergent for the quadrupole, the case is less clear for the hexadecapole, and the even parity estimators show non-convergence in both cases, with the $\alpha=4$ term being larger than that with $\alpha=2$. This may be rationalized by noting that convergence condition in \S\ref{subsec: mid-pk-even} is stricter than that of \S\ref{subsec: mid-2pcf-series} ($2\theta_\max\ll1$ compared with $\theta_\max\ll1$), leading to the poorer behavior. Overall, the plot indicates that our series are only marginally useful for a survey of BOSS width. 

\begin{figure}
    \centering
    \includegraphics[width=0.8\textwidth]{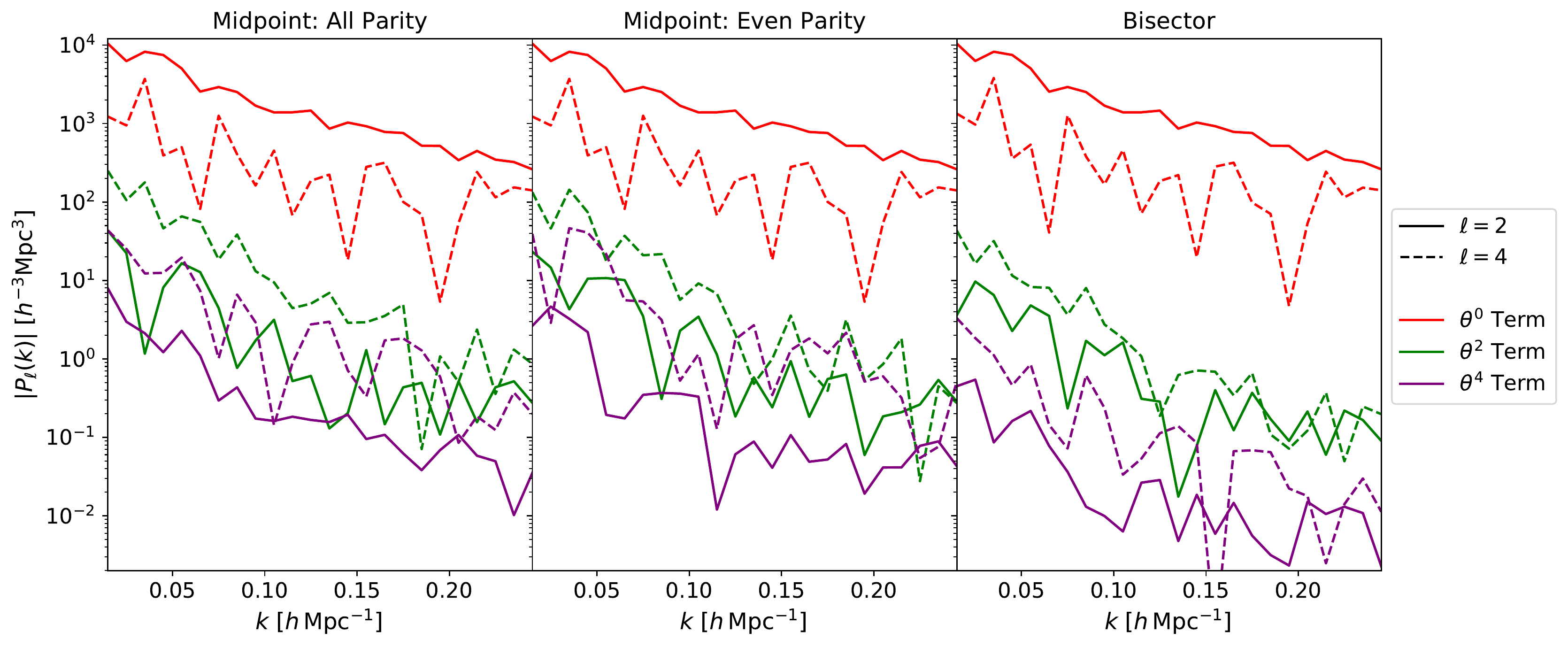}\\
    \includegraphics[width=0.8\textwidth]{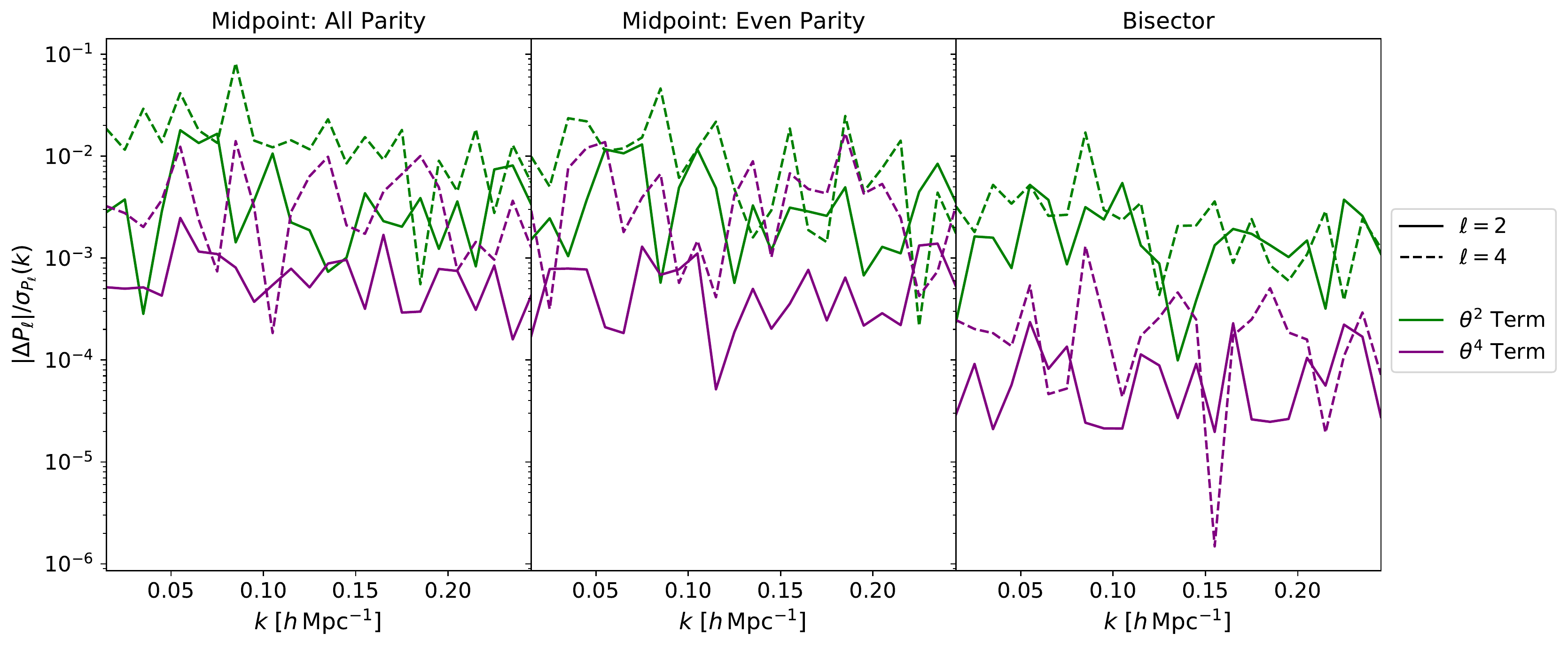}
    \caption{As Fig.\,\ref{fig: pk-corrections} but for a BOSS-like galaxy sample shifted radially outwards by an additional $2000\Mpch$ to reduce the $\theta_\max$ parameter and enforce greater convergence. In this instance, we use the mean of 5 \textsc{patchy} catalogs. As expected, the higher-order contributions are suppressed relative to Fig.\,\ref{fig: pk-corrections}, and the expansion is significantly more convergent (in the sense that the $\mathcal{O}(\theta_\max^4)$ term is subdominant to the $\mathcal{O}(\theta_\max^2)$ piece).}
    \label{fig: shift-pk-corrections}
\end{figure}

To better understand this, we consider a simple test; shifting the \textsc{patchy} mock data radially outwards by $2000\Mpch$, \resub{from its initial position, centered at $\sim$\,$1000\Mpch$.}. This increases the mean galaxy distance by a factor $\sim$\,$3$, thus reducing $\theta_\max$ by the same factor. \resub{This test is of relevance for future surveys such as DESI and Euclid which generally focus on higher redshifts than BOSS.} Resulting power spectrum contributions are shown in Fig.\,\ref{fig: shift-pk-corrections}, and match our expectations; we see a greater difference between the $\mathcal{O}(\theta_\max^2)$ and $\mathcal{O}(\theta_\max^4)$ contributions than before, and significantly improved convergence for all estimators. Clearly then, our estimators will perform better in regimes where the characteristic angle $\theta_\max$ is smaller. Even in this example, the systematic errors are at the percent level (and again artificially smaller for the bisector formalism, due to the resummation of higher orders into the zeroth-order term), which may become important for future surveys with \resub{far smaller statistical errors}.

\subsubsection{2PCF}

\begin{figure}
    \centering
    \includegraphics[width=0.8\textwidth]{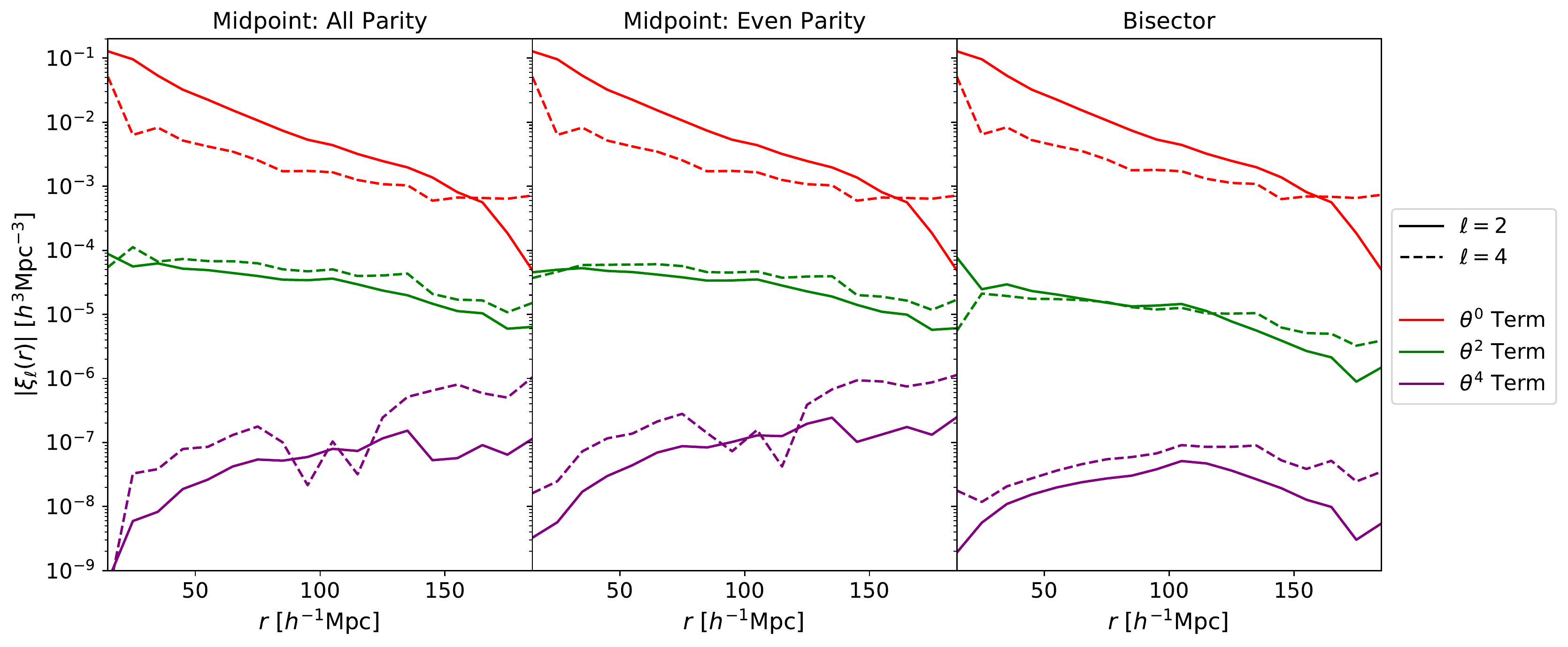}\\
    \includegraphics[width=0.8\textwidth]{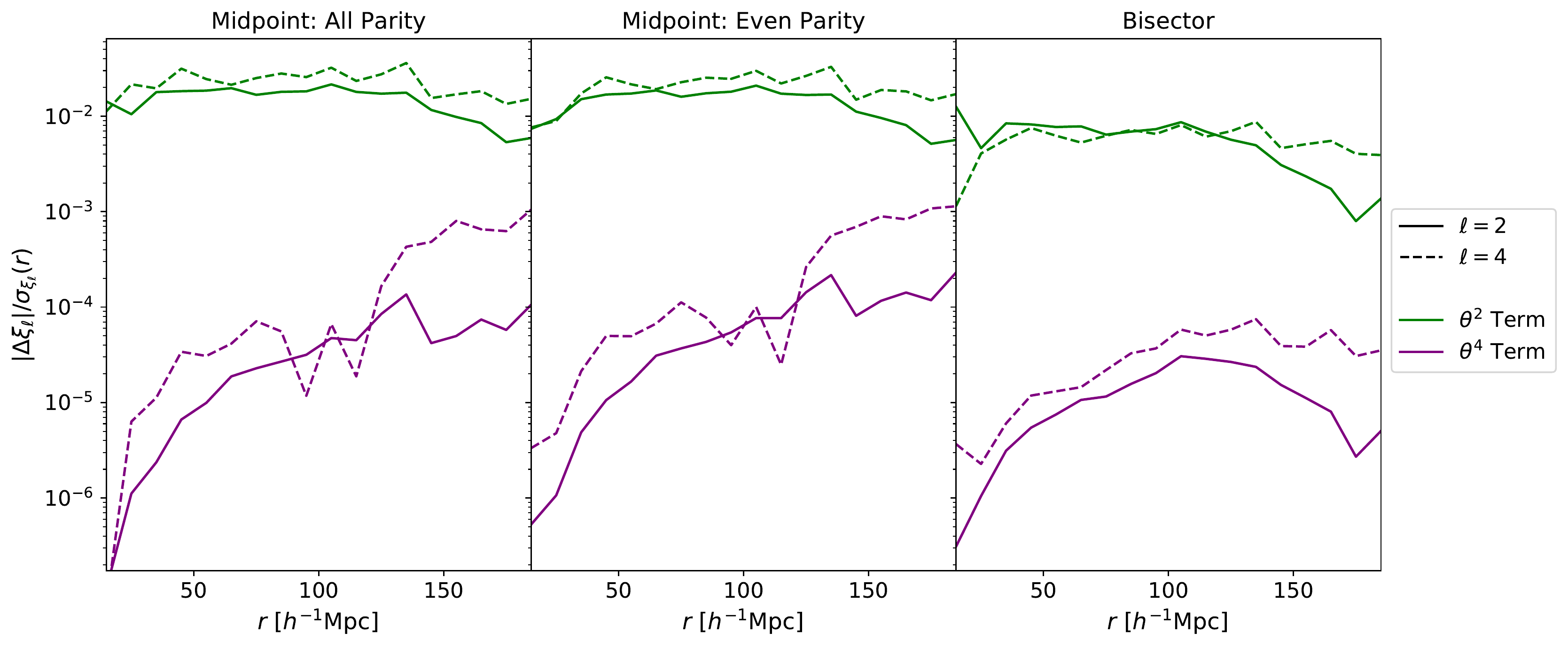}
    \caption{As Fig.\,\ref{fig: pk-corrections} but for the two-point correlation function multipoles. In this case, the hierarchy of terms is much clearer, with higher-order effects becoming important only at large $r$. For BOSS volumes, the quadratic-order correction is $\sim$\,$1\%$ of the error bar for both the monopole and quadrupole, though this will increase with the survey volume. We find good convergence of both the midpoint and bisector formalism on all scales tested. Note that we do not perform any window-function correction on these 2PCF measurements.}
    \label{fig: xi-corrections}
\end{figure}

Fig.\,\ref{fig: xi-corrections} displays the analogous results for the 2PCF multipoles of the (unshifted) \textsc{patchy} mocks. We reiterate that these do not include window-function corrections for simplicity.\footnote{The method of \citet{2016MNRAS.455L..31S} gives a straightforward approach by which to include these.} The 2PCF presents a very different story to the power spectrum. We see a clear demarcation of the expansion orders, with a high degree of convergence seen for the quadrupole and hexadecapole on all scales tested for both midpoint and bisector estimates. At larger radii the fractional contribution of the higher-order terms increases, and, extrapolating by eye, one would expect the expansions to break down around $300-400\Mpch$. These results clearly demonstrate that our expansions are performing as expected. As for the power spectrum, the post-Yamamoto corrections are small; $\sim$\,$1\%$ of the error bars on all scales, with a slight reduction at high radius (whereupon cosmic variance dominates).

The marked difference between the 2PCF and power spectrum begs the question: why does our power spectrum algorithm fare so much worse? Given that we observe similar behavior in the midpoint and bisector formalisms (with a very different expansion scheme), it is unlikely to stem from some algebraic fault. Our justification for this relies on the following observations: (1) the fractional contribution of the higher-order terms to the 2PCF increases as a function of $r$, becoming unity on scales comparable with the survey width, and (2) the power spectrum is an (Bessel function weighted) integral over the 2PCF. The scales on which the small-$\theta$ approximation is most tenuous thus have an impact on all the $k$-modes of the power spectrum. This is clearly seen from the (more convergent) bisector formalism; whilst the ratio of quartic to quadratic contribution strongly increases as a function of scale for the 2PCF, it is roughly constant for $P_\ell(k)$. As previously noted, for a convergent series expansion, the power spectrum requires the \textit{survey} opening angle be small, rather than just that of the maximum radial scale considered. This being said, we conclude that the algorithms developed herein are of particular use for the 2PCF, and may be applied also to the power spectrum, though the BOSS width lies in the limit of their convergence.

\subsection{Comparison of Methods}\label{subsec: method-comparison}
Given that the bisector and midpoint are both valid LoS definitions at $\mathcal{O}(\theta_\max^2)$ it is important to compare their utility in the context of the algorithms presented in this work. Below we list several differences, both in terms of efficiency and accuracy of their associated power spectrum and 2PCFs.
\begin{itemize}
    \item \textbf{Memory Requirements}: A na\"ive implementation of the bisector formalism requires holding in memory all possible $\ft{Y_J^M\delta}(\vk)$ fields up to a maximum multipole $\ell_\mathrm{max}+\beta_\mathrm{max}$ (such that we can take their outer product); a total of 21 for the $\mathcal{O}(\theta_\max^2)$ correction with $\ell_\mathrm{max}=4$. For a modest FFT grid of size $512^3$ in double precision, this requires $60\,\mathrm{GB}$ of memory, which is infeasible on many computer architectures. Whilst one could compute each function `on-the-fly', this will be significantly slower. For the same order of approximation, we require 28 $g^\alpha_{JM}$ functions (using even multipoles up to $\ell_\mathrm{max}+\alpha$), but, since no outer product is required, only one needs to be held in memory at any time. The most efficient CPU-wise implementation of the midpoint algorithm thus requires significantly less memory.
    \item \textbf{Computation Time}: An efficient metric with which to judge the runtime of the two approaches is the number of FFTs required at a given order. Assuming $\ell_\mathrm{max}=4$ and quadratic corrections, the midpoint $P_\ell(k)$ algorithm requires $70$ FFTs (two for each $g^\alpha_{JM}$ function, one for $\delta^*(\vk)$ and an addition 13 for transforming into $\vk$-space), whilst the bisector approach needs just $28$ (one per $\ft{Y_J^M\delta}(\vk)$ evaluation). The bisector method is thus somewhat faster, though this holds only if high-memory computational resources are available. For the 2PCF, the conclusion is similar, with the midpoint approach requiring $56$ FFTs, and the bisector $29$ (if memory is no concern).
    \item \textbf{Convergence}: As demonstrated in previous sections, the bisector $P_\ell(k)$ algorithms exhibit stronger convergence than those using the midpoint LoS. This is particularly true for the even-parity approach, which breaks down on BOSS survey scales (also indicating that the Yamamoto approximation works poorly there). For the 2PCF, all approaches converge well on scales of interest.
    \item \textbf{Grid Size}: As discussed in \S\ref{subsec: mid-pk-impl}, the midpoint power spectrum algorithm requires the particles to be placed on an FFT grid at least twice as wide as the survey itself to avoid errors. This requires a finer cell-size for the same Nyquist frequency, and thus slower (and more expensive) computaiton. This is not a concern for the bisector formalism or the 2PCF algorithms. 
    \item \textbf{Zeroth-Order Contribution}: The midpoint formalisms have the useful property that the zeroth-order ($\alpha=0$) contribution is equal to the well-known Yamamoto approximation. In contrast, the $\beta=0$ bisector algorithms discussed herein have a more complicated form. This is additionally slower to compute, involving four pairs of spherical harmonics for each $Y_\ell^{m*}(\hk)$ function rather than one.
    \item \textbf{Legendre Expansion}: The midpoint 2PCF algorithm can be simply expressed in terms of Legendre multipoles \eqref{eq: xi-midpoint-tmp} unlike the bisector approach, which must use spherical harmonics.\footnote{An analogous expression for the bisector is possible (and discussed in \citealt{2015arXiv151004809S}), but requires a different series expansion.} This is far simpler to interpret, and requires significantly fewer coupling coefficients. 
\end{itemize}

Overall, it is clear that both approaches can be implemented in an efficient manner, and give sensible results if the characteristic size is not too large. The optimal choice of LoS is left to the user, following the above considerations. In general, both approaches agree at $\mathcal{O}(\theta_\max^2)$ thus this choice is not of particular importance, particularly since any single LoS definition necessarily incurs an $\mathcal{O}(\theta_\max^4)$ error.

\section{Summary and Outlook}\label{sec: conclusion}
This work has presented efficient and practical algorithms for computing the two-point correlators using pairwise lines-of-sight; an extension to the usual single-particle Yamamoto approximation. Considering both the bisector and midpoint angle definitions, we have shown how convergent series expansions may be used to write the 2PCF and power spectrum estimators in a form allowing for implementation via FFTs. Any such correction is a function of the characteristic size, $\theta_\max$, which we have treated as a perturbation variable. Utilizing newly derived spherical harmonic and Legendre polynomial shift theorems, we have computed the midpoint corrections at arbitrary order in $\theta_\max$, paying close attention to the existence (and excision) of odd-parity terms. For the bisector LoS definition, the work of \citet{2018MNRAS.476.4403C} has been extended, including higher-order corrections and an efficient 2PCF algorithm. To demonstrate our approach, the methodology has been applied to a set of realistic galaxy catalogs, and shown to give good results for the 2PCF. For the power spectrum, the bisector approach is similarly effective, though the midpoint algorithms suffer somewhat with convergence issues as the expansion variable becomes large. Such series expansions perform best when the angular survey size is relatively small. In the opposing limit, convergence is difficult to achieve, yet also the Yamamoto approximation itself is poor.

For BOSS, the size of the post-Yamamoto effects is small; $\sim$\,$1\%$ of the statistical error. For future surveys such as DESI, the errorbars shrink, though the mean redshift also increases. Following \citet[Fig.\,6]{2018MNRAS.476.4403C}, we forecast that the estimator-induced wide angle effects discussed herein can be marginally important, especially on larger scales than considered herein. Since much information regarding primordial non-Gaussianity is found at low-$k$, implementing estimators beyond the Yamamoto approximation should be seriously considered. As shown above, there is freedom in how best to do this, since both the midpoint and bisector approaches fix the $\mathcal{O}(\theta_\max^2)$ systematic, but differ at $\mathcal{O}(\theta_\max^4)$. Here, we find little preference for one over the other; for efficient computation they require somewhat different algorithms, and the bisector approach fares a little better for the power spectrum, yet its memory consumption is significant.

Pairwise lines-of-sight are not the perfect solution however. To fully extract information from the large-scale modes, we ought to parametrize the 2PCF by \textit{two} lines-of-sight \citep[e.g.,][]{2008MNRAS.389..292P,2015MNRAS.447.1789Y}, though this presents difficulties since the power spectrum becomes ill-defined and the dimensionality increases. For surveys below $\sim$\,$10\degree$, the pairwise approximations are valid \citep{2015MNRAS.452.3704S}, yet we should bear in mind their sub-optimality for future surveys such as SPHEREx \citep{2014arXiv1412.4872D}. Even with improved estimators, it is necessary to model the wide-angle effects arising from other sources such as the galaxy selection function, and much work has been done to achieve this goal. Indeed, it may prove more straightforward to forward-model also the post-Yamamoto effects, and use a simpler estimator; given the results of this work, there is little justification for this on efficiency grounds. On a more philosophical note, it is interesting to see how even on the largest scales, where the Universe is most linear, the two-point function is still highly non-trivial. Despite being the most basic cosmological observable, the two-point function still has secrets up its sleeve.

\section*{Acknowledgements}
We thank Emanuele Castorina, Karolina Garcia, David Spergel, and \resub{Martin White} for insightful discussions. \resub{We are additionally grateful to the anonymous referee for an insightful report.} OHEP acknowledges funding from the WFIRST program through NNG26PJ30C and NNN12AA01C. The authors are pleased to acknowledge that the work reported on in this paper was substantially performed using the Princeton Research Computing resources at Princeton University which is consortium of groups led by the Princeton Institute for Computational Science and Engineering (PICSciE) and Office of Information Technology's Research Computing.

%%%%%%%%%%%%%%%%%%%%%%%%%%%%%%%%%%%%%%%%%%%%%%%%%%

%%%%%%%%%%%%%%%%% APPENDICES %%%%%%%%%%%%%%%%%%%%%

\appendix

\section{Useful Mathematical Relations}\label{appen: math-101}
We present a number of mathematical results used in this work. Firstly, we give the explicit form for $L_\ell(\mu)$ as a finite polynomial in $\mu$ (with $|\mu|\leq 1$):
\beq\label{eq: Leg-polynomial-def}
        L_\ell(\mu) &=& \sum_{n=0}^{\floor{\ell/2}} c_\ell^n\times\mu^{\ell-2n},\quad\,c_\ell^n = \frac{(-1)^n}{2^\ell}\binom{\ell}{n}\binom{2\ell-2n}{\ell} \equiv \frac{(-1)^n (2\ell-2n)!}{2^\ell n!(\ell-n)!(\ell-2n)!}
\eeq
\citep[Eq.\,22.3.8]{abramowitz+stegun}, where the $2\times1$ vectors are binomial coefficients and $\floor{\ell/2}$ indicates the largest integer $\leq \ell/2$. The inverse relation also proves useful:
\beq\label{eq: Leg-polynomial-inv-def}
    \mu^n = \sum_{\ell=n\downarrow2}\tilde{c}^\ell_n\times L_\ell(\mu),\quad\,\tilde{c}_n^\ell = \frac{(2\ell+1)}{2^{n+1}}\frac{\sqrt{\pi}\,\Gamma(1+n)}{\Gamma\left((n-\ell)/2+1\right)\Gamma\left((n+\ell+3)/2\right)}\equiv \frac{(2 \ell+1) n! (\ell+n+2)!!}{(\ell+n+2)!(n-\ell)!!}
\eeq
(derived from \citealt[Eq.\,7.126.1]{GandR} invoking Legendre polynomial orthogonality), where the summation is over all $\ell$ downwards from $n$ in steps of two, and we use the double factorial $n!! \equiv n(n-2)(n-4)...$ for positive integer $n$. Note that Legendre polynomials of even (odd) order depend only on even (odd) powers of $\mu$ (and \textit{vice versa}). To derive the simplified coefficients on the right-hand-side we have assumed $n$ to be integral, and noted that the summation rules imply that $\ell$, $n$ have the same sign. Inserting \eqref{eq: Leg-polynomial-inv-def} into \eqref{eq: Leg-polynomial-def} and using orthogonality gives the relation
\beq\label{eq: leg-forward-reverse-relation}
    &&\,L_\ell(\mu) = \sum_{n=0}^{\floor{\ell/2}}\sum_{L=(\ell-2n)\downarrow2}c_\ell^n\tilde{c}^{L}_{\ell-2n}L_L(\mu) \,\Rightarrow\, \sum_{n=0}^{\floor{\ell/2}}c_\ell^n\tilde{c}^L_{\ell-2n} = \delta^\mathrm{K}_{\ell L},
\eeq
where $\delta^\mathrm{K}$ is the Kronecker delta.

Legendre polynomials have the generating function
\beq\label{eq: leg-generator}
    \left[1-2st+t^2\right]^{-1/2} = \sum_{N=0}^\infty L_L(s) t^N
\eeq
\citep[Eq.\,8.921]{GandR}, and may related to spherical harmonics via the \textit{addition theorem}:
\beq\label{eq: addition-theorem}
    L_\ell(\hx\cdot\hy) = \frac{4\pi}{2\ell+1}\sum_{m=-\ell}^\ell Y_\ell^m(\hx)Y_\ell^{m*}(\hy)
\eeq
\citep[Eq.\,14.30.9]{nist_dlmf}, where $^*$ indicates a complex conjugate and $Y_\ell^m$ is the spherical harmonic of order $(\ell,m)$, which obeys the symmetries $Y_\ell^m(-\hx) = (-1)^\ell Y_\ell^m(\hx)$, $Y_\ell^{m*}(\hx) = (-1)^m Y_\ell^{-m}(\hx)$ and $Y_0^0 = (4\pi)^{-1/2}$. Furthermore, the spherical harmonics are orthonormal:
\beq\label{eq: spherical-harmonic-orthog}
    \int d\Omega_x Y_\ell^{m}(\hx)Y_{\ell'}^{m'*}(\hx) = \delta^\mathrm{K}_{\ell\ell'}\delta^\mathrm{K}_{mm'}
\eeq
\cite[Eq.\,14.30.8]{nist_dlmf}. An important relation is the \textit{product-to-sum} rule for spherical harmonics:
\beq\label{eq: spherical-harmonic-prod-to-sum}
    Y_{L_1}^{M_1}(\hx)Y_{L_2}^{M_2}(\hx) = \sum_{L_3=|L_1-L_2|}^{L_1+L_2}\sum_{M_3=-L_3}^{L_3}\mathcal{G}_{L_1L_2L_3}^{M_1M_2(-M_3)}(-1)^{M_3}Y_{L_3}^{M_3}(\hx),
\eeq
where $\mathcal{G}$ is the Gaunt integral obeying several selection rules including $M_1+M_2+M_3=0$. This can be derived from spherical harmonic orthogonality \eqref{eq: spherical-harmonic-orthog} and the definition of $\mathcal{G}$ as the integral over three spherical harmonics \citep[Eq.\,34.3.22]{nist_dlmf}. The Gaunt factor may be written explicitly in terms of Wigner 3-$j$ symbols as
\beq\label{eq: Gaunt-3j-def}
    \mathcal{G}^{\ell_1\ell_2\ell_3}_{m_1m_2m_3} = \left(\frac{(2\ell_1+1)(2\ell_2+1)(2\ell_3+1)}{4\pi}\right)^{1/2}\tj{\ell_1}{\ell_2}{\ell_3}{m_1}{m_2}{m_3}\tjo{\ell_1}{\ell_2}{\ell_3},
\eeq
\citep[Eq.\,34.3.22]{nist_dlmf}.

Closely related to spherical harmonics are the regular \textit{solid harmonics}, defined by
\beq\label{eq: solid-harmonic-def}
    R_\ell^m(\vr) = \sqrt{\frac{4\pi}{2\ell+1}}r^\ell Y_\ell^m(\hr).
\eeq
These obey an addition theorem:
\beq\label{eq: solid-harmonic-addition}
    R_\ell^m(\vr+\va) = \sum_{\lambda=0}^\ell \sum_{\mu=-\lambda}^\lambda \binom{\ell+m}{\lambda+\mu}^{1/2}\binom{\ell-m}{\lambda-\mu}^{1/2} R_{\lambda}^\mu(\vr)R_{\ell-\lambda}^{m-\mu}(\va)
\eeq
\citep{Tough_1977}, which is a finite series.

The Rayleigh plane wave expansion gives
\beq\label{eq: plane-wave}
    e^{i\vx\cdot\vy} = \sum_{\ell=0}^\infty i^\ell(2\ell+1)j_\ell(xy)L_\ell(\hx\cdot\hy) = 4\pi\sum_{\ell=0}^\infty\sum_{m=-\ell}^\ell i^\ell j_\ell(xy)Y_\ell^m(\hx)Y_\ell^{m*}(\hy)
\eeq
\citep[Eq.\,16.63]{arfken2013mathematical}, where we have used \eqref{eq: addition-theorem} to arrive at the second equality. Using this, we can prove the following identity:
\beq\label{eq: Omega-k-Leg-int}
    \int d\Omega_x\,e^{i\vx\cdot\vy}L_\ell(\vx\cdot\vz) &=& 4\pi \sum_{\ell'=0}\sum_{m'=-\ell'}^{\ell'}i^{\ell'}j_{\ell'}(xy)\int d\Omega_x\,Y_{\ell'}^{m'}(\hx)Y_{\ell'}^{m'*}(\hy)\frac{4\pi}{2\ell+1}\sum_{m=-\ell}^\ell Y_\ell^m(\hat{\vec z})Y_\ell^{m*}(\hx)\\\nonumber
    &=& 4\pi\,i^\ell j_\ell(xy)\frac{4\pi}{2\ell+1}\sum_{m=-\ell}^\ell Y_{\ell}^{m*}(\hy)Y_{\ell}^{m*}(\hat{\vec z}) \equiv 4\pi\,i^\ell j_\ell(xy)L_\ell(\hy\cdot\hat{\vec z}),
\eeq
using \eqref{eq: addition-theorem}\,\&\,\eqref{eq: plane-wave} in the first line, and \eqref{eq: spherical-harmonic-orthog} to obtain the second.

An additional class of orthogonal polynomials are Gegenbauer (or ultra-spherical) polynomials, with the generating function
\beq\label{eq: gegenbauer-generator}
    \left[1-2st+t^2\right]^{-\alpha} = \sum_{N=0}^\infty C_{N}^{(\alpha)}(s)\,t^N
\eeq
\citep[Eq.\,8.930]{GandR}, where $C_{N}^{(\alpha)}$ is the Gegenbauer polynomial (for integer order $N$), and $\alpha=1/2$ recovers the Legendre polynomials. These have the explicit form
\beq\label{eq: gegenbauer-explicit}
    C_N^{(\alpha)}(s) = \sum_{k=0}^{\floor{N/2}}C_{Nk}^{\alpha}\times(2s)^{N-2k}, \quad\, C_{Nk}^{\alpha} = (-1)^k\frac{\Gamma(\alpha+N-k)}{\Gamma(\alpha)k!(N-2k)!}
\eeq
\citep[Eq.\,22.3.4]{abramowitz+stegun}, with the special case of $C^0_{Nk}=\delta^\mathrm{K}_{N0}$. Note that this is a sum of even (odd) powers of $s$ for even (odd) $N$, as for the Legendre polynomials.

\section{Spherical Harmonic Shift Theorem}\label{appen: spherical-harmonic-shift-theorem}
\subsection{Derivation}\label{appen: sph-shift-deriv}
We present the derivation of a useful series expansion for spherical harmonics.\footnote{This is equivalent to that introduced in \citet{2020arXiv201103503G} for the three-point correlation function except with one position vector set to zero, affording significant simplification.} First, we consider the function $Y_\ell^m(\widehat{\va+\vA})$ for $\epsilon\equiv a/A\ll 1$. Converting this into the solid harmonic $R_\ell^m(\va+\vA)$ then using the addition theorem \eqref{eq: solid-harmonic-addition}, we obtain
\beq\label{eq: solid-harm-application}
    Y_{\ell}^m(\widehat{\va+\vA}) &=& \sum_{\lambda=0}^\ell \sum_{\mu=-\lambda}^\lambda \sqrt{\frac{4\pi(2\ell+1)}{(2\lambda+1)(2\ell-2\lambda+1)}} \binom{\ell+m}{\lambda+\mu}^{1/2}\binom{\ell-m}{\lambda-\mu}^{1/2}\frac{a^\lambda A^{\ell-\lambda}}{|\va+\vA|^{\ell}} Y_{\lambda}^\mu(\ha)Y_{\ell-\lambda}^{m-\mu}(\hA)\\\nonumber
    &\equiv&\sum_{\lambda=0}^\ell\sum_{\mu=-\lambda}^\lambda A_{\ell m}^{\lambda \mu}\frac{\epsilon^\lambda}{|\hA+\epsilon \ha|^\ell}Y_{\lambda}^\mu(\ha)Y_{\ell-\lambda}^{m-\mu}(\hA),
\eeq
defining the coefficients $A_{\ell m}^{\lambda \mu}$ in the second line. This is a finite expansion, allowing the composite spherical harmonic to be expressed in terms of the harmonics of $\ha$ and $\hA$. However, due to the angular dependence of $|\hA+\epsilon\ha|^\ell$, we require an infinite expansion to separate $\va$ and $\vA$ fully. To proceed, we recognize that the denominator is the generating function for a Gegenbauer series \eqref{eq: gegenbauer-generator} with $\alpha=\ell/2$, $t=\epsilon$, $s=-\ha\cdot\hA$:
\beq\label{eq: gegen-application}
    \frac{1}{|\hA+\epsilon\ha|^{\ell}} = \frac{1}{\left(1+2\epsilon \ha\cdot\hA + \epsilon^2\right)^{\ell/2}} = \sum_{N=0}^\infty C_{N}^{(\ell/2)}(-\ha\cdot\hA)\epsilon^N = \sum_{N=0}^\infty \sum_{k=0}^{\floor{N/2}} C_{Nk}^{\ell/2}\,\times\,(-2\ha\cdot\hA)^{N-2k}\epsilon^N,
\eeq
where the coefficients $C_{Nk}^{\ell/2}$ are defined in \eqref{eq: gegenbauer-explicit}. Inserting into \eqref{eq: solid-harm-application} gives:
\beq
    Y_\ell^m(\widehat{\va+\vA}) = \sum_{\lambda=0}^\ell\sum_{\mu=-\lambda}^\lambda A_{\ell m}^{\lambda \mu}\sum_{N=0}^\infty \sum_{k=0}^{\floor{N/2}} C_{Nk}^{\ell/2}(-2)^{N-2k}\times \epsilon^{\lambda+N} Y_{\lambda}^\mu(\ha)Y_{\ell-\lambda}^{m-\mu}(\hA)(\ha\cdot\hA)^{N-2k}.
\eeq
To simplify this further, we express $(\ha\cdot\hA)^{N-2k}$ in spherical harmonics, using \eqref{eq: Leg-polynomial-inv-def}:
\beq
    (\ha\cdot\hA)^{N-2k} = \sum_{L'=(N-2k)\downarrow2}\tilde{c}^{L'}_{N-2k}L_{L'}(\ha\cdot\hA) = \sum_{L'=(N-2k)\downarrow2}\frac{4\pi}{2L'+1}\sum_{M'=-L'}^{L'}\tilde{c}^{L'}_{N-2k}Y_{L'}^{M'}(\ha)Y_{L'}^{M'*}(\hA),
\eeq
expanding the Legendre polynomial into spherical harmonics via the addition theorem \eqref{eq: addition-theorem}. Finally, we can combine the two spherical harmonics in $\ha$ and $\hA$ using the product-to-sum relation \eqref{eq: spherical-harmonic-prod-to-sum}:
\beq
    Y_{\lambda}^\mu(\ha)Y_{L'}^{M'}(\ha) &=& \sum_{J_1=|\lambda-L'|}^{\lambda+L'}\sum_{M_1=-J_1}^{J_1}\mathcal{G}_{\lambda L' J_1}^{\mu M'(-M_1)}(-1)^{M_1}Y_{J_1}^{M_1}(\ha)\\\nonumber
    Y_{\ell-\lambda}^{m-\mu}(\hA)Y_{L'}^{M'*}(\hA) &=& \sum_{J_2=|\ell-\lambda-L'|}^{\ell-\lambda+L'}\sum_{M_2=-J_2}^{J_2}\mathcal{G}_{(\ell-\lambda) L' J_2}^{(m-\mu) (-M')(-M_2)}(-1)^{M'+M_2}Y_{J_2}^{M_2}(\hA).
\eeq
Combining results, we obtain the shift theorem for spherical harmonics:
\beq\label{eq: Ylm-shift-full}
    Y_\ell^m(\widehat{\va+\vA}) &=& \sum_{\lambda=0}^\ell \sum_{\mu=-\lambda}^\lambda A_{\ell m}^{\lambda \mu}\sum_{N=0}^\infty \sum_{k=0}^{\floor{N/2}} C_{Nk}^{\ell/2}(-2)^{N-2k}\sum_{L'=(N-2k)\downarrow2}\tilde{c}^{L'}_{N-2k}\frac{4\pi}{2L'+1}\\\nonumber
    &&\,\times\,\sum_{J_1=|\lambda-L'|}^{\lambda+L'}\sum_{M=-J_1}^{J_1}\mathcal{G}_{\lambda L' J_1}^{\mu (M-\mu)(-M)}
    \sum_{J_2=|\ell-\lambda-L'|}^{\ell-\lambda+L'}\mathcal{G}_{(\ell-\lambda) L' J_2}^{(m-\mu) (\mu-M)(M-m)}(-1)^{m+M-\mu}\\\nonumber
    &&\,\times\,\quad\,\epsilon^{\lambda+N} Y_{J_1}^{M}(\ha)Y_{J_2}^{m-M}(\hA),
\eeq
where we have noted that the Gaunt integrals imply $M_1=\mu+M'$, $M_2=m-\mu-M'=m-M_1$ and relabelled $M'\rightarrow M$. Defining the coefficients $\varphi^{\alpha,\ell m}_{J_1J_2M}$ where $\alpha = \lambda +N$, this can be written more succinctly as
% \beq
%     \boxed{Y_\ell^m(\widehat{\va+\vA}) = \sum_{\alpha J_1J_2M}\varphi^{\alpha,\ell m}_{J_1J_2M}\left(\frac{a}{A}\right)^{\alpha} Y_{J_1}^{M}(\ha)Y_{J_2}^{m-M}(\hA).}
% \eeq
\beq\label{eq: Ylm-shift-theorem}
    \boxed{Y_\ell^m(\widehat{\va+\vA}) = \sum_{\alpha=0}^\infty \sum_{J_1=0}^\alpha \sum_{J_2=\mathrm{max}(0,\ell-\alpha)}^{\ell+\alpha}\sum_{M=-J_1}^{J_1}\varphi^{\alpha,\ell m}_{J_1J_2M}\left(\frac{a}{A}\right)^{\alpha} Y_{J_1}^{M}(\ha)Y_{J_2}^{m-M}(\hA),}
\eeq
where the summation limits will be elaborated upon in the next section. Truncating \eqref{eq: Ylm-shift-theorem} at $\alpha = K$ incurs an error of $\mathcal{O}(\epsilon^{K+1})$, thus, for $\epsilon<1$ this is a valid perturbative expansion.

\subsection{Coefficient Properties}\label{appen: shift-coeff-prop}
From \eqref{eq: Ylm-shift-full}, the shift coefficients $\varphi$ are given by 
\begin{empheq}[box=\fbox]{align}\label{eq: shift-coeff-def}
    \varphi^{\alpha,\ell m}_{J_1J_2M} &= \sum_{\lambda=0}^\ell \sum_{\mu=-\lambda}^\lambda A_{\ell m}^{\lambda \mu}\sum_{N=0}^\infty \delta^\mathrm{K}_{\alpha(\lambda+N)} \sum_{k=0}^{\floor{N/2}} C_{Nk}^{\ell/2}(-2)^{N-2k}\sum_{L'=(N-2k)\downarrow2}\tilde{c}^{L'}_{N-2k}\frac{4\pi}{2L'+1}\\\nonumber
    &\,\times\,\mathcal{G}_{\lambda L' J_1}^{\mu (M-\mu)(-M)}
    \mathcal{G}_{(\ell-\lambda) L' J_2}^{(m-\mu) (\mu-M)(M-m)}(-1)^{m+M-\mu},
\end{empheq}
where the Kronecker delta in the first line enforces that $\lambda+N=\alpha$. Including the various coefficient definitions gives the explicit form
\beq\label{eq: shift-coeff-def-explicit}
    \varphi^{\alpha,\ell m}_{J_1J_2M} &=& \sqrt{4\pi(2\ell+1)(2J_1+1)(2J_2+1)}\sum_{\lambda=0}^{\mathrm{min}(\alpha,\ell)}\sum_{\mu=-\lambda}^\lambda \binom{\ell+m}{\lambda+\mu}^{1/2}\binom{\ell-m}{\lambda-\mu}^{1/2}\sum_{k=0}^{\floor{\alpha-\lambda/2}}\sum_{L'=(\alpha-\lambda-2k)\downarrow 2}\\\nonumber
    &&\,\times\,(-1)^{k+\alpha+\lambda+m+M+\mu}\frac{(2L'+1)\sqrt{\pi}\,\Gamma\left(\frac{\ell}{2}+\alpha-\lambda-k\right)}{2\,\Gamma\left(\ell/2\right)\Gamma\left((\alpha-\lambda-2k-L')/2+1\right)\Gamma\left((\alpha-\lambda-2k+L'+3)/2\right)k!}\\\nonumber
    &&\,\times\,\tj{\lambda}{L'}{J_1}{\mu}{M-\mu}{-M}\tj{\ell-\lambda}{L'}{J_2}{m-\mu}{\mu-M}{M-m}\tjo{\lambda}{L'}{J_1}\tjo{\ell-\lambda}{L'}{J_2}.
\eeq    

We consider two special cases. For $\alpha=0$, we require $\lambda=N=0$, and thus, by the summation limits, $\mu=k=L'=0$. This gives
\beq
    \varphi^{0,\ell m}_{J_1J_2M} = (4\pi)^{3/2}\mathcal{G}^{0M(-M)}_{00J_1}\mathcal{G}^{m(-M)M}_{\ell 0 J_2}(-1)^{m+M},
\eeq
using $A_{\ell m}^{00} = \sqrt{4\pi}$, $\tilde{c}^0_0=1$ and $C_{00}^{\ell/2} = 1$ \eqref{eq: Leg-polynomial-inv-def}. The Gaunt integrals obey 3-$j$ symmetries, such that $\mathcal{G}_{L_1L_2L_3}^{M_1M_2M_3}$ is zero unless $|L_1-L_2|<L_3<L_1+L_2$, implying $J_1=0$ (and thus $M=0$), $J_2=\ell$. In full, we obtain
\beq\label{eq: shift-coeff-alpha-zero}
    \varphi^{0,\ell m}_{J_1J_2M} = \delta^\mathrm{K}_{J_10}\delta^\mathrm{K}_{J_2\ell}\delta^\mathrm{K}_{M0}\times\sqrt{4\pi}.
\eeq
This gives $\left.Y_\ell^m(\widehat{\va+\vA})\right|_{\alpha=0} = Y_{\ell}^m(\hA)$, as expected.

Secondly, we consider $\ell=m=0$. The summation limits enforce $\lambda=\mu=0$, and, since $C_{Nk}^{0} = \delta^\mathrm{K}_{N0}$ \eqref{eq: gegenbauer-explicit}, $N=k=L'=0$ (and thus $\alpha=\lambda+N=0$), giving
\beq
    \varphi^{\alpha,00}_{J_1J_2M} = (4\pi)^{3/2}\mathcal{G}^{0M(-M)}_{00J_1}\mathcal{G}^{0(-M)M}_{00J_2}(-1)^M.
\eeq
In this case, the triangle conditions imply $J_1=J_2=0$, thus $M=0$, and 
\beq
    \varphi^{\alpha,00}_{J_1J_2M} = \delta^\mathrm{K}_{J_10}\delta^\mathrm{K}_{J_20}\delta^\mathrm{K}_{M0}\delta^\mathrm{K}_{\alpha0}\times\sqrt{4\pi},
\eeq
giving $Y_{0}^0(\widehat{\va+\vA}) = (4\pi)^{-1/2}$, as expected.

It is useful to consider properties relating to parity. Firstly, consider the parity transformation $\va\rightarrow-\va$, $\vA\rightarrow-\vA$:
\beq
    Y_{\ell}^m(\widehat{\va+\vA}) &\rightarrow& (-1)^\ell Y_\ell^m(\widehat{\va+\vA})\\\nonumber
    Y_{J_1}^{M}(\ha)Y_{J_2}^{m-M}(\hA) &\rightarrow& (-1)^{J_1+J_2} Y_{J_1}^{M}(\ha)Y_{J_2}^{m-M}(\hA).
\eeq
By orthogonality of the spherical harmonics, this implies $\varphi^{\alpha,\ell m}_{J_1J_2M} = (-1)^{\ell+J_1+J_2}\varphi^{\alpha,\ell m}_{J_1J_2M}$, and hence that the coefficients are vanishing if $\ell+J_1+J_2$ is odd. Furthermore, the Gaunt coefficients $\mathcal{G}_{L_1L_2L_3}^{M_1M_2M_3}$ contain 3-$j$ symbols with all $M_i$ fixed to zero; these vanish if $L_1+L_2+L_3$ is odd, implying that $\lambda+L'+J_1$ and $\ell-\lambda+L'+J_2$ must be even \eqref{eq: shift-coeff-def}. Finally, the summation on $L'$ indicates that $L'$ and $N$ must have the same sign, hence $\ell+N+J_1$ is even, thus so is $\alpha+J_1$. Our conclusion is that $\varphi^{\alpha,\ell m}_{J_1J_2M}$ must be vanishing unless both $\alpha+J_1$ and $\ell+\alpha+J_2$ are even. This result proves useful when forming the even parity expansion in \S\ref{subsec: mid-pk-even}.

Finally, we discuss the summation limits in \eqref{eq: Ylm-shift-theorem}. From the $L'$ summation, $0\leq L'\leq N$ (since $0\leq 2k \leq N$), and from the Gaunt integrals, $|\lambda-L'|\leq J_1\leq \lambda+L'$, $|\ell-\lambda-L'|\leq J_2\leq \ell-\lambda+L'$. Combining the two implies that $0\leq J_1\leq \alpha$, $\mathrm{max}(0,\ell-\alpha)\leq J_2\leq \ell+\alpha$, $|M|\leq J_1$, noting that the moduli enforce $J_1$, $J_2$ to be positive. For the $\alpha=1$ piece, we therefore need consider only $J_1=1$ (as $\alpha+J_1$ is even) and $J_2=\ell\pm1$, or, for $\alpha=2$, $J_1\in\{0,2\}$, $J_2\in\{\ell,\ell\pm 2\}$.

\section{Generalized Spherical Harmonic Shift Theorem}\label{appen: generalized-shift-theorem}
Below, we prove a useful generalization of the spherical harmonic shift theorem discussed in Appendix \ref{appen: spherical-harmonic-shift-theorem}. To begin, consider the expression
\beq
    \left(\frac{|\va|}{|\vA+\va|}\right)^\alpha Y_{J_1}^{M}(\ha)Y_{J_2}^{m-M}(\widehat{\va+\vA}) = \frac{1}{\left[1+2\epsilon\ha\cdot\hA+\epsilon^2\right]^{\alpha/2}} Y_{J_1}^{M}(\ha)Y_{J_2}^{m-M}(\widehat{\va+\vA}),
\eeq
for some $\ha$, $\hA$ with $\epsilon=a/A\ll 1$. The spherical harmonic in $\va+\vA$ can be expanded identically to Appendix \ref{appen: sph-shift-deriv}; with the prefactor (which is again a generator for Gegenbauer polynomials), we obtain the same expression except that $C_{Nk}^{\ell/2}$ is replaced with $C_{Nk}^{(\ell+\alpha)/2}$ in \eqref{eq: gegen-application} and the succeeding. In full, we obtain
\beq\label{eq: generalized-Ylm-tmp}
    \left(\frac{|\va|}{|\vA+\va|}\right)^\alpha Y_{J_1}^{M}(\ha)Y_{J_2}^{m-M}(\widehat{\va+\vA}) = \sum_{\beta J_1'J_2'M'}\omega^{\beta,J_2(m-M)\alpha}_{J_1'J_2'M'}\epsilon^{\alpha+\beta}Y_{J_1}^M(\ha)Y_{J_1'}^{M'}(\ha)Y_{J_2'}^{m-M-M'}(\hA),
\eeq
where $\omega^{\beta,\ell m\alpha}_{J_1J_2M}$ is equal to $\varphi^{\beta, \ell m}_{J_1J_2M}$ \eqref{eq: shift-coeff-def}, except that $C_{Nk}^{\ell/2}\rightarrow C_{Nk}^{(\ell+\alpha)/2}$, implying that $\omega^{\beta,\ell m0}_{J_1J_2M} = \varphi^{\beta, \ell m}_{J_1J_2M}$. Next, we can combine the two spherical harmonics in $\va$ using the product-to-sum relation \eqref{eq: spherical-harmonic-prod-to-sum} to yield
\beq\label{eq: generalized-shift-theorem}
     \left(\frac{|\va|}{|\vA+\va|}\right)^\alpha Y_{J_1}^{M}(\ha)Y_{J_2}^{m-M}(\widehat{\va+\vA}) &=& 
     \sum_{\beta S_1S_2T} \left(\sum_{J'} \omega^{\beta,J_2(m-M)\alpha}_{J'S_2(T-M)}\mathcal{G}_{J_1J'S_1}^{M(T-M)(-T)}(-1)^T\right) \epsilon^{\alpha+\beta}Y_{S_1}^T(\ha)Y_{S_2}^{m-T}(\hA)\nonumber
\eeq
\begin{eqnarray}
     \Rightarrow \boxed{\left(\frac{|\va|}{|\vA+\va|}\right)^\alpha Y_{J_1}^{M}(\ha)Y_{J_2}^{m-M}(\widehat{\va+\vA}) \equiv \sum_{\beta S_1S_2T}\Omega^{\beta,\alpha J_1J_2mM}_{S_1S_2T}\epsilon^{\alpha+\beta}Y_{S_1}^T(\ha)Y_{S_2}^{m-T}(\hA),}
\end{eqnarray}
liberally relabeling variables and defining the new coefficient set $\Omega$. The expression is now in the form of a power series combined with two spherical harmonics, just as in \eqref{eq: Ylm-shift-theorem}.

By analogous arguments to before, the $\omega_{J'S_2(T-M)}^{\beta,J_2(m-M)\alpha}$ coefficient requires $0\leq J'\leq \beta$ and $\mathrm{max}(0,J_2-\beta)\leq S_2\leq J_2+\beta$. Coupled with the triangle conditions on the additional Gaunt integral, we obtain $\mathrm{max}(0,J_1-\beta)\leq S_1\leq J_1+\beta$, $\mathrm{max}(0,J_2-\beta)\leq S_2\leq J_2+\beta$, and $|T|<S_1$. Furthermore, we require $\beta+J'$ and $J_2+\beta+S_2$ to be even, and, from the Gaunt factor, the same applies for $\beta+J_1+S_1$. The case $\beta = 0$ will also be of use. As in \eqref{eq: shift-coeff-alpha-zero}, $\omega^{0,J_2(m-M)\alpha}_{J'S_2(T-M)} = \sqrt{4\pi}\delta^\mathrm{K}_{J'0}\delta^\mathrm{K}_{S_2J_2}\delta^\mathrm{K}_{TM}$, thus $\Omega^{0,\alpha J_1J_2mM}_{S_1S_2T} = \sqrt{4\pi}\delta^\mathrm{K}_{S_2J_2}\delta^\mathrm{K}_{TM}\mathcal{G}^{M0(-M)}_{J_10S_1}(-1)^M$. The Gaunt integral requires $S_1=J_1$, thus we obtain
\beq\label{eq: Omega-alpha-0-coeff}
    \Omega^{0,\alpha J_1J_2mM}_{S_1S_2T} = \delta^\mathrm{K}_{S_1J_1}\delta^\mathrm{K}_{S_2J_2}\delta^\mathrm{K}_{TM}.
\eeq
%\sum_{\beta J_1'J_2'M'}\omega^{\beta,\ell m\alpha}_{J_1'J_2'M'}\sum_{\mathcal{J}\mathcal{M}}\mathcal{G}_{J_1J_1'\mathcal{J}}^{MM'(-\mathcal{M})}(-1)^{\mathcal{M}}\times \epsilon^{\alpha+\beta}Y_{\mathcal{J}_1}^\mathcal{M}(\ha)Y_{J_2'}^{m-M-M'}(\hA),

\section{Alternative Derivation for the Parity Even Expansion}\label{appen: parity-even-alternate}
We briefly present an alternative derivation of the parity-even expansion for spherical harmonics, which, whilst less conceptual, results in an easier-to-implement formalism. To begin, consider the function
\beq
    z^{\alpha,m}_{\ell} \equiv \frac{1}{2}\left[\epsilon_1^\alpha Y_\ell^m(\hr_1)+(-1)^\ell \epsilon_2^\alpha Y_\ell^m(\hr_2)\right].
\eeq
Writing $\vr_1=\vR-\vD$, $\vr_2=\vR+\vD$, each term may be expanded using a part of the generalized shift theorem of Appendix \ref{appen: generalized-shift-theorem}:
\beq
    z^{\alpha,m}_\ell &\equiv& \frac{1}{2}\left[\left(\frac{\Delta}{|\vR-\vD|}\right)^\alpha Y_\ell^m(\widehat{\vR-\vD})+(-1)^\ell\left(\frac{\Delta}{|\vR+\vD|}\right)^\alpha Y_\ell^m(\widehat{\vR+\vD})\right]\\\nonumber
    &=& \sum_{\beta J_1J_2M}\omega^{\beta,\ell m\alpha}_{J_1J_2M}\theta^{\alpha+\beta}Y_{J_1}^{M}(\hD)Y_{J_2}^{m-M}(\hR)\left[\frac{1+(-1)^{\ell+J_1}}{2}\right] .
\eeq
using \eqref{eq: generalized-Ylm-tmp} and recalling $\theta\equiv\Delta/R$. As before, $\omega^{\beta,\ell m\alpha}_{J_1J_2M}$ is non-zero unless $J_1+\beta$ is even, thus the sum extends only over even $\beta$ (assuming even $\ell$). Extracting the $\beta=0$ piece gives
\beq
    z^{\alpha,m}_\ell &\equiv& \theta^\alpha Y_{\ell}^{m}(\hR) + \sum_{\mathrm{even}\,\beta>0}\sum_{J_1J_2M}\omega^{\beta,\ell m\alpha}_{J_1J_2M}Y_{J_1}^{M}(\hD)\times \theta^{\alpha+\beta}Y_{J_2}^{m-M}(\hR).
\eeq
Assuming small $\theta$, this may be inverted order-by-order to find $\theta^\alpha Y_\ell^m$ in terms of $z_\ell^{\alpha,m}$ (which is in the form required for the convolutional power spectrum estimators). This gives
\beq
    \theta^\alpha Y_\ell^m(\hR) &=& z_\ell^{\alpha,m} - \sum_{\mathrm{even}\,\beta>0}\sum_{J_1J_2M}\omega^{\beta,\ell m\alpha}_{J_1J_2M}Y_{J_1}^{M}(\hD)\times \theta^{\alpha+\beta}Y_{J_2}^{m-M}(\hR)\\\nonumber
    &=& z_\ell^{\alpha,m} - \sum_{\mathrm{even}\,\beta>0}\sum_{J_1J_2M}\omega^{\beta,\ell m\alpha}_{J_1J_2M}Y_{J_1}^{M}(\hD)\times z_{J_2}^{(\alpha+\beta),(m-M)}\\\nonumber
    &&\,+\,\sum_{\mathrm{even}\,\beta>0}\sum_{J_1J_2M}\sum_{\mathrm{even}\,\gamma>0}\sum_{S_1S_2T}\omega^{\beta,\ell m\alpha}_{J_1J_2M}\omega^{\gamma,J_2(m_M)(\alpha+\beta)}_{S_1S_2T}Y_{J_1}^{M}(\hD)Y_{S_1}^{T}(\hD)\times z_{S_2}^{(\alpha+\beta+\gamma),(m-M-T)}+...
\eeq
where the first, second and third terms start at zeroth-, second- and fourth-order in $\theta$ respectively. Setting $\alpha=0$, we obtain a concise form for the parity-even expansion correct to fourth order:
\beq
    2\,Y_\ell^m(\hR) &=& \left[Y_\ell^m(\hr_1)+Y_\ell^m(\hr_2)\right]\\\nonumber
    &&\,-\,\sum_{\mathrm{even}\,\alpha>0}\sum_{J_1J_2M}\varphi^{\alpha,\ell m}_{J_1J_2M}Y_{J_1}^{M}(\hD) \left[\epsilon_1^\alpha Y_{J_2}^{m-M}(\hr_1)+\epsilon_2^\alpha Y_{J_2}^{m-M}(\hr_2)\right]\\\nonumber
    &&\,+\,\sum_{\mathrm{even}\,\alpha>0}\sum_{J_1J_2M}\varphi^{\alpha,\ell m}_{J_1J_2M}\sum_{\mathrm{even}\,\beta>0}\sum_{S_1S_2T}\Omega^{\beta,J_1J_2mM\alpha}_{S_1S_2T}Y_{S_1}^{T}(\hD)\left[\epsilon_1^{\alpha+\beta}Y_{S_2}^{m-M-T}(\hr_1)+\epsilon_2^{\alpha+\beta}Y_{S_2}^{m-M-T}(\hr_2)\right]\,+\,...
    %\omega^{\beta,J_2(m-M)\alpha}_{S_1S_2T}\sum_{J'T'}\mathcal{G}^{MT(-M-T)}_{J_1S_1J'}(-1)^{M+T}Y_{J'}^{M+T}(\hD)\\\nonumber
    %&&\quad\times\quad\left[\epsilon_1^{\alpha+\beta}Y_{S_2}^{m-M-T}(\hr_1)+\epsilon_2^{\alpha+\beta}Y_{S_2}^{m-M-T}(\hr_2)\right]\\\nonumber
    %&&\,+\,...
\eeq
assuming even $\ell$, recalling that $\omega^{0,\ell m\alpha}_{J_1J_2M} = \varphi^{\alpha,\ell m}_{J_1J_2M}$ and contracting the spherical harmonics via \eqref{eq: spherical-harmonic-prod-to-sum}. We have additionally inserted the definition of $\Omega$ from \eqref{eq: generalized-shift-theorem}. As before, this can be written as a simple summation over even $\alpha$:
\begin{eqnarray}
    \boxed{2\,Y_\ell^m(\hR) = \sum_{\mathrm{even}\,\alpha}\sum_{J_1J_2M}\Phi^{\alpha,\ell m}_{J_1J_2M}Y_{J_1}^{M}(\hD)\left[\epsilon_1^\alpha Y_{J_2}^{m-M}(\hr_1)+\epsilon_2^\alpha Y_{J_2}^{m-M}(\hr_2)\right],}
\end{eqnarray}
with 
\beq
    \Phi^{0,\ell m}_{J_1J_2M} &=& \sqrt{4\pi}\delta^\mathrm{K}_{J_10}\delta^\mathrm{K}_{J_2\ell}\delta^\mathrm{K}_{M0}\\\nonumber
    \Phi^{2,\ell m}_{J_1J_2M} &=& -\varphi^{2,\ell m}_{J_1J_2M}\\\nonumber
    \Phi^{4,\ell m}_{J_1J_2M} &=& -\varphi^{4,\ell m}_{J_1J_2M}+\sum_{J_1'J_2'M'}\varphi^{2,\ell m}_{J_1'J_2'M'}\Omega^{2,2J_1'J_2'mM'}_{J_1J_2M}.%\sum_{J'}\omega^{2,J_2'(m-M')2}_{J'J_2(M-M')}\mathcal{G}^{M'(M-M')(-M)}_{J_1'J'J_1}
\eeq
These are equivalent to the $\Phi$ coefficients of \S\ref{subsec: mid-pk-even}, through derived in a different manner. This is straightforward to extend to higher order, and somewhat faster to compute than the former procedure, since summations are only over even $\alpha$.

\section{Legendre Polynomial Shift Theorem}\label{appen: leg-shift-theorem}
\subsection{Derivation}\label{appen: leg-shift-theorem-deriv}
A similar relation to that of Appendix \ref{appen: spherical-harmonic-shift-theorem} may be derived for the Legendre polynomial $L_\ell(\ha\cdot\widehat{\va+\vA})$. This can be derived in one of two ways; either by expanding in powers of $\ha\cdot\widehat{\va+\vA}$ directly, or via the previously proved spherical harmonic shift theorem. For consistency with the above, we adopt the latter approach, though we note that the two yield consistent results.

To perform such an expansion, we first rewrite the Legendre polynomial in terms of spherical harmonics using the addition theorem \eqref{eq: addition-theorem} and apply \eqref{eq: Ylm-shift-theorem}, yielding
\beq
    L_\ell(\ha\cdot\widehat{\va+\vA}) &=& \frac{4\pi}{2\ell+1}\sum_{m=-\ell}^\ell (-1)^m Y_{\ell}^{m}(\ha)Y_{\ell}^{m}(\widehat{\va+\vA})\\\nonumber
    &=& \frac{4\pi}{2\ell+1}\sum_{m=-\ell}^\ell\sum_{\alpha J_1J_2M} (-1)^m \varphi^{\alpha,\ell m}_{J_1J_2M}\left(\frac{a}{A}\right)^\alpha  Y_{J_1}^M(\ha)Y_{\ell}^{m}(\ha)Y_{J_2}^{m-M}(\hA),
\eeq
additionally using $Y_{\ell}^{m*} = (-1)^mY_\ell^{-m}$. The two spherical harmonics in $\ha$ may be contracted using the product-to-sum relation \eqref{eq: spherical-harmonic-prod-to-sum}, giving
\beq
    L_\ell(\ha\cdot\widehat{\va+\vA}) &=& \frac{4\pi}{2\ell+1}\sum_{\alpha JJ_2M'}\left[\sum_{mJ_1}(-1)^m\varphi^{\alpha,\ell m}_{J_1J_2(m-M')}\mathcal{G}^{(-m)(m-M')M'}_{\ell J_1J}\right]\left(\frac{a}{A}\right)^\alpha Y_{J}^{M'*}(\ha)Y_{J_2}^{M'}(\hA)\\\nonumber
    &\equiv& \frac{4\pi}{2\ell+1}\sum_{\alpha JJ_2M'}f^{\alpha,\ell}_{JJ_2M'}\left(\frac{a}{A}\right)^\alpha Y_{J}^{M'*}(\ha)Y_{J_2}^{M'}(\hA),
\eeq
where we have relabelled $M' = m-M$, and defined the new coefficient $f^{\alpha,\ell}_{JJ_2M'}$. Whilst this may seem complicated, one can in fact show that $f^{\alpha,\ell}_{JJ_2}$ is independent of $M'$ and non-zero only for $J=J_2$. Defining reduced coefficients $f^{\alpha,\ell}_{J}$ via
\beq
    f^{\alpha,\ell}_{JJ_2M'} \equiv \delta^\mathrm{K}_{JJ_2}\times f^{\alpha,\ell}_{J},
\eeq
and using \eqref{eq: addition-theorem}, this leads to
\beq\label{eq: Legendre-shift-theorem-all-parity}
    \boxed{L_\ell(\ha\cdot\widehat{\va+\vA}) = \sum_{\alpha=0}^\infty \sum_{J=\mathrm{max}(\ell-\alpha,0)}^{\ell+\alpha}\frac{2J+1}{2\ell+1}f^{\alpha,\ell}_{J}\left(\frac{a}{A}\right)^\alpha L_{J}(\ha\cdot\hA);}
\eeq
a far more tractable expansion. Whilst the $(2J+1)/(2\ell+1)$ factor could be absorbed into $f_J^{\alpha,\ell}$, we retain it here for later convenience.

To see this, we first write out $f_{JJ_2M'}^{\alpha,\ell}$ explicitly:
\beq\label{eq: falpha-deriv-tmp}
    f_{JJ_2M'}^{\alpha,\ell} &=& \sum_{m J_1 \lambda \mu N k L'} A_{\ell m}^{\lambda \mu}\delta^\mathrm{K}_{\alpha(\lambda+N)} C_{Nk}^{\ell/2}(-2)^{N-2k}\tilde{c}^{L'}_{N-2k}\frac{4\pi}{2L'+1}\\\nonumber
    &&\,\times\,\mathcal{G}_{\lambda L' J_1}^{\mu (m-M'-\mu)(M'-m)}
    \mathcal{G}_{(\ell-\lambda) L' J_2}^{(m-\mu) (\mu-m+M')(-M')}\mathcal{G}^{(-m)(m-M')M'}_{\ell J_1J}(-1)^{m-M'-\mu}\\\nonumber
    &=& \sum_{\lambda N k L'}(2\ell+1)^{3/2}\sqrt{(2J'+1)(2J_2+1)}\binom{2\ell}{2\lambda}^{1/2}(-1)^\ell \delta^\mathrm{K}_{\alpha(\lambda+N)} C_{Nk}^{\ell/2}(-2)^{N-2k}\tilde{c}^{L'}_{N-2k}\\\nonumber
    &&\,\times\,\sum_{J_1}(2J_1+1)\tjo{\lambda}{L'}{J_1}\tjo{\ell-\lambda}{L'}{J_2}\tjo{\ell}{J_1}{J}\\\nonumber
    &&\,\times\,\left[\sum_{m\mu}(-1)^{M'-\mu}\tj{\lambda}{L'}{J_1}{\mu}{m-M'-\mu}{M'-m}\tj{\ell-\lambda}{L'}{J_2}{m-\mu}{\mu-m+M'}{-M'}\tj{\ell}{J_1}{J}{-m}{m-M'}{M'}\tj{\lambda}{\ell}{\ell-\lambda}{\mu}{-m}{m-\mu}\right],
\eeq
where, in the second line, we have written the Gaunt integrals in terms of Wigner 3-$j$ symbols \eqref{eq: Gaunt-3j-def}, and noted that 
\beq
    A_{\ell m}^{\lambda\mu} = (-1)^{\ell-m}\sqrt{\frac{4\pi(2\ell+1)^2}{(2\lambda+1)(2\ell-2\lambda+1)}}\binom{2\ell}{2\lambda}^{1/2}\tj{\lambda}{\ell}{\ell-\lambda}{\mu}{-m}{m-\mu}.
\eeq
To simplify, we first employ a summation identity for four 3-$j$ symbols, given in \citet[Eq.\,A.24]{lohmann2008angle}:
\beq
    &&\sum_{m_1m_2m_4m_5m_6}(-1)^{j_4+j_5+j_6+m_4+m_5+m_6}\tj{j_4}{j_5}{j_3}{m_4}{-m_5}{m_3}\tj{j_5}{j_6}{j_1}{m_5}{-m_6}{m_1}\tj{j_6}{j_4}{j_2}{m_6}{-m_4}{m_2}\tj{j_1}{j_2}{j_3'}{m_1}{m_2}{m_3'}\nonumber\\
    &&\quad = \quad\delta^\mathrm{K}_{j_3j_3'}\delta^\mathrm{K}_{m_3m_3'}\frac{1}{2j_3+1}\begin{Bmatrix}j_1 & j_2 & j_3\\ j_4 & j_5 & j_6\end{Bmatrix},
\eeq
where the quantity in curly parentheses is a Wigner 6-$j$ symbol (see \citealt[\S34.4]{nist_dlmf}). Notably, the RHS of the above equation is independent of $m_3$ and enforces $j_3=j_3'$. Applying this to the final line of \eqref{eq: falpha-deriv-tmp}, denoted by $[...]$, with $\{j_1,j_2,j_3,j_3',j_4,j_5,j_6\} = \{\ell,J_1,J_2,J,L',\ell-\lambda,\lambda\}$ with some massaging via the 3-$j$ symmetries gives
\beq
    \left[^{ }_{ }...\right] &=& (-1)^{J_1+\lambda+\ell}\delta^\mathrm{K}_{J_2J}\frac{1}{2J+1}\begin{Bmatrix}\ell & J_1 & J_2\\ L' & \ell-\lambda & \lambda\end{Bmatrix},
\eeq
enforcing $J_2=J$ and with no dependence on $M'$, as previously mentioned. We may further perform the $J_1$ summation analytically, using the 6-$j$ summation relation:
\beq
    \sum_{j_6}(-1)^{j_1+j_2-j_3+j_4+j_5+j_6-m_1-m_4}(2j_6+1)\begin{Bmatrix}j_1 & j_2 & j_3\\ j_4 & j_5 & j_6\end{Bmatrix}\tj{j_5}{j_1}{j_6}{m_5}{m_1}{-m_6}\tj{j_2}{j_4}{j_6}{m_2}{m_4}{m_6}
\eeq
\citep[Eq.\,A.26b]{lohmann2008angle}, which, following application of 6-$j$ symmetries, can be used to show
\beq
    &&\sum_{J_1}(2J_1+1)(-1)^{J_1}\tjo{\ell}{L'}{J_1}\tjo{\ell-\lambda}{L'}{J}\tjo{\ell}{J_1}{J}\begin{Bmatrix}\ell & J_1 & J\\ L' & \ell-\lambda & \lambda\end{Bmatrix}\\\nonumber
    &&\,=\,(-1)^{L'+J}\tjo{\ell-\lambda}{L'}{J}\tjo{\ell}{\lambda}{\ell-\lambda}\tjo{L'}{J}{\ell-\lambda}.
\eeq
Inserting into \eqref{eq: falpha-deriv-tmp} gives the final form for the reduced coefficients $f^{\alpha,\ell}_{J}$:
\begin{empheq}[box=\widebox]{align}\label{eq: f-alpha-def}
    f^{\alpha,\ell}_{J} &= (-1)^\ell(2\ell+1)^{3/2}\sum_{\lambda=0}^{\mathrm{min}(\alpha,\ell)}\binom{2\ell}{2\lambda}^{1/2}\tjo{\ell}{\lambda}{\ell-\lambda}\sum_{k=0}^{\floor{(\alpha-\lambda)/2}}C^{\ell/2}_{(\alpha-\lambda)k}(-2)^{\alpha-\lambda-2k}\\\nonumber
    &\,\times\,\sum_{L'=(\alpha-\lambda-2k)\downarrow2}\tilde{c}^{L'}_{\alpha-\lambda-2k}\tjo{\ell-\lambda}{L'}{J}^2.
\end{empheq}

As before, this satisfies $f^{0,\ell}_{J} = \delta^\mathrm{K}_{\ell J}$, and $f^{\alpha,0}_{J} = \delta^\mathrm{K}_{\alpha 0}\delta^\mathrm{K}_{J0}$, which, when inserted into \eqref{eq: Legendre-shift-theorem-all-parity} gives $L_\ell(\ha\cdot\widehat{\va+\vA}) = L_\ell(\ha\cdot\hA)$ at leading order and $L_0(\ha\cdot\widehat{\va+\vA}) = 1$. Furthermore, we have the parity-rule $\alpha+\ell+J=\mathrm{even}$ for non-zero $f^{\alpha,\ell}_{J}$ coefficients. We also note a curious property; the sum of $f^{\alpha,\ell}_J$ over all $J$ is equal to zero unless $\alpha=0$, as demonstrated in \eqref{eq: L2-all}\,\&\,\eqref{eq: L24-even}. To prove this, we set $\vA = k\va$ in \eqref{eq: Legendre-shift-theorem-all-parity} and recall $L_\ell(1) = 1$, giving
\beq
    1 = \sum_{\alpha}\sum_{J}f^{\alpha,\ell}_J \left(\frac{1}{k}\right)^{\alpha} = 1 + \sum_{\alpha>0}k^{-\alpha}\sum_{J}f^{\alpha,\ell}_J
\eeq
For this to be true for arbitrary $k>1$, we require $\sum_J f^{\alpha,\ell}_J = 0$ for $\alpha>0$.

\subsection{Parity-Even Form}\label{appen: leg-shift-theorem-even}
For the 2PCF estimators considered in \S\ref{subsec: mid-pk-even}, we use the Legendre polynomial shift theorem \eqref{eq: Legendre-shift-theorem-all-parity} in the following symmetrized form:
\beq
    L_\ell(\hD\cdot\widehat{\vr_1+\vr_2}) = \sum_{\alpha=0}^\infty\sum_{J=\mathrm{max}(\ell-\alpha,0)}^{\ell+\alpha}\frac{2J+1}{2\ell+1}f_J^{\alpha,\ell}\left[\epsilon_1^\alpha L_J(\hD\cdot\hr_1)+(-1)^{\ell+J}\epsilon_2^\alpha L_J(\hD\cdot\hr_2)\right],
\eeq
for $\epsilon_i \equiv \Delta/(2r_i)$. Just as for the spherical harmonic expansion (\S\ref{subsec: mid-pk-even}), these may be recast in a manifestly parity-even form, \textit{i.e.} one only involving even $J$ and $\alpha$ (assuming even $\ell$). The derivation of this is analogous to the above, and leads to 
\begin{eqnarray}\label{eq: Legendre-shift-theorem-even-parity}
    \boxed{L_\ell(\hD\cdot\widehat{\vr_1+\vr_2}) = \sum_{\mathrm{even}\,\alpha=0}^\infty \sum_{J=\mathrm{max}(\ell-\alpha,0)}^{\ell+\alpha}\frac{2J+1}{2\ell+1}F^{\alpha,\ell}_{J}\left(\frac{a}{A}\right)^{\alpha}L_{J}(\ha\cdot\hA),}
\end{eqnarray}
for $2\epsilon_i\ll 1$, where the parity-even coefficients are defined recursively via
\beq\label{eq: F-coeff-def}
    F^{\alpha,\ell}_{J} = f^{\alpha,\ell}_{J} - \frac{1}{2}\sum_{\mathrm{odd}\,\beta<\alpha}2^{\alpha-\beta}\sum_{S}f^{\beta,\ell}_{S}h^{(\alpha-\beta),\beta S}_{J}+\frac{1}{4}\sum_{\mathrm{odd}\,\beta<\alpha}2^{\alpha-\beta}\sum_{S}f^{\beta,\ell}_{S}\sum_{\mathrm{even}\,\gamma>0,\gamma<\alpha-\beta}\sum_{T}h^{\gamma,\beta S}_{T}h^{(\alpha-\beta-\gamma),(\beta+\gamma)T}_{J}+...
\eeq
using the method of \S\ref{subsec: mid-pk-even}, or equivalently
\beq
    F^{0,\ell}_{J} &=& \delta^\mathrm{K}_{\ell J}\\\nonumber
    F^{2,\ell}_{J} &=& -f^{2,\ell}_{J}\\\nonumber
    F^{4,\ell}_{J} &=&  - f^{4,\ell}_{J}+\sum_S f^{2,\ell}_S h^{2,2S}_{J}
\eeq
\textit{et cetera}, using that of Appendix \ref{appen: parity-even-alternate}. Here, the $h^{\beta,\alpha\ell}_{J}$ functions are identical to $f^{\alpha,\ell}_{J}$ \eqref{eq: f-alpha-def}, but with $\ell/2\rightarrow(\ell+\alpha)/2$ in the $C_{Nk}^{\ell/2}$ coefficient.

\section{Bisector Series Expansion}\label{appen: cw-math}
To derive the series expansion of \S\ref{subsec: pk-bis-series}, we start from the bisector vector definition $\vd = t\vr_1+(1-t)\vr_2$ where $t  = r_2/(r_1+r_2)$, and perform the solid-harmonic expansion \eqref{eq: solid-harmonic-addition}:
\beq\label{eq: Castorina-tmp}
    Y_\ell^m(\hat{\vec d}) = \sum_{\lambda=0}^\ell \sum_{\mu=-\lambda}^\lambda A_{\ell m}^{\lambda\mu} \left(\frac{tr_1}{d}\right)^\ell Y_{\lambda}^\mu(\hr_1)Y_{\ell-\lambda}^{m-\mu}(\hr_2),
\eeq
using $tr_1=(1-t)r_2$, and defining $A_{\ell m}^{\lambda \mu}$ as in \eqref{eq: solid-harm-application}. Whilst this is not immediately separable in $\vr_1$, $\vr_2$ due to the $d = |\vd|$ denominator, we follow \citet{2018MNRAS.476.4403C} and note that
\beq
    d^2 = \frac{r_1^2r_2^2}{(r_1+r_2)^2}|\hr_1+\hr_2|^2 = 2r_1^2t^2\left(1+\cos{\phi}\right),
\eeq
where $\cos\phi = \hr_1\cdot\hr_2$. Combining with the above:
\beq
    Y_\ell^m(\hat{\vec d}) = \left[2(1+\cos\phi)\right]^{-\ell/2}\sum_{\lambda=0}^\ell \sum_{\mu=-\lambda}^\lambda A_{\ell m}^{\lambda\mu} Y_{\lambda}^\mu(\hr_1)Y_{\ell-\lambda}^{m-\mu}(\hr_2).
\eeq
The former work proceeded to expand the prefactor to second order in $\cos\phi$ around $\cos\phi=1$; here, we note that
\beq
    \left[2(1+\cos\phi)\right]^{-\ell/2} = 2^{-\ell}\left[1+\frac{\cos\phi-1}{2}\right]^{-\ell/2} =  2^{-\ell}\sum_{\beta=0}^\infty \binom{-\ell/2}{\beta}2^{-\beta}\left(\cos\phi-1\right)^\beta,
\eeq
via Newton's generalized binomial theorem around $\cos\phi=1$ (where the $2\times1$ vector is a generalized binomial coefficient with $\binom{a}{b} = (a)_b/b!$ for falling-factorial Pochhammer symbol $(a)_b$, defined in \citealt{nist_dlmf}, \S5.2). Using the standard binomial theorem, this gives
\beq
    \left[2(1+\cos\phi)\right]^{-\ell/2} = 2^{-\ell}\sum_{\beta=0}^\infty \binom{-\ell/2}{\beta}2^{-\beta}\sum_{n=0}^\beta\binom{\beta}{n}(-1)^{\beta-n}\cos^n\phi.
\eeq
Finally, we may convert $\cos^n\phi$ into a sum over Legendre polynomials via \eqref{eq: Leg-polynomial-inv-def} and thus spherical harmonics using \eqref{eq: addition-theorem}. This yields
\beq
    \left[2(1+\cos\phi)\right]^{-\ell/2} = 2^{-\ell}\sum_{\beta=0}^\infty\binom{-\ell/2}{\beta}2^{-\beta}\sum_{n=0}^\beta\binom{\beta}{n}(-1)^{\beta-n}\sum_{L=n\downarrow2}\tilde{c}^L_n\frac{4\pi}{2L+1}\sum_{M=-L}^L Y_L^M(\hr_1)Y_L^{M*}(\hr_2).
\eeq
Combining the spherical harmonics with those in \eqref{eq: Castorina-tmp} via \eqref{eq: spherical-harmonic-prod-to-sum} gives the final form for the expansion:
\beq\label{eq: Castorina-expan}
    Y_\ell^m(\hat{\vec d}) &=& 2^{-\ell}\sum_{\lambda=0}^\ell\sum_{\mu=-\lambda}^\lambda A_{\ell m}^{\lambda \mu}\sum_{\beta=0}^\infty  \binom{-\ell/2}{\beta}2^{-\beta}\sum_{n=0}^\beta\binom{\beta}{n}(-1)^{\beta-n}\sum_{L=n\downarrow2}\tilde{c}^L_n\frac{4\pi}{2L+1}\sum_{M=-L}^L\\\nonumber
    &&\,\sum_{J_1=|\lambda-L|}^{\lambda+L}(-1)^{M-m}\mathcal{G}_{\lambda L J_1}^{\mu M(-\mu-M)}\sum_{J_2=|\ell-\lambda-L|}^{J_2=\ell-\lambda+L}\mathcal{G}_{(\ell-\lambda)LJ_2}^{(m-\mu)(-M)(M+\mu-m)}Y_{J_1}^{\mu+M}(\hr_1)Y_{J_2}^{m-\mu-M}(\hr_2)
\eeq
\beq
    \Rightarrow \boxed{Y_\ell^m(\hat{\vec d}) \equiv \sum_{\beta=0}^\infty \sum_{J_1=0}^{\ell+\beta}\sum_{J_2=0}^{\ell+\beta}\sum_{M=-J_1}^{J_1}B^{\beta, \ell m}_{J_1J_2M}Y_{J_1}^M(\hr_1)Y_{J_2}^{m-M}(\hr_2)\nonumber,}
\eeq
defining the coefficients
\beq\label{eq: B-def}
    \boxed{B^{\beta, \ell m}_{J_1J_2M} = 2^{-\ell}\sum_{\lambda=0}^\ell \sum_{\mu=-\lambda}^\lambda A_{\ell m}^{\lambda \mu}\binom{-\ell/2}{\beta}2^{-\beta}\sum_{n=0}^\beta\binom{\beta}{n}(-1)^{\beta-n}\sum_{L=n\downarrow2}\tilde{c}_n^{L}\frac{4\pi}{2L+1}(-1)^{M-\mu-m}\mathcal{G}^{\mu (M-\mu)(-M)}_{\lambda L J_1}\mathcal{G}^{(m-\mu)(\mu-M)(M-m)}_{(\ell-\lambda)LJ_2}.}
\eeq
The summation limits on $J_1,J_2$ come from the consideration of the Gaunt integral triangle conditions, and the constraints $L\leq n\leq \beta$, as in Appendix \ref{appen: shift-coeff-prop}. Furthermore, the Gaunt integrals require even $\lambda+L+J_1$ and $\ell-\lambda+L+J_2$, thus $\ell+J_1+J_2$ is even, but we do not require $J_1$ and $J_2$ themselves to be even. 

For the special case of $\beta = 0$, we obtain
\beq\label{eq: B-coeff-k=0}
    B_{J_1J_2M}^{0,\ell m} &=& 2^{-\ell}\sum_{\lambda=0}^{\ell}\sum_{\mu=-\lambda}^\lambda A_{\ell m}^{\lambda\mu} (4\pi)(-1)^{M-\mu-m}\mathcal{G}^{\mu(M-\mu)(-M)}_{\lambda 0 J_1}\mathcal{G}^{(m-\mu)(\mu-M)(M-m)}_{(\ell-\lambda)0J_2}\\\nonumber
    &=& \delta^\mathrm{K}_{J_2(\ell-J_1)}\times 2^{-\ell}A_{\ell m}^{J_1M}.
\eeq
using $\binom{-\ell/2}{0} = 1$, $\tilde{c}^0_0 = 1$ and the explicit forms of the Gaunt integral, assuming $J_1,J_2\leq \ell$ and $|M|\leq J_1$. Notably, $Y_{\ell}^m(\hat{\vec d})$ still involves a summation over $J_1,J_2,M$ at zeroth-order in $\beta$. Considering $\ell=m=0$, the coefficient becomes
\beq\label{eq: B-coeff-ell=0}
    B_{J_1J_2M}^{\beta,00} &=& A_{00}^{00}2^{-\beta}\sum_{n=0}^\beta\binom{\beta}{n}(-1)^{\beta-n}\sum_{L=n\downarrow2}\tilde{c}_n^{L}\frac{4\pi}{2L+1}(-1)^{M}\mathcal{G}^{0M(-M)}_{0LJ_1}\mathcal{G}^{0(-M)M}_{0LJ_2}\\\nonumber\\\nonumber
    &=& \delta^\mathrm{K}_{J_10}\delta^\mathrm{K}_{J_20}\delta^\mathrm{K}_{M0}\times \sqrt{4\pi}
\eeq
using $A_{00}^{00} = \sqrt{4\pi}$ and $\binom{0}{\beta} = \delta^\mathrm{K}_{\beta0}$. As expected, this recovers $Y_0^0(\hat{\vec d}) = 1/\sqrt{4\pi}$.

\bibliographystyle{mnras}
\bibliography{adslib,otherlib} % if your bibtex file is called example.bib

% Don't change these lines
\bsp	% typesetting comment
\label{lastpage}

\end{document}

% End of mnras_template.tex